\documentclass[smallextended]{svjour3}     % onecolumn (second format) (Shifflett chose this one)
\begin{document}
%\large
\newcommand{\ExpandDerivations}{0}

\title{
%$\Lambda$-renormalized Einstein-Schr\"{o}dinger theory
A modification of Einstein-Schr\"{o}dinger theory that contains both general
relativity and electrodynamics
}

\titlerunning{$\Lambda$-renormalized Einstein-Schr\"{o}dinger theory}  % if too long for running head

\author{J. A. Shifflett}

\authorrunning{J. A. Shifflett} % if too long for running head

\institute{J. A. Shifflett\\
           Washington University, Department of Physics,\\
           1 Brookings Drive\\
           St.~Louis, Missouri 63130\\
\email{shifflet@hbar.wustl.edu}
}

\date{September 19, 2007} %\today or an explicit date
%\date{Received: 11 April 2006 / Accepted: 25 October 2006\\
%\date{\today}

\maketitle

\begin{abstract}
We modify the Einstein-Schr\"{o}dinger theory to
include a cosmological constant $\Lambda_z$ which multiplies the
symmetric metric, and we show how the theory can be easily coupled to additional fields.
The cosmological constant $\Lambda_z$ is assumed to be nearly cancelled by
Schr\"{o}dinger's cosmological constant $\Lambda_b$ which multiplies
the nonsymmetric fundamental tensor, such that the total
$\Lambda\!\nobreak=\nobreak\!\Lambda_z\!+\!\Lambda_b$ matches measurement.
The resulting theory becomes exactly Einstein-Maxwell theory
in the limit as $|\Lambda_z|\!\rightarrow\!\infty$.
For $|\Lambda_z|\!\sim\!1/({\rm Planck~length}\!)^2$
the field equations match the ordinary Einstein and Maxwell equations except for
extra terms which are $<\!10^{-16}$ of the usual terms for
worst-case field strengths and rates-of-change accessible to measurement.
%The theory also avoids ghosts in an unusual way.
Additional fields can be included in the Lagrangian,
and these fields may couple to the symmetric
metric and the electromagnetic vector potential, just as in Einstein-Maxwell theory.
The ordinary Lorentz force equation is obtained by taking the divergence
of the Einstein equations when sources are included.
The Einstein-Infeld-Hoffmann (EIH)
equations of motion match the equations of motion for
Einstein-Maxwell theory to Newtonian/Coulombian order,
which proves the existence of a Lorentz force without requiring sources.
This fixes a problem of the original Einstein-Schr\"{o}dinger theory,
which failed to predict a Lorentz force.
An exact charged solution matches the Reissner-Nordstr\"{o}m solution except for
additional terms which are $\sim\!10^{-66}$ of the usual terms for
worst-case radii accessible to measurement.
An exact electromagnetic plane-wave
solution is identical to its counterpart in Einstein-Maxwell theory.
%Using the Lorentz force equation we find the equations of motion
%for charged or neutral particles around the charged solution.
%Periastron advance, deflection of light
%and time delay of light have a fractional difference of $<\!10^{-56}$
%compared to Einstein-Maxwell theory for worst-case parameters.
%When a spin-1/2 field is included, the theory gives
%the ordinary Dirac equation, and the charged solution
%results in fractional shifts of $<\!10^{-50}$ in Hydrogen atom energy levels.
%Lastly, we discuss the merits of our Lagrangian density
%compared to the Einstein-Maxwell Lagrangian density.
% Einstein Maxwell spacetimes (04.40.Nr)
% cosmological constant (98.80.Es)
% dark matter (95.35.+d)
% exact solutions of general relativity (04.20.Jb)
% unified field theories and models (12.10.-g)
% relativity and gravitation (95.30.Sf) - probably covered by (04.40.Nr)
% alternative theories of gravity (04.50.+h)
% No more than 4 of these are supposed to be used
%\pacs{04.40.Nr,98.80.Es,12.10.-g,04.50.+h}% PACS codes
%\PACS
%{04.40.Nr\and 98.80.Es\and 12.10.-g\and 04.50.+h} % PACS codes
%{PACS 04.40.Nr\and 98.80.Es\and 12.10.-g\and 04.50.+h}
%Use showkeys class option
%\keywords{Foliations \and Stability \and Motion law}
\end{abstract}

\newcommand{\rmt}{\sqrt{2}\,i}
\newcommand{\ca}{c_1}
\newcommand{\cb}{c_2}
\newcommand{\cc}{c_3}
\newcommand{\sR}{{{^*}R}}
\newcommand{\Nbar}{W}
\newcommand{\hR}{\mathcal R}
\newcommand{\tR}{\tilde\hR}
\newcommand{\hG}{\mathcal G}
\newcommand{\tG}{\tilde G}
\newcommand{\tB}{R}
\newcommand{\tT}{\tilde T}
\newcommand{\tS}{\tilde S}
\newcommand{\cUps}{\check\Upsilon}
\newcommand{\bUps}{\bar\Upsilon}
\newcommand{\tGam}{\tilde\Gamma}
\newcommand{\nGam}{\widehat\Gamma}
\newcommand{\sGam}{{^*\Gamma}}
\newcommand{\ftilde}{\tilde f}
\newcommand{\dual}{\vartheta}
\newcommand{\dualtilde}{\tilde\vartheta}
\newcommand{\Fdash}{\raise1pt\hbox{\rlap\textendash} F}
\newcommand{\uacute}{\acute{u}}
\newcommand{\hf}{\hat f}
\newcommand{\hj}{\hat j}
\newcommand{\hbj}{\hat\mathbf{j}}
\newcommand{\rmg}{\sqrt{-g}}
\newcommand{\rmN}{\sqrt{\!-N}}
\newcommand{\ehat}{Q}
\newcommand{\ff}{\,\ell}
\newcommand{\Aphi}{A}
\newcommand{\Gk}{G}
\newcommand{\sinht}{\check{s}}
\newcommand{\cosht}{\check{c}}
\newcommand{\extra}{V}
\newcommand{\Fdashoverf}{\raise1pt\hbox{\rlap\textendash} I}
\newcommand{\mum}{\mu}
\newcommand{\Ddag}{\overleftarrow D}
\newcommand{\ord}{{\mathcal O}}
\newcommand{\HR}{\Re}

\def\Stacksymbols #1#2#3#4{\def\theguybelow{#2}
   \def\verticalposition{\lower#3pt}
   \def\spacingwithinsymbol{\baselineskip0pt\lineskip#4pt}
   \mathrel{\mathpalette\intermediary#1}}
\def\intermediary#1#2{\verticalposition\vbox{\spacingwithinsymbol
     \everycr={}\tabskip0pt
     \halign{$\mathsurround0pt#1\hfil##\hfil$\crcr#2\crcr
            \theguybelow\crcr}}}

\section{\label{Introduction}Introduction}
The Einstein-Schr\"{o}dinger theory is a generalization of
vacuum general relativity which allows non-symmetric fields.
The theory without a cosmological constant
was first proposed by Einstein and
Straus\cite{EinsteinStraus,Einstein3,EinsteinBianchi,EinsteinKaufman,EinsteinMOR}.
Schr\"{o}dinger later showed that it could be derived from a very simple
Lagrangian density if a cosmological constant was included\cite{SchrodingerI,SchrodingerIII,SchrodingerSTS}.
Einstein and Schr\"{o}dinger suspected that the theory might include electrodynamics,
%One problem is that
but no Lorentz force was found\cite{Callaway,Infeld}
when using the Einstein-Infeld-Hoffmann (EIH) method\cite{EinsteinInfeld,Wallace}.
Here we show that a simple modification of the Einstein-Schr\"{o}dinger theory
closely approximates Einstein-Maxwell theory,
and the Lorentz force does result from the EIH method,
and in fact the ordinary Lorentz force equation results when sources
are included.
The modification is the addition of a second cosmological term $\Lambda_z g_{\mu\nu}$,
where $g_{\mu\nu}$ is the symmetric metric.
We assume this term nearly cancels
Schr\"{o}dinger's ``bare'' cosmological term $\Lambda_b N_{\mu\nu}$, where $N_{\mu\nu}$ is the
nonsymmetric fundamental tensor. The total cosmological constant
$\Lambda =\Lambda_b+\Lambda_z$ can then match
cosmological measurements of the accelerating universe.
Our theory is related to one in \cite{Kursunoglu}, but it is roughly the electromagnetic dual of that theory,
and it allows coupling to additional fields (sources), and it allows $\Lambda\ne0$.

%At the present development of the theory,
The origin of our $\Lambda_z$ is unknown.
One possibility is that
$\Lambda_z$ could arise from vacuum fluctuations,
an idea that has been discussed by many
authors\cite{Zeldovich,Sahni,Peskin,Carroll}. Zero-point
fluctuations are essential to both QED and the Standard-Model, and
are the cause of the Casimir force\cite{Sahni} and other effects.
Another possibility is that $\Lambda_z$ arises dynamically,
related to the minimum of a potential of some additional field in
the theory.  Speculation about the origin of this second
cosmological constant is beyond the scope of this paper. Our main goal
here is to demonstrate that the theory
closely approximates Einstein-Maxwell theory.

Like Einstein-Maxwell theory, our theory can be coupled to additional fields using
a symmetric metric $g_{\mu\nu}$ and vector potential $A_\mu$,
and it is invariant under a $U\!(1)$ gauge transformation.
The theory does not enlarge the invariance group.
When coupled to the Standard Model, the combined Lagrangian is invariant under
the usual $U\!(1)\otimes SU\!(2)\otimes SU\!(3)$ gauge group.
The usual $U\!(1)$ gauge term $F^{\mu\nu}\!F_{\mu\nu}$ is incorporated together with
the geometry, and is not explicitly in the Lagrangian.
Whether this is a step backwards from Einstein-Maxwell theory coupled to the Standard Model,
or whether the $SU\!(2)$ and $SU\!(3)$ gauge terms could also be incorporated
using non-Abelian fields as in \cite{Borchsenius76,Borchsenius78},
or by using higher space-time dimensions,
is speculation beyond the scope of this paper.

This paper is organized as follows. In \S\ref{LagrangianDensity} we discuss the Lagrangian density.
In \S\ref{EinsteinEquations}-\S\ref{Connections} we derive the field equations
and quantify how closely they approximate the field equations of Einstein-Maxwell theory.
In \S\ref{LorentzForce} we derive the ordinary Lorentz force equation by
taking the divergence of the Einstein equations when sources are included.
In \S\ref{EIHEquationsOfMotion} we derive the Lorentz force
using the EIH method, which requires no sources in the Lagrangian.
In \S\ref{Monopole} we give an exact charged solution
and show that it closely approximates the Reissner-Nordstr\"{o}m solution.
In \S\ref{ppwaves} we give an exact electromagnetic plane-wave solution which
is identical to its counterpart in Einstein-Maxwell theory.

\section{\label{LagrangianDensity}The Lagrangian density}
Einstein-Maxwell theory can be derived from a Palatini Lagrangian density,
\begin{eqnarray}
\label{GR}
{\mathcal L}(\Gamma^{\lambda}_{\rho\tau},g_{\rho\tau},A_\nu)
&=&-\frac{\lower2pt\hbox{$1$}}{16\pi}\rmg\left[\,g^{\mu\nu}
R_{\nu\mu}({\Gamma})+(n\!-\!2)\Lambda_b\,\right]\nonumber\\
&&+\frac{\lower2pt\hbox{$1$}}{4\pi}\rmg A_{[\alpha,\rho]}g^{\alpha\mu}g^{\rho\nu}\!A_{[\mu,\nu]}
+{\mathcal L}_m(u^\nu,\psi,g_{\mu\nu},A_\nu\cdots).~~~
\end{eqnarray}
Here $\Lambda_b$ is a bare cosmological constant.
The ${\mathcal L}_m$ term couples the
metric $g_{\mu\nu}$ and electromagnetic potential
$A_\mu$ to additional fields, such as a hydrodynamic velocity vector $u^\nu$,
spin-1/2 wavefunction $\psi$, or perhaps
the other fields of the Standard Model.
The original Einstein-Schr\"{o}dinger theory
allows a nonsymmetric $N_{\mu\nu}$ and $\nGam^{\lambda}_{\!\rho\tau}$
in place of the symmetric $g_{\mu\nu}$ and $\Gamma^{\lambda}_{\rho\tau}$, and
excludes the $\rmg A_{[\alpha,\rho]}g^{\alpha\mu}\!g^{\rho\nu}\!A_{[\mu,\nu]}$ term.
Our ``$\Lambda$-renormalized'' Einstein-Schr\"{o}dinger theory
introduces an additional cosmological term $\rmg\Lambda_z$,
\begin{eqnarray}
\label{JSlag1}
{\mathcal L}(\nGam^{\lambda}_{\!\rho\tau},N_{\rho\tau})
&=&-\frac{\lower1pt\hbox{$1$}}{16\pi}\rmN\left[N^{\dashv\mu\nu}
\hR_{\nu\mu}({\nGam})+(n\!-\!2)\Lambda_b\,\right]\nonumber\\
&&\!-\frac{\lower1pt\hbox{$1$}}{16\pi}\rmg\,(n\!-\!2)\Lambda_z
+{\mathcal L}_m(u^\nu,\psi,g_{\mu\nu},A_\nu\dots),
\end{eqnarray}
where $\Lambda_b\!\approx\!-\Lambda_z$ so that the
total $\Lambda$ matches astronomical measurements\cite{Astier},
\begin{eqnarray}
\label{Lambdadef}
\Lambda=\Lambda_b+\Lambda_z\approx&10^{-56}{\rm cm}^{-2},~~~~
\end{eqnarray}
and the physical metric and electromagnetic potential are defined to be
\begin{eqnarray}
\label{gdef}
\rmg\,g^{\mu\nu}=\rmN N^{\dashv(\mu\nu)},~~~~~~~~
\label{A}
%\Aphi_\nu=\frac{\lower1pt\hbox{$1$}}{(n\!-\!1)\sqrt{\!-2\Lambda_b}}\,{\nGam}{^\sigma_{\![\nu\sigma]}}.
\Aphi_\nu={\nGam}{^\sigma_{\![\nu\sigma]}}/[(n\!-\!1)\sqrt{\!-2\Lambda_b}\,].
\end{eqnarray}
Eq. (\ref{gdef}) defines $g^{\mu\nu}$ unambiguously because $\rmg\!=\![-det(\rmg\,g^{\mu\nu})]^{1/(n-2)}$.
Here and throughout this paper we use geometrized units with $c\!=\!G\!=\nobreak\!1$,
the symbols $\raise2pt\hbox{$_{(~)}$}$ and $\raise2pt\hbox{$_{[~]}$}$ around indices
indicate symmetrization and antisymmetrization, $g\!=\!det(g_{\mu\nu})$,
$N\!=\!det(N_{\mu\nu})$, and $N^{\dashv\sigma\nu}$ is the inverse of $N_{\nu\mu}$ such that
$N^{\dashv\sigma\nu}\!N_{\nu\mu}\!=\nobreak\!\delta^\sigma_\mu$.
The dimension is assumed to be n=4, but ``n'' is retained in the equations to show how easily the
theory can be generalized.
The ${\mathcal L}_m$ term is not to include a
$\rmg A_{[\alpha,\beta]}g^{\alpha\mu}g^{\beta\nu}\!A_{[\mu,\nu]}$
% $\rmg\,F^{\mu\nu}\!F_{\mu\nu}$
part but may contain the rest of the Standard Model.
In (\ref{JSlag1}), $\hR_{\nu\mu}(\nGam)$
is a form of Hermitianized Ricci tensor\cite{EinsteinStraus},
\begin{eqnarray}
\label{HermitianizedRicci0}
\!\!\!\!\hR_{\nu\mu}(\nGam)
\label{HermitianizedRicci}
=\nGam^\alpha_{\nu\mu,\alpha}
-\nGam^\alpha_{\!(\alpha(\nu),\mu)}
+\nGam^\sigma_{\nu\mu}\nGam^\alpha_{\!(\alpha\sigma)}
-\nGam^\sigma_{\nu\alpha}\nGam^\alpha_{\sigma\mu}
\!-\nGam^\tau_{\![\tau\nu]}\nGam^\alpha_{\![\alpha\mu]}/(n\!-\!1).
\end{eqnarray}
This tensor reduces to the ordinary Ricci tensor when
$\Gamma^\alpha_{\![\nu\mu]}\!=\!0$ and
%$\Gamma^\alpha_{\alpha[\nu,\mu]}
%\!=\!R^\alpha_{~\alpha\mu\nu}/2\!=\!0$.
%\!=\!(ln\rmg\,)_{,[\nu,\mu]}\!=\!0$.
%$\Gamma^\alpha_{\alpha\nu}\!=\!(ln\rmg\,)_{,\nu}\Rightarrow
$\Gamma^\alpha_{\alpha[\nu,\mu]}\!=\!0$, as occurs in ordinary general relativity.

It is helpful to decompose ${\nGam}{^\alpha_{\nu\mu}}$
into a new connection ${\tGam}{^\alpha_{\nu\mu}}$, and $A_\sigma$ from~(\ref{A}),
\begin{eqnarray}
\label{gamma_natural}
{\nGam}{^\alpha_{\nu\mu}}&=&\tGam{^\alpha_{\nu\mu}}
+(\delta^\alpha_\mu\Aphi_\nu
\!-\delta^\alpha_\nu \Aphi_\mu)\sqrt{\!-2\Lambda_b},\\
\label{gamma_tilde}
{\rm where}~~~\tGam{^\alpha_{\nu\mu}}
&=&{\nGam}{^\alpha_{\nu\mu}}\!+
%\frac{\lower1pt\hbox{$1$}}{(n\!-\!1)}
(\delta^\alpha_\mu{\nGam}{^\sigma_{\![\sigma\nu]}}
-\delta^\alpha_\nu{\nGam}{^\sigma_{\![\sigma\mu]}})/(n\!-\!1).
\end{eqnarray}
By contracting (\ref{gamma_tilde}) on the right and left we see that
$\tGam{^\alpha_{\nu\mu}}$ has the symmetry
\begin{eqnarray}
\label{JScontractionsymmetric}
\tGam^\alpha_{\nu\alpha}
\!=\!\nGam^\alpha_{\!(\nu\alpha)}
\!=\!\tGam^\alpha_{\alpha\nu},
\end{eqnarray}
so it has only $n^3\!-\!n$ independent components.
Substituting (\ref{gamma_natural}) into (\ref{HermitianizedRicci}) gives
\begin{eqnarray}
\label{RnGam0}
{\mathcal R}_{\nu\mu}(\nGam)={\mathcal R}_{\nu\mu}(\tGam)+2A_{[\nu,\mu]}\sqrt{\!-2\Lambda_b}.
\end{eqnarray}
Using (\ref{RnGam0}), the Lagrangian density (\ref{JSlag1}) can be written in terms of
${\tGam}{^\alpha_{\nu\mu}}$ and $A_\sigma$,
\begin{eqnarray}
\label{JSlag3}
{\mathcal L}(\nGam^{\lambda}_{\!\rho\tau},N_{\rho\tau})
&=&\!-\frac{\lower2pt\hbox{$1$}}{16\pi}\rmN\left[N^{\dashv\mu\nu}({\tR}_{\nu\mu}
%+\ca\tGam^\alpha_{\alpha[\nu,\mu]}
+2\Aphi_{[\nu,\mu]}\sqrt{\!-2\Lambda_b}\,)+(n\!-\!2)\Lambda_b\,\right]\nonumber\\
&&\!-\frac{\lower2pt\hbox{$1$}}{16\pi}\rmg\,(n\!-\!2)\Lambda_z+{\mathcal L}_m(u^\nu,\psi,g_{\mu\nu},A_\sigma\dots).
\end{eqnarray}
Here $\tR_{\nu\mu}\!=\!\hR_{\nu\mu}(\tGam)$,
and from (\ref{JScontractionsymmetric},\ref{HermitianizedRicci}) we have
\begin{eqnarray}
\label{HermitianizedRiccit}
\tR_{\nu\mu}
&=&\tGam^\alpha_{\nu\mu,\alpha}
-\tGam^\alpha_{\alpha(\nu,\mu)}
+\tGam^\sigma_{\nu\mu}\tGam^\alpha_{\sigma\alpha}
-\tGam^\sigma_{\nu\alpha}\tGam^\alpha_{\sigma\mu}.
\end{eqnarray}
From (\ref{gamma_natural},\ref{JScontractionsymmetric}), $\tGam^\alpha_{\nu\mu}$ and $A_\nu$ fully
parameterize $\nGam^\alpha_{\nu\mu}$ and can be treated as independent variables.
%So when we set $\delta{\mathcal L}/\delta\tGam^\alpha_{\nu\mu}\!=0$
%and $\delta{\mathcal L}/\delta A_\nu\!=0$,
%the same field equations must result as with
%$\delta{\mathcal L}/\delta\nGam^\alpha_{\nu\mu}\!=0$.
%It is simpler to calculate the field equations using
%$\tGam^\alpha_{\nu\mu}$ and $A_\nu$ instead of
%$\nGam^\alpha_{\nu\mu}$, so we will follow this method.
It is simpler to calculate the field equations by setting
$\delta{\mathcal L}/\delta\tGam^\alpha_{\nu\mu}\!=0$ and $\delta{\mathcal L}/\delta A_\nu\!=0$
%instead of using $\nGam^\alpha_{\nu\mu}$,
instead of setting $\delta{\mathcal L}/\delta\nGam^\alpha_{\nu\mu}\!=0$,
so we will follow this method.

To do quantitative comparisons of this theory to Einstein-Maxwell
theory we will need to use some value for $\Lambda_z$. One possibility
is that $\Lambda_z$ results from zero-point fluctuations\cite{Zeldovich,Sahni,Peskin,Carroll},
in which case using (\ref{Lambdadef}) we get
%\begin{eqnarray}
%\label{cutoff}
%\end{eqnarray}
%where
%\begin{eqnarray}
%\label{Planck}
%$l_P\!=\!({\rm Planck~length})\!=\!\sqrt{\hbar G/c^3}\!=\!1.6\times 10^{-33}cm$.
%\end{eqnarray}
%Then from (\ref{Lambdadef},\ref{cutoff}) and assuming all of the known
%fundamental particles we have\cite{Sahni},
\begin{eqnarray}
\label{Lambdab}
\Lambda_b&\approx&-\Lambda_z\sim C_z \omega_c^4 l_P^2
\sim\!10^{66}{\rm cm}^{-2},\\
\label{cutoff}
%\omega_c&=&\left({{\lower2pt\hbox{cutoff}}\atop{\raise2pt\hbox{frequency}}}\right)\!\sim\!1/l_P,~~~~
\omega_c&=&({\rm cutoff~frequency})\!\sim\!1/l_P,~~~~\\
\label{Cz}
C_z&=&\frac{\lower2pt\hbox{1}}{2\pi}\!\left({{\lower2pt\hbox{fermion}}\atop{\raise2pt\hbox{spin~states}}}
\!-\!{{\lower2pt\hbox{boson}}\atop{\raise2pt\hbox{spin~states}}}\right)
\sim \frac{\lower2pt\hbox{60}}{2\pi}~~~~~
%~~~~ C_z={\rm(fermions\!-\!bosons)}/2\pi\sim 60/2\pi
\end{eqnarray}
where $l_P\!=\!({\rm Planck~length})\!=\!1.6\times 10^{-33}cm$.
%and from astronomical measurements\cite{Astier}
%\begin{eqnarray}
%\label{Lambda}
%\Lambda &\approx&1.4\times 10^{-56}{\rm cm}^{-2},~~~~
%\label{LambdaoverLambdab}
%\Lambda/\Lambda_b\sim 10^{-122}.
%\end{eqnarray}
We will also consider the limit $\omega_c\!\rightarrow\!\infty$,
$|\Lambda_z|\!\rightarrow\!\infty$, $\Lambda_b\!\rightarrow\!\infty$
as in QED, and we will prove that
\begin{eqnarray}
\label{LRESlimit}
{{\lower2pt\hbox{lim}}\atop{\raise2pt\hbox{$|\Lambda_z|\!\rightarrow\!\infty$}}}
\left(\lower2pt\hbox{$\Lambda$-renormalized}\atop{\raise2pt\hbox{Einstein-Schr\"{o}dinger theory}}\right)
\!=\!\left(\lower2pt\hbox{Einstein-Maxwell}\atop{\raise2pt\hbox{theory}}\right).
\end{eqnarray}

The Hermitianized Ricci tensor (\ref{HermitianizedRicci}) has the following invariance properties
\begin{eqnarray}
\label{transpositionsymmetric}
&&\hR_{\nu\mu}(\nGam^T)=\hR_{\mu\nu}(\nGam)~~~~~~~~~~~~~~~~~({\rm T=transpose}),\\
\label{gaugesymmetric}
&&\hR_{\nu\mu}(\nGam^\alpha_{\rho\tau}\!+\delta^\alpha_{[\rho}\varphi_{,\tau]})=
\hR_{\nu\mu}(\nGam^\alpha_{\rho\tau})~~~{\rm for~an~arbitrary}~\varphi(x^\sigma).
\end{eqnarray}
From (\ref{transpositionsymmetric},\ref{gaugesymmetric}),
the Lagrangians (\ref{JSlag1},\ref{JSlag3}) are invariant under charge conjugation,
\begin{eqnarray}
\label{transposition}
Q\!\rightarrow\!-Q,~
A_\sigma\!\rightarrow\!-A_\sigma,~
\tGam^\alpha_{\nu\mu}\!\rightarrow\!\tGam^\alpha_{\mu\nu},~
\nGam^\alpha_{\nu\mu}\!\rightarrow\!\nGam^\alpha_{\mu\nu},~
N_{\nu\mu}\!\!\rightarrow\!N_{\mu\nu},~
N^{\dashv\nu\mu}\!\rightarrow\!N^{\dashv\mu\nu}\!,~~~
\end{eqnarray}
and also under an electromagnetic gauge transformation
\begin{eqnarray}
\label{gauge}
\!\!\!\!\psi\!\rightarrow\!\psi e^{i\phi},~~
A_\alpha\!\rightarrow\! A_\alpha\!-\!\frac{\lower1pt\hbox{$\hbar$}}{Q}\,\phi_{,\alpha},~~
\tGam^\alpha_{\rho\tau}\!\rightarrow\!\tGam^\alpha_{\rho\tau},~~
\nGam^\alpha_{\rho\tau}\!\rightarrow\!\nGam^\alpha_{\rho\tau}\!+\!\frac{\lower1pt\hbox{$2\hbar$}}{Q}\,\delta^\alpha_{[\rho}\phi_{,\tau]}\sqrt{\!-2\Lambda_b},~~
\end{eqnarray}
assuming that ${\mathcal L}_m$ is invariant.
With $\Lambda_b\!>0,~\Lambda_z\!<0$ as in (\ref{Lambdab}) then
$\tGam^\alpha_{\nu\mu}$, $\nGam^\alpha_{\nu\mu}$, $N_{\nu\mu}$ and $N^{\dashv\nu\mu}$ are all Hermitian,
$\tR_{\nu\mu}$ and $\hR_{\nu\mu}(\nGam)$ are Hermitian from (\ref{transpositionsymmetric}),
and $g_{\nu\mu}$, $A_\sigma$ and ${\mathcal L}$ are real from (\ref{gdef},\ref{JSlag1},\ref{JSlag3}).

In this theory the metric (\ref{gdef}) is used for measuring space-time intervals, for calculating geodesics,
and for raising and lowering of indices. The covariant derivative ``;'' is always
done using the Christoffel connection formed from $g_{\mu\nu}$,
\begin{eqnarray}
\label{Christoffel}
\Gamma^\alpha_{\nu\mu}&=&\frac{\lower2pt\hbox{$1$}}{2}\,g^{\alpha\sigma}(g_{\mu\sigma,\nu}
+g_{\sigma\nu,\mu}-g_{\nu\mu,\sigma}).
\end{eqnarray}
%If ${\mathcal L}_m\!=\!0$, the metric (\ref{gdef}) is the unique choice for which the divergence
%of the Einstein equations vanishes when using (\ref{Christoffel}) for the covariant derivative.
%With the metric (\ref{gdef}),
We will see that taking the divergence of the Einstein equations using (\ref{Christoffel},\ref{gdef})
% vanishes if ${\mathcal L}_m\!=\nobreak\!0$, and it
gives the ordinary Lorentz force equation.
% if ${\mathcal L}_m\!\ne\!0$.
%where (\ref{Christoffel}) is used for the covariant derivative.
%And when $N_{\mu\nu}$ and $\nGam^\alpha_{\mu\nu}$ are symmetric,
%the definition (\ref{gdef}) requires $g_{\mu\nu}\!=\!N_{\mu\nu}$,
%the definition (\ref{A}) requires $A_\sigma\!=\!0$,
%and the theory reduces to ordinary general relativity without electromagnetism.
The electromagnetic field is defined in terms of the potential (\ref{A})
\begin{eqnarray}
\label{Fdef}
F_{\mu\nu}=A_{\nu,\mu}-A_{\mu,\nu}.
\end{eqnarray}
However, we will also define another field $f^{\mu\nu}$
\begin{eqnarray}
\label{fdef}
\rmg\,f^{\mu\nu}=\rmN N^{\dashv[\nu\mu]}\Lambda_b^{\!1/2}/\rmt.
\end{eqnarray}
Then from (\ref{gdef}), $g^{\mu\nu}$ and $f^{\mu\nu}\rmt\Lambda_b^{\!-1/2}$ are
%the symmetric and antisymmetric
parts of a total field,
\begin{eqnarray}
\label{Wdef}
(\rmN/\rmg\,)N^{\dashv\nu\mu}=g^{\mu\nu}\!+\!f^{\mu\nu}\rmt\Lambda_b^{\!-1/2}.
\end{eqnarray}
We will see that the field equations require $f_{\mu\nu}\!\approx\!F_{\mu\nu}$
to a very high precision.
%so it is mainly just a matter of terminology which one
%is called the electromagnetic field.
The definitions (\ref{gdef}) of $g_{\mu\nu}$ and $A_\nu$ in terms
of the ``fundamental" fields $N_{\rho\tau},\nGam^{\lambda}_{\!\rho\tau}$
%in (\ref{JSlag1}) are only connected via (\ref{gdef},\ref{A}) with
may seem unnatural from an empirical viewpoint.
On the other hand, our Lagrangian density (\ref{JSlag1}) seems simpler than (\ref{GR})
of Einstein-Maxwell theory, it contains fewer fields, and these fields have no symmetry restrictions.
However, these are all very subjective considerations.
It is much more important that our theory closely matches Einstein-Maxwell theory, and hence measurement.

Note that there are many nonsymmetric generalizations
of the Ricci tensor besides
the Hermitianized Ricci tensor $\hR_{\nu\mu}(\nGam)$
from (\ref{HermitianizedRicci})
and the ordinary Ricci tensor $R_{\nu\mu}(\nGam)$.
For example, we could form any weighted average of
$R_{\nu\mu}(\nGam)$,
$R_{\mu\nu}(\nGam)$,
$R_{\nu\mu}(\nGam^T)$
and $R_{\mu\nu}(\nGam^T)$,
and then add any linear combination of the tensors
%$R^\alpha{_{\alpha\nu\mu}}(\nGam)$,
%$R^\alpha{_{\alpha\nu\mu}}(\nGam^T)$,
$\nGam^\alpha_{\alpha[\nu,\mu]}$,
$\nGam^\alpha_{\![\nu|\alpha,|\mu]}$,
%$\nGam^\alpha_{\![\alpha\nu],\mu}\!\!+\nGam^\alpha_{\mu\sigma}\nGam^\sigma_{\![\alpha\nu]}$,
$\nGam^\alpha_{\![\nu\mu]}\nGam^\sigma_{\![\sigma\alpha]}$,
$\nGam^\alpha_{\![\nu\sigma]}\nGam^\sigma_{\![\mu\alpha]}$,
and $\nGam^\alpha_{\![\alpha\nu]}\nGam^\sigma_{\![\sigma\mu]}$.
All of these generalized Ricci tensors would be linear in $\nGam^\alpha_{\nu\mu,\sigma}$, quadratic in $\nGam^\alpha_{\nu\mu}$,
and would reduce to the ordinary Ricci tensor when
$\Gamma^\alpha_{\![\nu\mu]}\!=\!0$ and $\Gamma^\alpha_{\alpha[\nu,\mu]}\!=\!0$
as occurs in ordinary general relativity.
Even if we limit the tensor to only four terms, there are still eight possibilities.
We assert that invariance properties like (\ref{transpositionsymmetric},\ref{gaugesymmetric})
are the most sensible way to choose among the different alternatives,
not criteria such as the number of terms in the expression.

Finally, let us discuss some notation issues.
We use the symbol $\Gamma^\alpha_{\!\nu\mu}$ for
the Christoffel connection (\ref{Christoffel}) whereas
Einstein and Schr\"{o}dinger used it for our $\tGam^\alpha_{\nu\mu}$
and $\nGam^\alpha_{\!\nu\mu}$ respectively.
We use the symbol $g_{\mu\nu}$ for the symmetric metric (\ref{gdef})
whereas Einstein and Schr\"{o}dinger used it
for our $N_{\mu\nu}$, the nonsymmetric fundamental tensor.
Also, to represent the inverse of $N_{\alpha\mu}$ we use $N^{\dashv\sigma\alpha}$
instead of the more conventional $N^{\alpha\sigma}$,
because this latter notation would be ambiguous when using
$g^{\mu\nu}$ to raise indices.
While our notation differs from previous literature on the
Einstein-Schr\"{o}dinger theory,
this change is required by our explicit metric definition, and
it is necessary to be consistent with the much
larger body of literature on Einstein-Maxwell theory.

\section{\label{EinsteinEquations}The Einstein equations}
To set $\delta{\mathcal L}/\delta(\rmN N^{\dashv\mu\nu})\!=0$
we need some initial results.
%From (\ref{gdef}) we get,
%\begin{eqnarray}
%\label{gcontravariantderiv}
%\!\!\!\frac{\partial\left(\rmg g^{\rho\tau}\right)}{\partial(\rmN N^{\dashv\mu\nu})}
%=\frac{\partial\left(\rmg g^{\rho\tau}\right)}{\partial(\rmg g^{\mu\nu})}
%&=&\delta^{(\rho}_\mu\delta^{\tau)}_\nu.
%\label{gcovariantderiv}
%\frac{\partial(g_{\tau\sigma}/\rmg)}{\partial(\rmN N^{\dashv\mu\nu})}
%&=&\frac{-g_{\tau(\nu}g_{\mu)\sigma}}{\rmg\rmg}
%~~\left({\rm because}~
%\frac{\partial\left(\rmg g^{\rho\tau}g_{\tau\sigma}/\rmg\right)}
%{\partial(\rmN N^{\dashv\mu\nu})}=0\right)\!.~~~~
%\end{eqnarray}
Using (\ref{gdef})
and the identities $det(sM^{})\!=s^n det(M^{})$,~$det(M^{-1}_{})\!=1/det(M^{})$ gives
\begin{eqnarray}
%\rmg\,g^{\rho\tau}&=&\rmN N^{\dashv(\rho\tau)},\\
%\rmg\,f^{\rho\tau}&=&-\rmN N^{\dashv[\rho\tau]},\\
\label{rmN2}
\rmN&=&(-det(\rmN N^{\dashv\mu\nu}))^{1/(n-2)},\\
\label{rmg2}
\rmg&=&(-det(\rmg\,g^{\mu\nu}))^{1/(n-2)}
=(-det(\rmN N^{\dashv(\mu\nu)}))^{1/(n-2)}.
\end{eqnarray}
Using (\ref{rmN2},\ref{rmg2},\ref{gdef}) and the identity
$\partial(det(M^{\cdot\cdot}))/\partial M^{\mu\nu}\!=M^{-1}_{\nu\mu}det(M^{\cdot\cdot})$ gives
\begin{eqnarray}
\label{rmNderiv}
\!\!\!\!\!\!\frac{\partial\rmN}{\partial(\!\rmN N^{\dashv\mu\nu})}
%\!&=&\!\frac{(-det(\rmN\,N^{\dashv..}))^{1/(n-2)-1+1}}{(n\!-\!2)}\frac{N_{\nu\mu}}{\rmN}
\!&=&\!\frac{N_{\nu\mu}}{(n\!-\!2)},~~~~~~~~~~
\label{rmgderiv}
\frac{\partial\rmg}{\partial(\!\rmN N^{\dashv\mu\nu})}
%\!&=&\!\frac{(-det(\rmg\,g^{..}))^{1/(n-2)-1+1}}{(n\!-\!2)}\frac{g_{\nu\mu}}{\rmg}
%=\frac{\partial\rmg}{\partial(\!\rmg g^{\mu\nu})}
=\frac{g_{\nu\mu}}{(n\!-\!2)}.
\end{eqnarray}
Setting $\delta{\mathcal L}/\delta(\rmN N^{\dashv\mu\nu})\!=0$ using (\ref{JSlag3},\ref{rmNderiv})
gives the field equations,
\begin{eqnarray}
\label{para}
\tR_{\nu\mu}
\!+2\Aphi_{[\nu,\mu]}\sqrt{\!-2\Lambda_b}
\!+\Lambda_b N_{\nu\mu}
\!+\Lambda_z g_{\nu\mu}
\!=8\pi S_{\nu\mu},
\end{eqnarray}
where $S_{\nu\mu}$ and the energy-momentum tensor $T_{\nu\mu}$ are defined by
\begin{eqnarray}
\label{Sdef}
S_{\nu\mu}\!&\equiv&2\frac{\delta{\mathcal L}_m}{\delta(\rmN N^{\dashv\mu\nu})}
=2\frac{\delta{\mathcal L}_m}{\delta(\rmg g^{\mu\nu})},\\
\label{Tdef}
T_{\nu\mu}&\equiv&S_{\nu\mu}\!-\frac{\lower1pt\hbox{$1$}}{2}\,g_{\nu\mu}S^\alpha_\alpha,~~~~~
S_{\nu\mu}=T_{\nu\mu}\!-\frac{1}{(n-2)}g_{\nu\mu}T^\alpha_\alpha.
\end{eqnarray}
The second equality in (\ref{Sdef}) results because ${\mathcal L}_m$ in (\ref{JSlag1})
contains only the metric $\rmg\,g^{\mu\nu}\!=\!\rmN N^{\dashv(\mu\nu)}$ from (\ref{gdef}),
and not $\rmN N^{\dashv[\mu\nu]}$.
Taking the symmetric and antisymmetric parts of (\ref{para}) and using (\ref{Fdef}) gives
\begin{eqnarray}
\label{JSsymmetric}
&&~~~~~ \tR_{(\nu\mu)}
+\Lambda_b N_{(\nu\mu)}+\Lambda_z g_{\nu\mu}
=8\pi\!\left(T_{\nu\mu}
-\frac{1}{(n-2)}g_{\nu\mu}T^\alpha_\alpha\right),\\
\label{JSantisymmetric}
&&~~~~~~N_{[\nu\mu]}=
F_{\nu\mu}\rmt\Lambda_b^{\!-1/2}
\!-\tR_{[\nu\mu]}\Lambda_b^{\!-1}.
\end{eqnarray}
Also from the curl of (\ref{JSantisymmetric}) we get
\begin{eqnarray}
\label{JScurl}
\tR_{[\nu\mu,\sigma]}+\Lambda_b N_{[\nu\mu,\sigma]}=0.
\end{eqnarray}

To put (\ref{JSsymmetric}) into a form which looks more like the ordinary Einstein equations,
we need some preliminary results.
The definitions (\ref{gdef},\ref{fdef}) of $g_{\nu\mu}$ and $f_{\nu\mu}$
can be inverted exactly to give $N_{\nu\mu}$ in terms of $g_{\nu\mu}$ and $f_{\nu\mu}$.
An expansion in powers of $\Lambda_b^{\!-1}$ will better serve our purposes,
and is derived in Appendix \ref{ApproximateFandg},
\begin{eqnarray}
\label{approximateNbar}
N_{(\nu\mu)}&\!\!=\!& g_{\nu\mu}-2\!\left({f_\nu}^\sigma f_{\sigma\mu}
-\frac{1}{2(n\!-\!2)}g_{\nu\mu}f^{\rho\sigma}\!f_{\sigma\rho}\right)\!\Lambda_b^{\!-1}
+(f^4)\Lambda_b^{\!-2}\dots\\
\label{approximateNhat}
N_{[\nu\mu]}&\!\!=\!& f_{\nu\mu}\rmt\Lambda_b^{\!-1/2}
+(f^3)\Lambda_b^{\!-3/2}\dots.
\end{eqnarray}
Here the notation $(f^3)$ and $(f^4)$ is for terms like
$f_{\nu\alpha}f^\alpha{\!_\sigma}f^\sigma{\!_\mu}$ and
$f_{\nu\alpha}f^\alpha{\!_\sigma}f^\sigma{\!_\rho}f^\rho{\!_\mu}$.
Let us consider the size of these higher order terms relative to the leading order term
for worst-case fields accessible to measurement.
In geometrized units an elementary charge has
\begin{eqnarray}
\label{redef}
Q_e=e\sqrt{\frac{\lower2pt\hbox{$G$}}{c^4}}
=\sqrt{\frac{\lower2pt\hbox{$e^2$}}{\hbar c}\frac{\lower2pt\hbox{$G\hbar$}}{c^3}}
=\sqrt{\alpha}\,l_P=1.38\times 10^{-34}cm
\end{eqnarray}
where $\alpha =e^2/\hbar c$ is the fine structure constant
and $l_P\!=\!\sqrt{G\hbar/c^3}$ is the Planck length.
%From \cite{cShifflett} we know there is an exact electric monopole solution
%for this theory which approximates a $f^1{_0}\!\sim\!Q/r^2$ field.
If we assume that charged particles retain $f^1{_0}\!\sim\!Q/r^2$
down to the smallest radii probed by high energy particle physics
experiments ($10^{-17}{\rm cm}$) we have from (\ref{redef},\ref{Lambdab}),
\begin{eqnarray}
\label{highenergyskew}
|f^1{_0}|^2/\Lambda_b\sim (Q_e/(10^{-17})^2)^2/\Lambda_b\sim 10^{-66}.
\end{eqnarray}
Here $|f^1{_0}|$ is assumed to be in some
standard spherical or cartesian coordinate system. If an equation has a tensor term which can
be neglected in one coordinate system, it can be neglected in any coordinate system,
so it is only necessary to prove it in one coordinate system.
The fields at $10^{-17}{\rm cm}$ from an elementary charge
would be larger than near any macroscopic charged object,
and would also be larger than the strongest plane-wave fields.
Therefore the higher order terms in (\ref{approximateNbar}-\ref{approximateNhat})
must be $<\!10^{-66}$ of the leading order terms, so they will be completely negligible for most purposes.

In \S\ref{Connections} we will calculate the connection equations resulting
from $\delta{\mathcal L}/\delta\tGam^\alpha_{\nu\mu}\!\nobreak=\nobreak\!0$. Solving these equations gives
(\ref{upsilonsymmetric},\ref{upsilonantisymmetric},\ref{tGminusG},\ref{antisymmetricpreliminary}),
which can be abbreviated as
\begin{eqnarray}
\tGam^\alpha_{(\nu\mu)}&=&\Gamma^\alpha_{\nu\mu}+\ord(\Lambda_b^{\!-1}),
\,~~~~~~\tGam^\alpha_{[\nu\mu]}=\ord(\Lambda_b^{\!-1/2}),\\
\label{tGapproxG}
\tG_{\nu\mu}&=&G_{\nu\mu}\!+\ord(\Lambda_b^{\!-1}),
\label{tRapprox0}
~~~~~~\tR_{[\nu\mu]}=\ord(\Lambda_b^{\!-1/2}),
\end{eqnarray}
where $\Gamma^\alpha_{\nu\mu}$ is the Christoffel connection (\ref{Christoffel}),
$\tR_{\nu\mu}\!=\!{\mathcal R}_{\nu\mu}(\tGam)$, $R_{\nu\mu}\!=\!R_{\nu\mu}(\Gamma)$~and
\begin{eqnarray}
\label{genEinstein}
~~~~~\tG_{\nu\mu}&=&\tR_{(\nu\mu)}-\frac{\lower1pt\hbox{$1$}}{2}\,g_{\nu\mu}\tR^\rho_\rho,
~~~~~~~G_{\nu\mu}=R_{\nu\mu}-\frac{\lower1pt\hbox{$1$}}{2}\,g_{\nu\mu}R.
\end{eqnarray}
In (\ref{tGapproxG}) the notation $\ord(\Lambda_b^{\!-1})$ and $\ord(\Lambda_b^{\!-1/2})$
indicates terms like $f^\sigma{_{\nu;\alpha}}f^\alpha{_{\mu;\sigma}}\Lambda_b^{\!-1}$
and $f_{[\nu\mu,\alpha];}{^\alpha}\Lambda_b^{\!-1/2}$.

From the antisymmetric part of the field equations (\ref{JSantisymmetric})
and (\ref{approximateNhat},\ref{tRapprox0}) we get
\begin{eqnarray}
\label{fapproxF}
f_{\nu\mu}&=&F_{\nu\mu}+\ord(\Lambda_b^{\!-1}).
\end{eqnarray}
So $f_{\nu\mu}$ and $F_{\nu\mu}$ only differ by terms with $\Lambda_b$ in the denominator,
and the two become identical in the limit as $\Lambda_b\!\rightarrow\!\infty$.
Combining (\ref{JSsymmetric}) with its contraction,
and substituting (\ref{genEinstein},\ref{approximateNbar},\ref{Lambdadef})
\ifnum\ExpandDerivations=1
\begin{eqnarray}
N_{(\nu\mu)}
-\frac{1}{2}\,g_{\nu\mu}N^\rho_\rho
\!&=&g_{\nu\mu}-2\left({f_\nu}^\sigma f_{\sigma\mu}
-\frac{1}{2(n\!-\!2)}g_{\nu\mu}f^{\rho\sigma}\!f_{\sigma\rho}\right)\!\Lambda_b^{\!-1}\nonumber\\
&&-\frac{1}{2}g_{\nu\mu}n+g_{\nu\mu}\!\left(f^{\rho\sigma}\!f_{\sigma\rho}
-\frac{1}{2(n\!-\!2)}nf^{\rho\sigma}\!f_{\sigma\rho}\right)\!\Lambda_b^{\!-1}
+(f^4)\Lambda_b^{\!-2}\dots\nonumber\\
\!&=&g_{\nu\mu}\left(1-\frac{n}{2}\right)
-2{f_\nu}^\sigma f_{\sigma\mu}\Lambda_b^{\!-1}\nonumber\\
&&+g_{\nu\mu}\!\left(\frac{1}{(n\!-\!2)}+\!1
\!-\frac{n}{2(n\!-\!2)}\right)f^{\rho\sigma}\!f_{\sigma\rho}\Lambda_b^{\!-1}
+(f^4)\Lambda_b^{\!-2}\dots\nonumber\\
\!&=&-2\left({f_\nu}^\sigma f_{\sigma\mu}
-\frac{1}{4}g_{\nu\mu}f^{\rho\sigma}\!f_{\sigma\rho}\right)\!\Lambda_b^{\!-1}
-\left(\frac{n}{2}-1\right)g_{\nu\mu}
+(f^4)\Lambda_b^{\!-2}\dots\nonumber
\end{eqnarray}
\fi
gives the Einstein equations
\begin{eqnarray}
\label{Einstein}
\!\!\!\tG_{\nu\mu}&=&8\pi T_{\nu\mu}
-\Lambda_b\!\left(N_{(\nu\mu)}
\!-\!\frac{1}{2}g_{\nu\mu}N^\rho_\rho\right)
\!+\!\Lambda_z\!\left(\frac{n}{2}-1\right)g_{\nu\mu},\\
\label{Einstein2}
&=&8\pi T_{\nu\mu}
+2\left({f_\nu}^\sigma\! f_{\sigma\mu}
\!-\!\frac{1}{4}g_{\nu\mu}f^{\rho\sigma}\!f_{\sigma\rho}\right)
+\Lambda\left(\frac{n}{2}-1\right)g_{\nu\mu}+(f^4)\Lambda_b^{\!-1}\dots.~~~
\end{eqnarray}
%As in (\ref{JSlag1}), we are assuming that the matter Lagrangian term ${\mathcal L}_m$ couples
%additional fields to the theory only through $g_{\nu\mu}$, not through $N_{\nu\mu}$.
From (\ref{Sdef},\ref{Tdef})
we see that $T_{\nu\mu}$ will be the same as in ordinary general relativity,
for example when we include classical hydrodynamics or spin-1/2 fields as in \cite{sShifflett,Birrell}.
Therefore from (\ref{fapproxF},\ref{tGapproxG}), equation (\ref{Einstein2}) differs from the
ordinary Einstein equations only by terms with $\Lambda_b$ in the denominator,
and it becomes identical to the ordinary Einstein equations
in the limit as $\Lambda_b\!\rightarrow\!\infty$
(with an observationally valid total $\Lambda$).
In \S\ref{Connections} we will examine how close the
approximation is for $\Lambda_b$ from (\ref{Lambdab}).

\section{\label{MaxwellEquations}Maxwell's equations}
Setting $\delta{\mathcal L}/\delta\Aphi_\tau\!=0$
and using (\ref{JSlag3},\ref{fdef}) gives
\begin{eqnarray}
0
%&=&\!\frac{4\pi}{\rmg}\!\left[\frac{\partial {\mathcal L}}{\partial \Aphi_\tau}
%-\left(\frac{\partial {\mathcal L}}{\partial \Aphi_{\tau,\omega}}\right)\!{_{,\,\omega}}\right]\!\\
\label{Ampere0}
&=&\frac{\rmt\Lambda_b^{\!1/2}}{2\rmg}\,(\rmN N^{\dashv[\omega\tau]})_{,\,\omega}-4\pi j^\tau
=\frac{(\rmg f^{\omega\tau})_{,\,\omega}}{\rmg}-4\pi j^\tau,
\end{eqnarray}
where
\begin{eqnarray}
\label{jdef}
j^\tau&=&\!\frac{-1}{\rmg}\!\left[\frac{\partial {\mathcal L}_m}{\partial \Aphi_\tau}
-\left(\frac{\partial {\mathcal L}_m}{\partial \Aphi_{\tau,\omega}}\right)\!{_{,\,\omega}}\right].
\end{eqnarray}
From (\ref{Ampere0},\ref{Fdef}) we get Maxwell's equations,
\begin{eqnarray}
\label{Ampere}
{f^{\omega\tau}}_{;\,\omega}&=&4\pi j^\tau,\\
\label{Faraday}
F_{[\nu\mu,\alpha]}&=&0.
\end{eqnarray}
where $f_{\nu\mu}=F_{\nu\mu}+\ord(\Lambda_b^{\!-1})$ from (\ref{fapproxF}).
%As in (\ref{JSlag1}), we assume that the matter Lagrangian term ${\mathcal L}_m$ couples additional
%fields to the theory only through $A_\mu$, not through other components of $\nGam^\alpha_{\mu\nu}$.
From (\ref{JSlag1},\ref{jdef}) we see that $j^{\mu}$ will be the same as in ordinary general relativity,
for example when we include classical hydrodynamics
or spin-1/2 fields as in \cite{sShifflett,Birrell}.
From (\ref{fapproxF}),
we see that equations (\ref{Ampere},\ref{Faraday}) differ from the ordinary Maxwell equations only by
terms with $\Lambda_b$ in the denominator, and these equations become identical to the ordinary Maxwell
equations in the limit as $\Lambda_b\!\rightarrow\!\infty$.
In \S\ref{Connections} we will examine how close the approximation is
for $\Lambda_b$ from (\ref{Lambdab}).

Because ${\mathcal L}_m$ in (\ref{JSlag1}) couples to additional fields only through $g_{\mu\nu}$ and $A_\mu$,
any equations associated with additional fields will be the same as in ordinary general relativity.
For example in the spin-1/2 case, setting $\delta{\mathcal L}/\delta\bar\psi\!=\!0$
will give the ordinary Dirac equation in curved space as in \cite{sShifflett,Birrell}.
It would be interesting to investigate what results if one includes
$f_{\mu\nu}$, $N_{\mu\nu}$ or $\tGam^\alpha_{\mu\nu}$ in ${\mathcal L}_m$,
although there does not appear to be any empirical reason for doing so.
A continuity equation follows from (\ref{Ampere}) regardless of the type of source,
\begin{eqnarray}
\label{continuity}
j^\rho{_{\!;\rho}}&=&\frac{1}{4\pi}f^{\tau\rho}{_{;[\tau;\rho]}}=0.
\end{eqnarray}
Note that the covariant derivative in (\ref{Ampere},\ref{continuity}) is done using the
Christoffel connection (\ref{Christoffel}) formed from the symmetric metric (\ref{gdef}).

\section{\label{Connections}The connection equations}
Setting $\delta{\mathcal L}/\delta\tGam^\alpha_{\nu\mu}\!=0$
with a Lagrange mulitiplier term $\Omega^\mu\tGam^\sigma_{\![\mu\sigma]}$
to enforce the symmetry (\ref{JScontractionsymmetric}),
and using (\ref{JSlag3},\ref{Ampere0}) gives
\begin{eqnarray}
\label{JSconnection}
&&(\rmN N^{\dashv\rho\tau})_{\!,\,\beta}
+\tGam^\tau_{\sigma\beta}\rmN N^{\dashv\rho\sigma}
+\tGam^\rho_{\beta\sigma}\rmN N^{\dashv\sigma\tau}
-\tGam^\alpha_{\beta\alpha}\rmN N^{\dashv\rho\tau}\nonumber\\
&&~~~~~~~~~~~~~~~~~~~~~~~~~~~~~~~~~~~~~~~~~~~~~~~~~~~~~~~
=\frac{8\pi\rmt}{(n\!-\!1)\Lambda_b^{1/2}}\rmg j^{[\rho}\delta^{\tau]}_\beta.~~~~
\end{eqnarray}
These are the connection equations, analogous to $g^{\rho\tau}{_{;\beta}}\!=\!0$
in the symmetric case.
Note that we can also derive Ampere's law (\ref{Ampere0}) by antisymmetrizing and contracting
these equations.
% and using the symmetry (\ref{JScontractionsymmetric}).
From the definition of matrix inverse,
$N^{\dashv\rho\tau}\!\!\nobreak=\nobreak\!(1/N)\partial N\!/\partial N_{\tau\rho}$ and
$N^{\dashv\rho\tau}\!N_{\tau\mu}\!\nobreak=\nobreak\!\delta^\rho_\mu$ we get the identity
\begin{eqnarray}
\label{sqrtdetcomma}
(\!\rmN\,)_{,\sigma}
=\frac{\partial\rmN}
{\partial N_{\tau\rho}}N_{\tau\rho,\sigma}
=\frac{\rmN}{2}N^{\dashv\rho\tau}N_{\tau\rho,\sigma}
\label{sqrtdetcomma2}
=-\frac{\rmN}{2}
{N^{\dashv\rho\tau}}_{,\sigma} N_{\tau\rho}.
\end{eqnarray}
Contracting (\ref{JSconnection}) with $N_{\tau\rho}$
using (\ref{JScontractionsymmetric},\ref{sqrtdetcomma}),
\ifnum\ExpandDerivations=1
\begin{eqnarray}
0=(n\!-\!2)((\rmN\,)_{,\,\beta}
-\tGam^\alpha_{\alpha\beta}\rmN\,)
+\frac{8\pi\rmt}{(n\!-\!1)\Lambda_b^{1/2}}\rmg j^\rho N_{[\rho\beta]},\nonumber
\end{eqnarray}
\fi
and dividing this by $(n\!-\!2)$ gives,
\begin{eqnarray}
\label{der0}
(\rmN\,)_{,\,\beta}-\tGam^\alpha_{\alpha\beta}\rmN
=-\frac{8\pi\rmt}{(n\!-\!1)(n\!-\!2)\Lambda_b^{1/2}}\rmg j^\rho N_{[\rho\beta]}.
\end{eqnarray}
%and from (\ref{der0}) we get
%\begin{eqnarray}
%\label{funnytensor}
%\tGam^\alpha_{\alpha[\nu,\mu]}
%-\frac{8\pi\rmt}{(n\!-\!1)(n\!-\!2)\Lambda_b^{1/2}}
%\left(\!\frac{\rmg}{\rmN}j^\rho N_{[\rho[\nu]}\!\right)\!\!{_{,\mu]}}
%=(ln\rmN\,)_{,[\nu,\mu]}=0.
%\end{eqnarray}
%Because of (\ref{funnytensor}), the $\tGam^\alpha_{\alpha[\nu,\mu]}$ part of
%(\ref{HermitianizedRicci}) only affects the field equations
%if ${\mathcal L}_m$ contains $A_\mu$, that is if $j^\alpha\!\ne\!0$.
Multiplying (\ref{JSconnection}) by
$-N_{\nu\rho}N_{\tau\mu}$ and using (\ref{der0}) gives
% the covariant connection equations,
\begin{eqnarray}
\label{JSconnection0}
\!\!\!\!\!\!N_{\nu\mu,\beta}\!-\!\tGam^\alpha_{\nu\beta}N_{\alpha\mu}
\!-\!\tGam^\alpha_{\beta\mu}N_{\nu\alpha}
\!=\!\frac{-8\pi\rmt}{(n\!-\!1)\Lambda_b^{\!1/2}}\frac{\rmg}{\rmN}\!\left(
N_{\nu[\alpha}N_{\beta]\mu}
\!+\!\frac{N_{[\alpha\beta]}N_{\nu\mu}}{(n\!-\!2)} \!\right)\!j^\alpha\!.
\end{eqnarray}
Equation (\ref{JSconnection0}) together with (\ref{JSsymmetric},\ref{JScurl},\ref{JScontractionsymmetric})
are often used to define the Einstein-Schr\"{o}dinger theory,
particularly when $T_{\nu\mu}\!=\!0$, $j^\alpha\!=\!0$.

Equations (\ref{JSconnection}) or (\ref{JSconnection0}) can be solved exactly\cite{Tonnelat,jShifflett},
similar to the way $g_{\rho\tau;\beta}\!=\nobreak\!0$ can be solved to get the
Christoffel connection.
An expansion in powers of $\Lambda_b^{\!-1}$ will better serve our purposes,
and such an expansion is derived in Appendix E of \cite{sShifflett},
%confirmed by tetrad methods in \cite{jShifflett},
and is also stated without derivation in \cite{Antoci3},
\begin{eqnarray}
\label{gammadecomposition}
\tGam^\alpha_{\nu\mu}&\!\!=\!&\Gamma^\alpha_{\nu\mu}
+\Upsilon^\alpha_{\nu\mu},\\
\Upsilon^\alpha_{\!(\nu\mu)}
\label{upsilonsymmetric}
&\!\!=\!&\!-2\left[f^\tau{_{\!\!(\nu}}f_{\mu)}{\!^\alpha}{_{\!;\tau}}
\!+f^{\alpha\tau}\!f_{\tau(\nu;\mu)}
\!+\!\frac{1}{4(n\!-\!2)}((f^{\rho\sigma}\!f_{\sigma\rho})_,{^\alpha}g_{\nu\mu}
\!-2(f^{\rho\sigma}\!f_{\sigma\rho})_{,(\nu}\delta^\alpha_{\mu)})\right.\nonumber\\
&&~~~~~~\left.+\frac{4\pi}{(n\!-\!2)}j^\rho\left(f^\alpha{_\rho}\,g_{\nu\mu}
+\frac{2}{(n\!-\!1)}f_{\rho(\nu}\delta^\alpha_{\mu)}\right)\right]\!\Lambda_b^{\!-1}
+(f^{4\prime})\Lambda_b^{\!-2}\dots,\\
\label{upsilonantisymmetric}
\Upsilon^\alpha_{\![\nu\mu]}
&\!\!=\!&\!\left[\frac{1}{2}(f_{\nu\mu;}{^\alpha}\!+\!f^\alpha{_{\mu;\nu}}\!-\!f^\alpha{_{\nu;\mu}})
+\frac{8\pi}{(n\!-\!1)}j_{[\nu}\delta^\alpha_{\mu]}\right]\!\rmt\Lambda_b^{\!-1/2}
\!+(f^{3\prime})\Lambda_b^{\!-3/2}\dots\,.
%\label{upsiloncontracted}
%\Upsilon^\alpha_{\alpha\nu}
%&\!=&2\left[\frac{1}{2(n\!-\!2)}(f^{\rho\sigma}\!f_{\sigma\rho})_{,\nu}
%-\frac{8\pi}{(n\!-\!1)(n\!-\!2)}j^\alpha f_{\alpha\nu}\right]\!\Lambda_b^{\!-1}
%+(f^{4\prime})\Lambda_b^{\!-2}\!\!\dots\,.
\end{eqnarray}
In (\ref{gammadecomposition}), $\Gamma^\alpha_{\nu\mu}$ is the Christoffel
connection (\ref{Christoffel}).
The notation $(f^{3\prime})$ and $(f^{4\prime})$ refers to terms like
$f^\alpha{_\tau}f^\tau{_\sigma}f^\sigma{_{[\nu;\mu]}}$ and
$f^\alpha{_\tau}f^\tau{_\sigma}f^\sigma{_\rho}f^\rho{_{(\nu;\mu)}}$.
%and $j^\tau$ can be considered as a substitute for $(c/4\pi){f^{\omega\tau}}_{;\,\omega}$
%because of Ampere's law (\ref{Ampere}).
As in (\ref{approximateNbar},\ref{approximateNhat}), we see from (\ref{highenergyskew}) that
the higher order terms in (\ref{upsilonsymmetric}-\ref{upsilonantisymmetric})
must be $<\!10^{-66}$ of the leading order terms, so they will be completely negligible for most purposes.

Extracting $\Upsilon^\tau_{\sigma\beta}$ of (\ref{gammadecomposition}) from
the Hermitianized Ricci tensor (\ref{HermitianizedRiccit}) gives,
\begin{eqnarray}
\!\hR_{(\nu\mu)}(\tGam)
\label{Ricciadditionsymmetric0}
\!&=& R_{\nu\mu}(\Gamma)+{\Upsilon}^\alpha_{\!(\nu\mu);\alpha}
\!-\!\Upsilon^\alpha_{\alpha(\nu;\mu)}
\!-\!{\Upsilon}^\sigma_{\!(\nu\alpha)}{\Upsilon}^\alpha_{\!(\sigma\mu)}
\!-\!{\Upsilon}^\sigma_{\![\nu\alpha]}{\Upsilon}^\alpha_{\![\sigma\mu]}
\!+\!{\Upsilon}^\sigma_{\!(\nu\mu)}\Upsilon^\alpha_{\sigma\alpha},~~~\\
\!\hR_{[\nu\mu]}(\tGam)
\label{Ricciadditionantisymmetric0}
\!&=&{\Upsilon}^\alpha_{\![\nu\mu];\alpha}
%\!-\!\Upsilon^\alpha_{\alpha[\nu;\mu]}
\!-\!{\Upsilon}^\sigma_{\!(\nu\alpha)}{\Upsilon}^\alpha_{\![\sigma\mu]}
\!-\!{\Upsilon}^\sigma_{\![\nu\alpha]}{\Upsilon}^\alpha_{\!(\sigma\mu)}
\!+\!{\Upsilon}^\sigma_{\![\nu\mu]}\Upsilon^\alpha_{\sigma\alpha}.~~~
\end{eqnarray}
%Here the covariant derivative is done using $\Gamma^\alpha_{\nu\mu}$ from (\ref{Christoffel}), and
%$R_{\nu\mu}\!=\!R_{\nu\mu}(\Gamma)$.
Substituting (\ref{gammadecomposition}-\ref{upsilonantisymmetric},\ref{Ampere})
into (\ref{Ricciadditionsymmetric0}),
\ifnum\ExpandDerivations=1
with $\ff=f^{\rho\sigma}\!f_{\sigma\rho}$,
\begin{eqnarray}
\tR_{(\nu\mu)}\!
&=& R_{\nu\mu}+{\Upsilon}^\alpha_{\!(\nu\mu);\,\alpha}
\!-\Upsilon^\alpha_{\alpha(\nu;\mu)}
\!-{\Upsilon}^\sigma_{\![\nu\alpha]}{\Upsilon}^\alpha_{\![\sigma\mu]}\dots\nonumber\\
&=&R_{\nu\mu}
-2\left[\left(f^\tau{_{(\nu}}f_{\mu)}{^\alpha}{_{;\,\tau}}
+f^{\alpha\tau}f_{\tau(\nu;\,\mu)}
+\frac{1}{4(n\!-\!2)}(\ff_,{^\alpha}g_{\nu\mu}
-2\ff_{,(\nu}\delta^\alpha_{\mu)})\right){_{;\,\alpha}}\right.\nonumber\\
&&~~~~~~~~~~~~~~+\frac{4\pi}{(n\!-\!2)}j^\rho{_{;\alpha}}\!\left(f^\alpha{_\rho}g_{\nu\mu}
+\frac{2}{(n\!-\!1)}f_{\rho(\nu}\delta^\alpha_{\mu)}\right)\nonumber\\
&&~~~~~~~~~~~~~~+\frac{4\pi}{(n\!-\!2)}j^\rho\!\left(f^\alpha{_\rho}g_{\nu\mu}
+\frac{2}{(n\!-\!1)}f_{\rho(\nu}\delta^\alpha_{\mu)}\right){_{\!;\alpha}}\nonumber\\
&&~~~~~~~~~~~~~~+\frac{1}{2(n\!-\!2)}\ff_{,(\nu;\,\mu)}
-\frac{8\pi}{(n\!-\!1)(n\!-\!2)}(j^\alpha f_{\alpha(\nu})_{;\mu)}\nonumber\\
&&~~~~~~~~~~~~~~-\frac{1}{4}\left(f_{\nu\alpha;}{^\sigma}\!+\!f^\sigma{_{\alpha;\nu}}
\!-\!f^\sigma{_{\nu;\,\alpha}}+\frac{16\pi}{(n\!-\!1)}j_{[\nu}\delta^\sigma_{\alpha]}\right)\nonumber\\
&&~~~~~~~~~~~~~~~~~\left.\times\left(f_{\sigma\mu;}{^\alpha}\!+\!f^\alpha{_{\mu;\,\sigma}}
\!-\!f^\alpha{_{\sigma;\,\mu}}+\frac{16\pi}{(n\!-\!1)}j_{[\sigma}\delta^\alpha_{\mu]}\right)\right]\!\Lambda_b^{\!-1}\dots\nonumber\\
\label{Riccitensor}
&=&R_{\nu\mu}
-\left[2f^\tau{_{(\nu}}f_{\mu)}{^\alpha}{_{;\tau;\alpha}}
+2f^{\alpha\tau}f_{\tau(\nu;\,\mu)}{_{;\,\alpha}}
+\frac{1}{2(n\!-\!2)}\ff_,{^\alpha}{_{;\alpha}}g_{\nu\mu}\right.\nonumber\\
&&-f^\sigma{_{\nu;\alpha}}f^\alpha{_{\mu;\sigma}}
+f^\sigma{_{\nu;\,\alpha}}f_{\sigma\mu;}{^\alpha}
+\frac{1}{2}f^\sigma{_{\alpha;\nu}}f^\alpha{_{\sigma;\,\mu}}\nonumber\\
&&\left.+8\pi j^\tau f_{\tau(\nu;\mu)}
\!-\!\frac{32\pi^2}{(n\!-\!1)}j_\nu j_\mu
\!+\!\frac{32\pi^2}{(n\!-\!2)}j^\rho j_\rho g_{\nu\mu}
\!+\!\frac{8\pi}{(n\!-\!2)}j^\rho{_{;\alpha}}f^\alpha{_\rho}g_{\nu\mu}\right]\!\Lambda_b^{\!-1}\dots,\nonumber\\
\tR^\rho_\rho &=&R
-\left[2f^{\tau\beta}f_{\beta}{^\alpha}{_{;\tau;\alpha}}
+\frac{n}{2(n\!-\!2)}\ff_,{^\alpha}{_{;\alpha}}
-f{^{\sigma\beta}}_{;\alpha}f^\alpha{_{\beta;\sigma}}
+\frac{1}{2}f{^{\sigma\beta}}_{;\,\alpha}f_{\sigma\beta;}{^\alpha}\right.\nonumber\\
&&\left.-8\pi f^{\alpha\tau}j_{\tau;\alpha}
\!-\!32\pi^2\left(1\!+\!\frac{1}{(n\!-\!1)}\!-\!\frac{n}{(n\!-\!2)}\right)j^\rho j_\rho
\!+\!\frac{8\pi n}{(n\!-\!2)}j^\rho{_{;\alpha}}f^\alpha{_\rho}\right]\!\Lambda_b^{\!-1}\dots\nonumber\\
\label{Ricciscalar}
&=&R
+\left[-2f^{\tau\beta}f_{\beta}{^\alpha}{_{;\tau;\alpha}}
-\frac{n}{2(n\!-\!2)}\ff_,{^\alpha}{_{;\alpha}}
-\frac{3}{2}f_{[\sigma\beta;\alpha]}f^{[\sigma\beta}{_;}{^{\alpha]}}\right.\nonumber\\
&&~~~~~~~~~\left.-\frac{32\pi^2n}{(n\!-\!1)(n\!-\!2)}j^\rho j_\rho
-\frac{16\pi}{(n\!-\!2)}f^{\alpha\tau}j_{\tau;\alpha}\right]\!\Lambda_b^{\!-1}\dots\,.\nonumber
\end{eqnarray}
\fi
and using (\ref{genEinstein}) gives
\begin{eqnarray}
\label{tGminusG}
&&(\tG_{\nu\mu}-G_{\nu\mu})=\nonumber\\
&&-\left(2f^\tau{_{(\nu}}f_{\mu)}{^\alpha}{_{;\tau;\alpha}}
+2f^{\alpha\tau}f_{\tau(\nu;\,\mu)}{_{;\alpha}}
\!-\!f^\sigma{_{\nu;\alpha}}f^\alpha{_{\mu;\sigma}}
\!+f^\sigma{_{\nu;\alpha}}f_{\sigma\mu;}{^\alpha}
+\frac{1}{2}f^\sigma{_{\alpha;\nu}}f^\alpha{_{\sigma;\,\mu}}\right.\nonumber\\
&&~~~~~-g_{\nu\mu}f^{\tau\beta}f_{\beta}{^\alpha}{_{;\tau;\alpha}}
\!-\!\frac{1}{4}g_{\nu\mu}(f^{\rho\sigma}\!f_{\sigma\rho})_,{^\alpha}{_{;\alpha}}
-\frac{3}{4}g_{\nu\mu}f_{[\sigma\beta;\alpha]}f^{[\sigma\beta}{_;}{^{\alpha]}}\nonumber\\
&&~~~~~\left.+8\pi j^\tau f_{\tau(\nu;\mu)}
-\frac{32\pi^2}{(n\!-\!1)}j_\nu j_\mu
+\frac{16\pi^2}{(n\!-\!1)}g_{\nu\mu}j^\rho j_\rho+(f^4)\right)\!\Lambda_b^{\!-1}\dots.
\end{eqnarray}
From (\ref{Einstein2}) we can define an ``effective'' energy momentum tensor $\tilde T_{\nu\mu}$
which applies when $G_{\nu\mu}$ is used in the Einstein equations and ${\mathcal L}_m\!=\!0$,
\begin{eqnarray}
\label{tildeT}
8\pi\tilde T_{\nu\mu}&=&2\left({f_\nu}^\sigma\! f_{\sigma\mu}\!-\!\frac{1}{4}\,g_{\nu\mu}f^{\rho\sigma}\!f_{\sigma\rho}\right)
-(\tG_{\nu\mu}-G_{\nu\mu}).
\end{eqnarray}
Substituting (\ref{upsilonantisymmetric},\ref{Ampere}) into (\ref{Ricciadditionantisymmetric0}) gives
\begin{eqnarray}
\!\!\!\tR_{[\nu\mu]}&=&
\ifnum\ExpandDerivations=1
{\Upsilon}^\alpha_{\![\nu\mu];\alpha}+\ord(\Lambda_b^{\!-3/2})\dots\nonumber\\
&=&\!\left(\frac{1}{2}(f_{\nu\mu;}{^\alpha}
\!+\!f^\alpha{_{\mu;\nu}}\!-\!f^\alpha{_{\nu;\mu}}){_{;\alpha}}
+\frac{8\pi}{(n\!-\!1)}j_{[\nu,\mu]}\right)\!\rmt\Lambda_b^{\!-1/2}\dots\nonumber\\
&=&\!\left(\!\frac{3}{2}f_{[\nu\mu,\alpha];}{^\alpha}
+f^\alpha{_{\mu;\nu;\alpha}}-f^\alpha{_{\nu;\mu;\alpha}}
+\frac{8\pi}{(n\!-\!1)}j_{[\nu,\mu]}\right)\!\rmt\Lambda_b^{\!-1/2}\dots\nonumber\\
&=&
\fi
\label{antisymmetricpreliminary}
\!\left(\!\frac{3}{2}f_{[\nu\mu,\alpha];}{^\alpha}
\!+\!2f^\alpha{_{\mu;[\nu;\alpha]}}\!-\!2f^\alpha{_{\nu;[\mu;\alpha]}}
\!-\!\frac{8\pi(n\!-\!2)}{(n\!-\!1)}\,j_{[\nu,\mu]}\right)\!\rmt\Lambda_b^{\!-1/2}\dots.~~~
\end{eqnarray}

As we have noted in \S\ref{EinsteinEquations} and \S\ref{MaxwellEquations},
the $\Lambda_b$ in the denominator of (\ref{tGminusG},\ref{antisymmetricpreliminary})
causes our Einstein and Maxwell equations (\ref{Einstein2},\ref{Ampere},\ref{Faraday})
to become the ordinary Einstein and Maxwell equations in the limit as
$\omega_c\!\rightarrow\!\infty$, $|\Lambda_z|\!\rightarrow\!\infty$, $\Lambda_b\!\rightarrow\!\infty$,
and it also causes the relation $f_{\nu\mu}\!\approx\!F_{\nu\mu}$ from (\ref{fapproxF})
to become exact in this limit.
Let us examine how close these approximations are when $\Lambda_b\sim 10^{66}cm^{-2}$ as in (\ref{Lambdab}).

We will start with the Einstein equations (\ref{Einstein2}).
Let us consider worst-case values of the ${\mathcal O}(\Lambda_b^{\!-1})$ terms in (\ref{tGminusG})
and compare these to the ordinary electromagnetic term in (\ref{tildeT}).
If we assume that charged particles retain $f^1{_0}\!\sim\!Q/r^2$
down to the smallest radii probed by high energy particle physics
experiments ($10^{-17}{\rm cm}$) we have,
\begin{eqnarray}
\label{highenergyderiv1}
|f^1{_{0;1}}/f^1{_0}|^2/\Lambda_b&\sim& 4/\Lambda_b\,(10^{-17})^2\sim 10^{-32},\\
\label{highenergyderiv2}
|f^1{_{0;1;1}}/f^1{_0}|/\Lambda_b&\sim& 6/\Lambda_b\,(10^{-17})^2\sim 10^{-32}.
\end{eqnarray}
So for electric monopole fields, terms like $f^\sigma{_{\nu;\alpha}}f^\alpha{_{\mu;\sigma}}\Lambda_b^{\!-1}$
and $f^{\alpha\tau}f_{\tau(\nu;\,\mu)}{_{;\alpha}}\Lambda_b^{\!-1}$ in (\ref{tGminusG})
must be $<\!10^{-32}$ of the ordinary electromagnetic term in (\ref{tildeT}).
And regarding $j^\tau$ as a substitute for $(1/4\pi){f^{\omega\tau}}_{;\,\omega}$ from (\ref{Ampere}),
the same is true for the $j^\tau$ terms.
For an electromagnetic plane-wave in a flat background space
\begin{eqnarray}
\label{planewaveA}
A_\mu&=&A\epsilon_\mu{\rm sin}(k_\alpha x^\alpha)
~~,~~\epsilon^\alpha\epsilon_\alpha=-1
~~,~~k^\alpha k_\alpha=k^\alpha\epsilon_\alpha=0,\\
\label{planewavef}
f_{\nu\mu}&=&2\Aphi_{[\mu,\nu]}
=2A\epsilon_{[\mu} k_{\nu]}{\rm cos}(k_\alpha x^\alpha),~~~~j^\sigma=0.
\end{eqnarray}
Here $A$ is the magnitude, $k^\alpha$ is the wavevector, and
$\epsilon^\alpha$ is the polarization.
Substituting (\ref{planewaveA},\ref{planewavef}) into (\ref{tGminusG}),
all of the terms vanish for a flat background space.
Also, for the highest energy gamma rays known in nature ($10^{20}$eV, $10^{34}$Hz)
we have from (\ref{Lambdab}),
\begin{eqnarray}
\label{gammaderiv1}
|f^1{_{0;1}}/f^1{_0}|^2/\Lambda_b&\sim& (E/\hbar c)^2/\Lambda_b\sim 10^{-16},\\
\label{gammaderiv2}
|f^1{_{0;1;1}}/f^1{_0}|/\Lambda_b&\sim& (E/\hbar c)^2/\Lambda_b\sim 10^{-16}.
\end{eqnarray}
So for electromagnetic plane-wave fields, even if some of the terms
in (\ref{tGminusG}) were non-zero because of spatial curvatures, they must still be
$<\!10^{-16}$ of the ordinary electromagnetic term in (\ref{tildeT}).
%For the purpose of looking at worst-case curvatures, electrons could perhaps be
%considered as tiny Schwarzschild solutions with an effective Schwarzschild radius of
%$r_s=2Gm_e/c^2\!=1.4\nobreak\times\nobreak\!10^{-55}~{\rm cm}$.
%Existing particle accelerators can only create energies high
%enough to probe radii of around $r\!\sim\!10^{-17}~{\rm cm}$, where
%\begin{eqnarray}
%\label{electroncurvature}
%\frac{\tilde C_{trtr}}{\Lambda_b}
%&\sim&\frac{r_s}{\Lambda_br^3}
%=\frac{1}{\Lambda_br^3}\left(\frac{2G m_e}{c^2}\right)\!\sim\!10^{-70}.
%\end{eqnarray}
Therefore even for the most extreme worst-case fields accessible to measurement, the extra terms
in the Einstein equations (\ref{Einstein2}) must all be $<\!10^{-16}$ of the ordinary electromagnetic term.

Now let us look at the approximation $f_{\nu\mu}\!\approx\!F_{\nu\mu}$ from (\ref{fapproxF}),
and Maxwell's equations (\ref{Ampere},\ref{Faraday}).
From the covariant derivative commutation rule,
the cyclic identity $2R_{\nu[\tau\alpha]\mu}=R_{\nu\mu\alpha\tau}$,
the definition of the Weyl tensor $C_{\nu\mu\alpha\tau}$, and
the Einstein equations $R_{\nu\mu}=-\Lambda g_{\nu\mu}+(f^2)\dots$ from
(\ref{Einstein2}) we get
\begin{eqnarray}
2f^\alpha{_{\nu;[\mu;\alpha]}}&=&
\ifnum\ExpandDerivations=1
R^\tau{_{\nu\mu\alpha}}f^\alpha{_\tau}
+R_\tau{^\alpha}{_{\mu\alpha}}f^\tau{_\nu}
=\frac{1}{2}R_{\nu\mu\alpha\tau}f^{\alpha\tau}
+R^\tau{_\mu}f_{\tau\nu}\nonumber\\
%\!&=&\frac{1}{2}\left(C_{\nu\mu\alpha\tau}+\!\frac{2}{(n\!-\!2)}
%(g_{\nu[\alpha}R_{\tau]\mu}-g_{\mu[\alpha}R_{\tau]\nu})
%-\frac{2}{(n\!-\!1)(n\!-\!2)}Rg_{\nu[\alpha}g_{\tau]\mu}\right)f^{\alpha\tau}\nonumber\\
%&&+R^\tau{_\mu}f_{\tau\nu}\\
\!&=&\frac{1}{2}\left(C_{\nu\mu}{^{\alpha\tau}}
+\!\frac{4}{(n\!-\!2)}\delta^{[\alpha}_{[\nu}R^{\tau]}_{\mu]}
-\frac{2}{(n\!-\!1)(n\!-\!2)}\delta^{[\alpha}_{[\nu}\delta^{\tau]}_{\mu]}R\right)f_{\alpha\tau}
-R^\tau{_\mu}f_{\nu\tau}\nonumber\\
%\!&=&\frac{1}{2}C_{\nu\mu\alpha\tau}f^{\alpha\tau}
%+\frac{2}{(n\!-\!2)}R^\tau{_{[\mu}}f_{\nu]\tau}
%-\frac{1}{(n\!-\!1)(n\!-\!2)}Rf_{\nu\mu}
%+R^\tau{_\mu}f_{\tau\nu}.
&=&
\fi
\label{Fterm0}
\frac{1}{2}f^{\alpha\tau}C_{\alpha\tau\nu\mu}
+\frac{(n\!-\!2)\Lambda}{(n\!-\!1)}f_{\nu\mu}+(f^3)\dots.
%\label{Rantisymmetric}
%\tR_{[\nu\mu]}\!=\!-\!\left(\theta_{[\tau,\alpha]}\varepsilon_{\nu\mu}{^{\tau\alpha}}
%\!+\!f^{\alpha\tau}C_{\alpha\tau\nu\mu}
%\!+\!\frac{2(n\!-\!2)\Lambda}{(n\!-\!1)}f_{\nu\mu}
%\!+\!\frac{8\pi(n\!-\!2)}{(n\!-\!1)}j_{[\nu,\mu]}\!+\!(f^3)\right)\!\frac{\rmt}{\Lambda_b^{\!1/2}}\dots
\end{eqnarray}
Substituting (\ref{approximateNhat})
into the field equations (\ref{JSantisymmetric})
%gives
%\begin{eqnarray}
%\label{antisymmetric0}
%f_{\nu\mu}
%\!&=&\!F_{\nu\mu}+\tR_{[\nu\mu]}\rmt\Lambda_b^{\!-1/2}/2+(f^3)\Lambda_b^{\!-1}\dots,
%\end{eqnarray}
and using (\ref{antisymmetricpreliminary},\ref{Fterm0}) we get
\begin{eqnarray}
\!\!\!\!f_{\nu\mu}
\label{threeparts}
\!&=&\!F_{\nu\mu}
\!+\!\!\left(\theta_{[\tau,\alpha]}\varepsilon_{\nu\mu}{^{\tau\alpha}}
\!+\!f^{\alpha\tau}C_{\alpha\tau\nu\mu}
\!+\!\frac{2(n\!-\!2)\Lambda}{(n\!-\!1)}f_{\nu\mu}\right.\nonumber\\
&&~~~~~~~~~~~~~~~~~~~~~~~~~~~~~~\left.\!+\!\frac{8\pi(n\!-\!2)}{(n\!-\!1)}j_{[\nu,\mu]}\!+\!(f^3)\right)\!\Lambda_b^{\!-1}\dots
\end{eqnarray}
where $\varepsilon_{\tau\nu\mu\alpha}={(\rm Levi\!-\!Civita~tensor)}$,~
$C_{\alpha\tau\nu\mu}=({\rm Weyl~tensor})$, and
\begin{eqnarray}
\label{thdef}
\theta_\tau &=&\frac{\lower1pt\hbox{$1$}}{4}f_{[\nu\mu,\alpha]}\varepsilon_\tau{^{\nu\mu\alpha}},~~
f_{[\nu\mu,\alpha]}=-\frac{\lower1pt\hbox{$2$}}{3}
\,\theta_\tau\varepsilon^\tau{_{\nu\mu\alpha}}.
%~~\varepsilon_{\tau\nu\mu\alpha}
%&=&{(\rm Levi\!-\!Civita~tensor)},
%&=&{[\,\nu\mu\tau\tau]}\rmg~~~~~~,~~
%\varepsilon^{\nu\mu\tau\tau}=-[\,\nu\mu\tau\tau]/\rmg~,\\
%{[\,\nu\mu\tau\tau]}
%&=&\left\{\matrix{
%+1~{\rm for~even~permutations~of~0123}\cr
%-1~{\rm for~odd~~permutations~of~0123}\cr
%~0~{\rm for~two~equal~indices~~~~~~~~~~~}
%}\right.\\
%\label{Ctilde}
%~~~~~C_{\alpha\tau\nu\mu}=({\rm Weyl~tensor}).
%\tC_{\nu\mu\alpha\tau}&=&R_{\nu\mu\alpha\tau}
%-g_{\nu[\alpha}R_{\tau]\mu}+g_{\mu[\alpha}R_{\tau]\nu},\\
%\label{Fdef}
%~~~~F_{\nu\mu}&=&2A_{[\mu,\nu]}.
\end{eqnarray}
The $\theta_{[\tau,\alpha]}\varepsilon_{\nu\mu}{^{\tau\alpha}}\Lambda_b^{\!-1}$
term in (\ref{threeparts}) is divergenceless so that it has no effect on Ampere's law (\ref{Ampere}).
The $f_{\nu\mu}\Lambda/\Lambda_b$ term is $\sim 10^{-122}$ of
$f_{\nu\mu}$ from (\ref{Lambdadef},\ref{Lambdab}).
The $(f^3)\Lambda_b^{\!-1}$ term is $<\!10^{-66}$ of $f_{\nu\mu}$ from (\ref{highenergyskew}).
The largest observable values of the Weyl tensor might be expected to occur
near the Schwarzschild radius, $r_s\!\nobreak=\nobreak\!2Gm/c^2$, of black
holes, where it takes on values around $r_s/r^3$.
%However, since the lightest black holes have the smallest Schwarzschild radius,
%they will create the largest value of
%$r_s/r_s^3=1/r_s^2$.
The largest value of $r_s/r^3$ would occur near the lightest black holes,
which would be of about one solar mass, where from (\ref{Lambdab}),
\begin{eqnarray}
\label{blackholecurvature}
\frac{C_{0101}}{\Lambda_b}
&\sim&\frac{1}{\Lambda_br^2_s}
=\frac{1}{\Lambda_b}\left(\frac{c^2}{2Gm_\odot}\right)^2\!\sim\!10^{-77}.
\end{eqnarray}
And regarding $j^\tau$ as a substitute for $(1/4\pi){f^{\omega\tau}}_{;\,\omega}$ from
(\ref{Ampere}), the $j_{[\nu,\mu]}\Lambda_b^{\!-1}$ term is
$<\!10^{-32}$ of $f_{\nu\mu}$ from (\ref{highenergyderiv2}).
Therefore, the last four terms in (\ref{threeparts}) must all be $<\!10^{-32}$ of $f_{\nu\mu}$.
%Taking the divergence of (\ref{threeparts}) using (\ref{Ampere}),
%%and ignoring $C_{\nu\mu\alpha\tau}f^{\alpha\tau}/\Lambda_b$
%%and $f_{\nu\mu}\Lambda/\Lambda_b$
%%due to (\ref{blackholecurvature},\ref{LambdaoverLambdab}),
%the divergenceless term $\dual_{[\tau,\alpha]}\varepsilon_{\nu\mu}{^{\tau\alpha}}$
%falls out and we get an extremely close approximation to Maxwell's equations,
%\begin{eqnarray}
%\label{Maxwell}
%F_{\nu\mu;}{^\nu}=4\pi j_\mu
%\!+\!\left[(f^{\tau\alpha}C_{\alpha\tau\nu\mu});{^\nu}
%\!-\!\frac{8\pi(n\!-\!2)\Lambda}{(n\!-\!1)}j_\mu
%\!-\!\frac{8\pi(n\!-\!2)}{(n\!-\!1)}\,j_{[\nu,\mu];}{^\nu}+(f^{3\prime})\right]\!\Lambda_b^{\!-1}\dots,\\
%\label{Faraday}
%F_{[\nu\mu,\sigma]}=0.
%\end{eqnarray}
%As usual, Faraday's law (\ref{Faraday}) is just an identity which follows from
%the definition (\ref{Fdef}).
Consequently, even for the most extreme worst-case
fields accessible to measurement, the extra terms in
Maxwell's equations (\ref{Ampere},\ref{Faraday}) must be
$<\nobreak\!10^{-32}$ of the ordinary terms.

The divergenceless term $\theta_{[\tau,\alpha]}\varepsilon_{\nu\mu}{^{\tau\alpha}}\Lambda_b^{\!-1}$ of
(\ref{threeparts}) should also be expected to be $<\!10^{-32}$ of $f_{\nu\mu}$ from
(\ref{highenergyderiv1},\ref{highenergyderiv2},\ref{thdef}).
However, we need to consider the possibility where $\theta_\tau$ changes extremely rapidly.
%so let us consider the ``dual'' part of (\ref{threeparts}).
%but where there are no extreme spatial curvatures so that
%the Weyl tensor term in (\ref{threeparts}) is still negligible.
Taking the curl of (\ref{threeparts}),
the $F_{\nu\mu}$ and $j_{[\nu,\mu]}$ terms drop out,
\ifnum\ExpandDerivations=1
\begin{eqnarray}
f_{[\nu\mu,\sigma]}
=\!\left(\theta_{\tau;\alpha;[\sigma}\,\varepsilon_{\nu\mu]}{^{\tau\alpha}}
+(f^{\alpha\tau}C_{\alpha\tau[\nu\mu}){_{,\sigma]}}
+\frac{2(n\!-\!2)\Lambda}{(n\!-\!1)}f_{[\nu\mu,\sigma]}
+(f^{3\prime})\right)\!\Lambda_b^{\!-1}\dots.\nonumber
\end{eqnarray}
Contracting this with $\Lambda_b\varepsilon^{\rho\sigma\nu\mu}/2$ and using (\ref{thdef}) gives,
\begin{eqnarray}
\label{rawProca}
2\Lambda_b\theta^\rho
%\!&=&\!-2\,\dual^\tau{\!_;}{^\alpha}{_{;\,\sigma}}\delta^{[\rho}_{\,\tau}\delta^{\sigma]}_\alpha\\
\!&=&\!-2\,\theta^{[\rho}{_;}{^{\sigma]}}{_{\!;\,\sigma}}
+\frac{1}{2}\varepsilon^{\rho\sigma\nu\mu}(f^{\alpha\tau}C_{\alpha\tau[\nu\mu}){_{,\sigma]}}
+\frac{4(n\!-\!2)\Lambda}{(n\!-\!1)}\,\theta^\rho
+(f^{3\prime})\dots\nonumber
\end{eqnarray}
Using $\theta^\sigma{_{;\sigma}}\!=0$ from (\ref{thdef}), the covariant derivative commutation rule,
and the Einstein equations $R_{\nu\mu}\!=\!-\Lambda g_{\nu\mu}+(f^2)\dots$ from (\ref{Einstein2}),
gives $\theta^\sigma{_{;\rho;\sigma}}\!=\!R_{\sigma\rho}\theta^\sigma\!=\!-\theta_\rho\Lambda+(f^{3\prime})\dots$,
\fi
and we get something similar to the Proca equation\cite{Proca,Jackson},
\begin{eqnarray}
\label{Proca}
\!\!\theta_\rho=\left(-\,\theta_\rho{_{;\,\sigma;}}{^\sigma}
+\frac{1}{2}\varepsilon_\rho{^{\sigma\nu\mu}}(f^{\alpha\tau}C_{\alpha\tau[\nu\mu}){_{,\sigma]}}
+\frac{(3n\!-\!7)\Lambda}{(n\!-\!1)}\,\theta_\rho
+(f^{3\prime})\right)\!\frac{1}{2\Lambda_b}\dots.~~~
\end{eqnarray}
Here the constraint $\theta^\nu_{;\nu}\!=\!0$ results from the definition (\ref{thdef})
and we are using a $(1,-1,-1,-1)$ metric signature.
Eq. (\ref{Proca}) suggests that $\theta_\rho$ Proca-wave solutions might exist in this theory.
Assuming that the magnitude of $C_{\alpha\tau\nu\mu}$ is roughly proportional to $\theta_\rho$
for such waves, and assuming that $f_{\mu\nu}$ goes according to (\ref{threeparts}) with $F_{\mu\nu}\!=\!0$,
the extra terms in (\ref{Proca}) could perhaps be neglected in the weak field approximation.
Using (\ref{Proca}) and $\Lambda_b\!\approx\!-\Lambda_z\!=\!C_z\omega_c^4l_P^2$
from (\ref{Lambdab}), such Proca-wave solutions would have an extremely high minimum frequency
\begin{eqnarray}
\label{omegac}
\omega_{Proca}\!=\!\sqrt{2\Lambda_b}\approx\!\sqrt{2C_z}\,\omega_c^2l_P\sim 10^{43} rad/s,
\end{eqnarray}
where the cutoff frequency $\omega_c$ and $C_z$ come from (\ref{cutoff},\ref{Cz}).

There are several points to make about (\ref{Proca},\ref{omegac}).
1) A particle associated with a $\theta_\rho$ field would have mass $\hbar\omega_{Proca}$,
which is much greater than could be produced by particle accelerators,
and so it would presumably not conflict with high energy physics experiments.
2) We have recently shown that $sin[\,kr\!-\!\omega t]$ Proca-wave solutions
do not exist in the theory, using an asyptotically flat Newman-Penrose $1/r$ expansion
similar to \cite{PersidesI,PersidesII}.
However, it is still possible that wave-packet solutions could exist.
3) Substituting the $k\!=\!0$ flat space Proca-wave solution
$\theta_\rho\!=\!(0,1,0,0)sin[\omega_{Proca}t]$ and $F_{\mu\nu}\!\nobreak=\nobreak\!0$
into (\ref{threeparts},\ref{tildeT},\ref{tGminusG}), and assuming a flat background space
gives $\tilde T_{00}\!=\!-2/\Lambda_b\!<\!0$.
This suggests that Proca-wave solutions might have negative energy,
but because $sin[\,kr\!-\!\omega t]$ solutions do not exist,
and because of the other approximations used,
this calculation is extremely uncertain.
4) With a cutoff frequency $\omega_c\!\nobreak\sim\nobreak\!1/l_P$
from (\ref{cutoff}) we have $\omega_{Proca}\!>\omega_c$
from (\ref{omegac},\ref{cutoff},\ref{Cz}), so Proca-waves would presumably be cut off.
More precisely, (\ref{omegac}) says that Proca-waves would be cut off if
$\omega_c\!>\!1/(l_P\sqrt{2C_z}\,)$.
Whether $\omega_c$ is caused by a discreteness, uncertainty or foaminess of
spacetime near the Planck length\cite{Sakharov,Garay,Padmanabhan,Padmanabhan2,Smolin},
or by some other effect, the same $\omega_c$ which cuts off $\Lambda_z$ in (\ref{Lambdab})
should also cut off very high frequency electromagnetic and gravitational waves,
and Proca-waves.
5) If wave-packet Proca-wave solutions do exist, and they have negative energy,
it is possible that $\theta_\rho$ could function
as a kind of built-in Pauli-Villars field. Pauli-Villars regularization in quantum electrodynamics
requires a negative energy Proca field with a mass $\hbar\omega_{Proca}$ that
goes to infinity as $\omega_c\!\rightarrow\!\infty$, as we have from (\ref{omegac}).
6) As mentioned initially, it might be more correct to take the limit of this theory as
$\omega_c\!\nobreak\rightarrow\nobreak\!\infty$, $|\Lambda_z|\!\nobreak\rightarrow\nobreak\!\infty$,
$\Lambda_b\!\nobreak\rightarrow\nobreak\!\infty$, as in quantum electrodynamics. In this limit
(\ref{Proca},\ref{omegac}) require that $\theta_\rho\!\rightarrow\!0$
or $\omega_{Proca}\!\nobreak\rightarrow\nobreak\!\infty$, and
the theory becomes exactly Einstein-Maxwell theory as in (\ref{LRESlimit}).
7) Finally, we should emphasize that Proca-wave solutions are only
a possibility suggested by equation (\ref{Proca}).
Their existence and their possible interpretation are just speculation at this point.
We are continuing to pursue these questions.

\section{\label{LorentzForce}The Lorentz force equation}
A generalized contracted Bianchi identity for this theory
can be derived using only the connection equations (\ref{JSconnection}) and the symmetry
(\ref{JScontractionsymmetric}) of $\tGam^\alpha_{\nu\mu}$,
\begin{eqnarray}
\label{niceform}
(\rmN N^{\dashv\nu\sigma}\tR{_{\sigma\lambda}}
+\rmN N^{\dashv\sigma\nu}\tR_{\lambda\sigma}){_{,\nu}}
-\rmN N^{\dashv\nu\sigma}\tR{_{\sigma\nu,\lambda}}=0.
\end{eqnarray}
This identity can also be written
%\begin{eqnarray}
%\label{niceform2}
%(\rmN N^{\dashv\nu\sigma}\tR{_{\sigma\lambda}}
%+\rmN N^{\dashv\sigma\nu}\tR_{\lambda\sigma}){_{;\nu}}
%-\rmN N^{\dashv\nu\sigma}\tR{_{\sigma\nu;\lambda}}=0,
%\end{eqnarray}
in terms of $g^{\rho\tau},f^{\rho\tau}$ and $\tG_{\nu\mu}$ from (\ref{gdef},\ref{fdef},\ref{genEinstein}),
\begin{eqnarray}
\label{contractedBianchi}
\tG^\sigma_{\nu;\,\sigma}
=\left(\frac{\lower2pt\hbox{$3$}}{2}f^{\sigma\rho}\,\tR_{[\sigma\rho,\nu]}
+f^{\alpha\sigma}{_{\!;\alpha}}\tR_{[\sigma\nu]}\right)\!\rmt\Lambda_b^{\!-1/2}.
\end{eqnarray}
The identity was originally derived\cite{EinsteinBianchi,SchrodingerIII}
assuming $j^\nu\!=\!0$ in (\ref{JSconnection}).
%and it was later expressed in terms of the metric (\ref{gdef}) by \cite{Kursunoglu,Hely,Treder57}.
The derivation for $j^\nu\!\ne\!0$ was first done\cite{Antoci3}
by applying an infinitesimal coordinate transformation to an invariant integral,
and it is also done in Appendix B of \cite{sShifflett} using a much different direct computation method.
Clearly (\ref{niceform},\ref{contractedBianchi}) are generalizations of
the ordinary contracted Bianchi identity
$2(\rmg\,R^\nu{_\lambda})_{,\nu}\!-\!\rmg\,g^{\nu\sigma}\!R_{\sigma\nu,\lambda}\!=\!0$
or $G^\sigma_{\nu;\sigma}\!=\!0$,
which is also valid in this theory.

Another useful identity\cite{Kursunoglu} can be derived
%is derived in Appendix \ref{UsefulIdentity}
using only the definitions (\ref{gdef},\ref{fdef})
% of $g_{\mu\nu}$ and $f_{\mu\nu}$,
\begin{eqnarray}
\label{usefulidentity}
\left(N^{(\mu}{_{\nu)}} \!-\!\frac{\lower2pt\hbox{$1$}}{2}\delta^\mu_\nu
N^\rho_\rho\right)\!{_{;\,\mu}}
=\left(\frac{\lower2pt\hbox{$3$}}{2}f^{\sigma\rho}N_{[\sigma\rho,\nu]}
+f^{\sigma\rho}{_{;\sigma}}N_{[\rho\nu]}\right)\!\rmt\Lambda_b^{\!-1/2}.
\end{eqnarray}

The ordinary Lorentz force equation results from
taking the divergence of the Einstein equations (\ref{Einstein}) using
(\ref{contractedBianchi},\ref{Ampere},\ref{JSantisymmetric},\ref{usefulidentity},\ref{Fdef})
\begin{eqnarray}
\label{divergence}
T^\sigma_{\nu;\,\sigma}
&=&\!\frac{1}{8\pi}\left[\tG^\sigma_{\nu;\,\sigma}
+\Lambda_b\!\left(N^{(\mu}{_{\nu)}}
\!-\!\frac{\lower2pt\hbox{$1$}}{2}\delta^\mu_\nu N^\rho_\rho\right)\!{_{;\,\mu}}\right]
%&=&\!\left(4\pi j^\sigma\tR_{[\sigma\nu]}
%\!-\!\Lambda_b\frac{\lower2pt\hbox{$3$}}{2}f^{\sigma\rho}N_{[\sigma\rho,\nu]}\right)\!\rmt\Lambda_b^{\!-1/2}
%\!+\!\Lambda_b\!\left(N^{(\mu}{_{\nu)}}
%\!-\!\frac{\lower2pt\hbox{$1$}}{2}\delta^\mu_\nu N^\rho_\rho\right)\!{_{;\,\mu}}~~~\\
%&=&(4\pi j^\sigma\tR_{[\sigma\nu]}
%+\Lambda_b f^{\rho\sigma}{_{;\rho}}N_{[\sigma\nu]})\rmt\Lambda_b^{\!-1/2}\\
%&=&4\pi j^\sigma(\tR_{[\sigma\nu]}
%+\Lambda_bN_{[\sigma\nu]})\rmt\Lambda_b^{\!-1/2}\\
%&=&16\pi j^\sigma A_{[\sigma,\nu]},\\
\label{Euler}
%\!\!\!\!T^\sigma_{\nu;\,\sigma}
=F_{\nu\sigma}j^\sigma.
\end{eqnarray}
Note that the covariant derivatives in
(\ref{contractedBianchi},\ref{usefulidentity},\ref{Euler}) are all done using the
Christoffel connection (\ref{Christoffel}) formed from the symmetric metric (\ref{gdef}).

\section{\label{EIHEquationsOfMotion}The Einstein-Infeld-Hoffmann Equations of motion}
%Here we derive the Lorentz force from the theory using the
%Einstein-Infeld-Hoffmann (EIH) method\cite{EinsteinInfeld}.
For Einstein-Maxwell theory, the EIH method allows the equations of
motion to be derived directly from the electro-vac field
equations. For neutral particles the method has been verified to
Post-Newtonian order\cite{EinsteinInfeld}, and in fact it was the
method first used to derive the Post-Newtonian equations of
motion\cite{EIH}. For charged particles the method has been
verified to Post-Coulombian
order\cite{Wallace,WallaceThesis,Gorbatenko}, meaning that it
gives the same result as the Darwin Lagrangian\cite{Jackson}.
%The EIH method is valuable because it does not require any additional
%assumptions, such as the postulate that neutral particles follow
%geodesics, or the {\it ad hoc} inclusion of matter terms in the
%Lagrangian density.
%When the EIH method was applied to the
%original Einstein-Schr\"{o}dinger theory, no Lorentz force was
%found between charged particles\cite{Callaway,Infeld}. The basic
%difference between our case and \cite{Callaway,Infeld} is that our
%Einstein equations (\ref{Einstein2}) contain the familiar term
%$f_\nu{^\sigma}f_{\sigma\mu}\!-\!(1/4)g_{\nu\mu}f^{\rho\sigma}f_{\sigma\rho}$.
%This term appears because we assumed $\Lambda_b\ne 0$, and because
%of our metric definition (\ref{gdef}) and (\ref{approximateNbar}).
%With this term, the EIH method predicts the same Lorentz force as
%it does for electro-vac Einstein-Maxwell theory. Also, it happens
%that the extra terms in our approximate Einstein and Maxwell
%equations due to (\ref{tGminusG},\ref{threeparts}) cause no
%contribution beyond the Lorentz force, to Newtonian/Coulombian
%order. The basic reason for the null result of
%\cite{Callaway,Infeld} is that they assumed $\Lambda_b\!=\!0$ and
%$g_{\mu\nu}\!=\!N_{(\mu\nu)}$, so that every term in their
%effective energy-momentum tensor has ``extra
%derivatives''\cite{Voros}. For the same reason that
%\cite{Callaway,Infeld} found no Lorentz force, the extra
%derivative terms in (\ref{tGminusG})
%cause no contribution to the equations of motion.
In \S\ref{LorentzForce} we derived the exact Lorentz force equation
for this theory by including source terms in the Lagrangian.
Here we derive the Lorentz force using
the EIH method because it requires no source terms,
and also to show definitely that the well known negative
result of \cite{Callaway,Infeld} for the unmodified
Einstein-Schr\"{o}dinger theory does not apply to the present
theory. We will only cover the bare essentials of the EIH method
which are necessary to derive the Lorentz force.
%and the references above should be consulted for a more complete explanation.
We will also only calculate the equations of motion
to Newtonian/Coulombian order, because this is the order where the
Lorentz force first appears.

\iffalse
With the EIH method, one does not just find equations of motion,
but rather one finds approximate solutions $g_{\mu\nu}$ and
$f_{\mu\nu}$ of the field equations which correspond to a system
of two or more particles. These approximate solutions will in
general contain $1/r^p$ singularities, and these are considered to
represent particles. It happens that acceptable solutions to the
field equations can only be found if the motions of these
singularities obey certain equations of motion.
%The assumption is that these approximate solutions for
%$g_{\mu\nu}$ and $f_{\mu\nu}$ should approach exact solutions
%asymptotically, and therefore the motions of the singularities
%should approximate the motions of exact solutions.
Any event horizon or other unusual feature of exact solutions at
small radii is irrelevant because the singularities are assumed to be
separated by much larger distances, and because the method relies
on surface integrals done at large distances from the
singularities. Some kind of exact Reissner-Nordstr\"{o}m-like
solution should probably exist in order for the EIH method to make
sense, and the electric monopole solution in \S\ref{Monopole}
fills this role in our case. However, exact solutions are really
only used indirectly to identify constants of integration.
\fi

The EIH method assumes the ``slow motion approximation'', meaning
that $v/c\!\ll\!1$. The fields are expanded in the
form\cite{EinsteinInfeld,Wallace,WallaceThesis,Gorbatenko},
\begin{eqnarray}
g_{\mu\nu}&=&\eta_{\mu\nu}+\gamma_{\mu\nu}\!-\eta_{\mu\nu}\eta^{\sigma\rho}\gamma_{\sigma\rho}/2,\\
\label{g00expansion}
\gamma_{00}&=&{_2}\gamma_{00}\lambda^2+{_4}\gamma_{00}\lambda^4\dots\\
\label{g0kexpansion}
\gamma_{0k}&=&{_3}\gamma_{0k}\lambda^3+{_5}\gamma_{0k}\lambda^5\dots\\
\label{gikexpansion}
\gamma_{ik}&=&{_4}\gamma_{ik}\lambda^4\dots\\
\label{A0expansion}
A_{0}&=&{_2}A_{0}\lambda^2+{_4}A_{0}\lambda^4\dots\\
\label{Akexpansion}
A_{k}&=&{_3}A_{k}\lambda^3+{_5}A_{k}\lambda^5\dots\\
\label{f0kexpansion}
f_{0k}&=&{_2}f_{0k}\lambda^2+{_4}f_{0k}\lambda^4\dots\\
\label{fikexpansion}
f_{ik}&=&{_3}f_{ik}\lambda^3+{_5}f_{ik}\lambda^5\dots
\end{eqnarray}
where $\lambda\sim v/c$ is the expansion parameter, the order of
each term is indicated with a left subscript\cite{Callaway},
$\eta_{\mu\nu}={\rm diag}(1,-1,-1,-1)$, and Latin indices run from
1-3. The field $\gamma_{\mu\nu}$ (often called $\bar h_{\mu\nu}$
in other contexts) is used instead of $g_{\mu\nu}$ only because it
simplifies the calculations.
%\begin{eqnarray}
%\gamma_{\mu\nu}=h_{\mu\nu}\!-\eta_{\mu\nu}\eta^{\sigma\rho}h_{\sigma\rho}/2,~~~~~
%g_{\mu\nu}=\eta_{\mu\nu}+h_{\mu\nu},
%~~~~~h_{\mu\nu}=\gamma_{\mu\nu}\!-\eta_{\mu\nu}\eta^{\sigma\rho}\gamma_{\sigma\rho}/2,\\
%g_{\mu\nu}=\eta_{\mu\nu}+\gamma_{\mu\nu}\!-\eta_{\mu\nu}\eta^{\sigma\rho}\gamma_{\sigma\rho}/2,\\
%\eta_{\mu\nu}=({\rm Minkowski~metric})={\rm diag}(1,-1,-1,-1).
%\end{eqnarray}
Because $\lambda\sim v/c$, when the expansions are substituted
into the Einstein and Maxwell equations, a time derivative counts
the same as one higher order in $\lambda$. The general procedure
is to substitute the expansions, and solve the resulting field
equations order by order in $\lambda$, continuing to higher orders
until a desired level of accuracy is achieved. At each order in
$\lambda$, one of the ${_l}\gamma_{\mu\nu}$ terms and one of the
${_l}f_{\mu\nu}$ terms will be unknowns, and the equations will
involve known results from previous orders because of the
nonlinearity of the Einstein equations.

The expansions (\ref{g00expansion}-\ref{fikexpansion}) use only
alternate powers of $\lambda$ essentially because the Einstein and
Maxwell equations are second order differential
equations\cite{EIH}, although for higher powers of $\lambda$, all
terms must be included to predict
radiation\cite{Wallace,WallaceThesis,Gorbatenko}. Because
$\lambda\!\sim\!v/c$, the expansions have the magnetic components
$A_{k}$ and $f_{ik}$ due to motion at one order higher in
$\lambda$ than the electric components $A_0$ and $f_{0i}$. As in
\cite{Wallace,WallaceThesis,Gorbatenko}, $f_{0k}$ and $f_{ik}$
have even and odd powers of $\lambda$ respectively. This is the
opposite of \cite{Callaway,Infeld} because we are assuming a
direct definition of the electromagnetic field
(\ref{fdef},\ref{approximateNhat},\ref{threeparts},\ref{Fdef})
instead of the dual definition
$f^{\alpha\rho}=\varepsilon^{\alpha\rho\sigma\mu}N_{[\sigma\mu]}/2$
assumed in \cite{Callaway,Infeld}.

The field equations are assumed to be of the standard form
\begin{eqnarray}
\label{EIHG0}
{\rm ~~~~~~} G_{\mu\nu}=8\pi T_{\mu\nu} ~~{\rm where}~~
G_{\mu\nu}=R_{\mu\nu}-\frac{\lower2pt\hbox{$1$}}{2}g_{\mu\nu}g^{\alpha\beta}R_{\alpha\beta},\\
%\end{eqnarray}
%or equivalently
%\begin{eqnarray}
\label{EIHR0}
{\rm or~~~~} R_{\mu\nu}=8\pi S_{\mu\nu} ~~{\rm where}~~
\,S_{\mu\nu}=\,T_{\mu\nu}-\frac{\lower2pt\hbox{$1$}}{2}g_{\mu\nu}g^{\alpha\beta}\,T_{\alpha\beta}.
\end{eqnarray}
However, with the EIH method we solve a sort of
quasi-Einstein equations,
\begin{eqnarray}
\label{quasi}
0&=&\breve G_{\mu\nu}-8\pi \breve T_{\mu\nu},
\end{eqnarray}
where
\begin{eqnarray}
\breve G_{\mu\nu}&=&R_{\mu\nu}-\frac{\lower2pt\hbox{$1$}}{2}\eta_{\mu\nu}\eta^{\alpha\beta}R_{\alpha\beta},
\label{EIHT}
~~~~~~~~~~~\breve T_{\mu\nu}=S_{\mu\nu}-\frac{\lower2pt\hbox{$1$}}{2}\eta_{\mu\nu}\eta^{\alpha\beta}S_{\alpha\beta}.
\end{eqnarray}
Here the use of $\eta_{\mu\nu}$ instead of $g_{\mu\nu}$ is not an
approximation because (\ref{EIHR0}) implies (\ref{quasi}) whether
$\breve G_{\mu\nu}$ and $\breve T_{\mu\nu}$ are defined with
$\eta_{\mu\nu}$ or $g_{\mu\nu}$. Note that the references use many
different notations in (\ref{quasi}): instead of $\breve G_{\mu\nu}$
others use $\Pi_{\mu\nu}/2\!+\!\Lambda_{\mu\nu}$,
$\Phi_{\mu\nu}/2\!+\!\Lambda_{\mu\nu}$ or $[{\rm LS}\!:\!\mu\nu]$
and instead of $8\pi\breve T_{\mu\nu}$ others use $-2S_{\mu\nu}$,
$-\Lambda'_{\mu\nu}$, $-\Lambda_{\mu\nu}$ or $[{\rm RS}\!:\!\mu\nu]$.

The equations of motion result as a condition that the field
equations (\ref{quasi}) have acceptable solutions. In the language
of the EIH method, acceptable solutions are those that contain
only ``pole'' terms and no ``dipole'' terms, and this can be
viewed as a requirement that the solutions should resemble
Reissner-Nordstr\"{o}m solutions asymptotically. To express the
condition of solvability we must consider the integral of the
field equations (\ref{quasi}) over 2D surfaces $S$ surrounding
each singularity,
\begin{eqnarray}\
\label{integral} {_l}C_\mu=\frac{1}{2\pi}\int^S(\,{_l}\breve G_{\mu k}-8\pi\,{_l}\breve T_{\mu k})n_kdS.
\end{eqnarray}
Here $n_k$ is the surface normal and $l$ is the order in
$\lambda$. Assuming that the divergence of the Einstein equations
(\ref{EIHG0}) vanishes, and that (\ref{quasi}) has been solved to
all previous orders, it can be shown\cite{EinsteinInfeld} that in
the current order
\begin{eqnarray}
\label{zerodivergenceEIH} (\,{_l}\breve G_{\mu k}-8\pi\,{_l}\breve T_{\mu k})_{|k}\!=\!0.
\end{eqnarray}
Here and throughout this section ``$|$'' represents ordinary
derivative\cite{EinsteinInfeld}. From Green's theorem,
(\ref{zerodivergenceEIH}) implies that ${_l}C_\mu$ in
(\ref{integral}) will be independent of surface size and
shape\cite{EinsteinInfeld}.
%because of Gauss's divergence theorem.
%However, $C_\mu$ will depend on the motion of any singularity
%at the center of the surface.
The condition for the existence of an acceptable solution for
${_4}\gamma_{ik}$ is simply
\begin{eqnarray}
\label{C} {_4}C_i=0,
\end{eqnarray}
and these are also our three $\ord(\lambda^4)$ equations of
motion\cite{EinsteinInfeld}. The $C_0$ component of
(\ref{integral}) causes no constraint on the
motion\cite{EinsteinInfeld}
so we only need to calculate $\breve G_{ik}$ and $\breve T_{ik}$.

At this point let us introduce a Lemma from \cite{EinsteinInfeld}
which is derived from Stokes's theorem. This Lemma states that
\begin{eqnarray}
\label{Lemma} \int^S \mathcal{F}_{(\cdots)kl|l}n_kdS=0 ~~~{\rm
if}~~~\mathcal{F}_{(\cdots)kl}=-\mathcal{F}_{(\cdots)lk},
\end{eqnarray}
where  $\mathcal{F}_{(\cdots)kl}$ is any antisymmetric function of
the coordinates, $n_k$ is the surface normal, and $S$ is any
closed 2D surface which may surround a singularity. The equation
${_4}C_i\!=\!0$ is a condition for the existence of a solution for
${_4}\gamma_{ik}$ because ${_4}\gamma_{ik}$ is found by solving
the $\ord(\lambda^4)$ field equations (\ref{quasi}), and ${_4}C_i$
is the integral (\ref{integral}) of these equations. However,
because of the Lemma (\ref{Lemma}), it happens that the
${_4}\gamma_{ik}$ terms in ${_4}\breve G_{ik}$ integrate to zero
in (\ref{integral}), so that ${_4}C_i$ is actually independent of
${_4}\gamma_{ik}$. In fact it is a general rule that $C_i$ for one
order can be calculated using only results from previous
orders\cite{EinsteinInfeld}, and this is a crucial aspect of the
EIH method. Therefore, the calculation of the $\ord(\lambda^4)$
equations of motion (\ref{C}) does not involve the calculation of
${_4}\gamma_{ik}$, and we will see below that it also does not
involve the calculation of ${_3}f_{ik}$ or ${_4}f_{0k}$.

The ${_4}\breve G_{ik}$ contribution to (\ref{integral}) is
derived in \cite{EinsteinInfeld}. For two particles with masses
$m_1$, $m_2$ and positions $\xi^i_1$, $\xi^i_2$, the
$\ord(\lambda^4)$ term from the integral over the first particle
is
\begin{eqnarray}
\label{neutral}
{^{\breve G}_{\,4}}C_i&=&\frac{1}{2\pi}\int^1{_4}\breve G_{ik}n_kdS
=-4\left\{m_1\ddot{\xi}_1^i
-m_1m_2\frac{\partial}{\partial\xi_1^i}\left(\frac{1}{r}\right)\right\},~~~~
\end{eqnarray}
where
\begin{eqnarray}
%{\rm where~~~~~~~}
r&=&\sqrt{\lower1pt\hbox{$(\xi_1^s-\xi_2^s)(\xi_1^s-\xi_2^s)$}}\,.
\end{eqnarray}
If there is no other contribution to (\ref{integral}), then
(\ref{C}) requires that ${^{\breve G}_{\,4}}C_i\!=\!0$ in
(\ref{neutral}), and the particle acceleration will be
proportional to a $\nabla(m_1m_2/r)$ Newtonian gravitational
force. These are the EIH equations of motion for vacuum general
relativity to $\ord(\lambda^4)$, or Newtonian order.

Because our effective energy momentum tensor (\ref{tildeT}) is
quadratic in $f_{\mu\nu}$, and the expansions
(\ref{g00expansion}-\ref{fikexpansion}) begin with $\lambda^2$
terms, the $\ord(\lambda^2)-\ord(\lambda^3)$ calculations leading
to (\ref{neutral}) are unaffected by the addition of
the electromagnetic terms to the vacuum field equations. However, the
$8\pi\,{_4}\breve T_{ik}$ contribution to (\ref{integral}) will
add to the ${_4}\breve G_{ik}$ contribution. To calculate this
contribution, we will assume that our singularities in
$f_{\nu\mu}$ are simple moving Coulomb potentials, and that
$\theta^\rho\!=\!0$, $\Lambda\!=\!0$. Then from
(\ref{threeparts},\ref{f0kexpansion}-\ref{fikexpansion}) we see
that ${_2}F_{0k}\!=\!{_2}f_{0k}$, and from inspection of the extra
terms in our Maxwell equations (\ref{Ampere},\ref{Faraday},\ref{threeparts}) and
Proca equation (\ref{Proca}), we see that these equations are both
solved to $\ord(\lambda^3)$. Because (\ref{tildeT})
is quadratic in $f_{\mu\nu}$, we see from
(\ref{f0kexpansion}-\ref{fikexpansion}) that only ${_2}f_{0k}$ can
affect the $\ord(\lambda^4)$ equations of motion. Including only
${_2}f_{0k}$, our $f_{\mu\nu}$ is then a sum of two Coulomb
potentials with charges $Q_1$, $Q_2$ and positions $\xi^i_1$,
$\xi^i_2$ of the form
\begin{eqnarray}
\label{singularity}
{_2}\Aphi_\mu&=&({_2}\varphi,0,0,0)~~,
~~~{_2}f_{0k}=2\,{_2}\Aphi_{[k|0]}=-\,{_2}\varphi_{|k},\\
%~~~{_4}f_{0k}={_4}\dot\Aphi_{,k},
%~~~{_3}f_{ik}=2\,{_3}\Aphi_{[k|i]},\\
\label{singularity2}
{_2}\varphi\!&=&\!\psi^1+\psi^2~~~~~~,~~~\psi^1=Q_1/r_1~~~,~~~\psi^2=Q_2/r_2,\\
%~~~{_2}\varphi\!=\!\sum_{p=1}^N\psi^p,
%~~~{_3}\Aphi_k\!=\!\sum_{p=1}^N\psi^p\dot\xi_p^k,~~~\psi^p=Q_p/r_p,\\
r_a\!&=&\sqrt{\lower1pt\hbox{$(x^s-\xi_a^s)(x^s-\xi_a^s)$}}~~~~~,~~~a=1...2\,.
\end{eqnarray}

Because (\ref{tildeT}) is
quadratic in both $f_{\mu\nu}$ and $g_{\mu\nu}$, and the
expansions (\ref{g00expansion}-\ref{fikexpansion}) start at
$\lambda^2$ in both of these quantitites, no
gravitational-electromagnetic interactions will occur at
$\ord(\lambda^4)$. This allows us to replace covariant derivatives
with ordinary derivatives, and $g_{\nu\mu}$ with $\eta_{\nu\mu}$
in (\ref{tildeT}). This also allows us to replace
$\breve T_{\mu\nu}$ from (\ref{quasi},\ref{EIHT}) with (\ref{tildeT}).
Keeping only $\ord(\lambda^4)$ terms
when (\ref{singularity}) is substituted, the spacial part of (\ref{tildeT}) gives,
\begin{eqnarray}
\label{Ttemp2}
\!\!\!\!\!\!\!\!8\pi\!\!\!\!&&\!\!\!{_4}\breve T_{sm}=2\left({f_s}^0 f_{0m}
-\frac{1}{2}\eta_{sm}f^{r0}f_{0r}\right)\nonumber\\
\!\!\!\!&+&\!\left(2f^{a0}\!f_{0(s|m)}{_{|a}}
\!+\!f^0{_{s|a}}f_{0m|}{^a}\!+\!f^0{_{a|s}}f^a{_{0|m}}
\!-\!\frac{\lower1pt\hbox{1}}{2}\,\eta_{sm}(f^{r0}\!f_{0r})_|{^a}{_{|a}}\right)\!\Lambda_b^{\!-\!1}.
%\!\!\!&=&\!-2\left(f_{0s}f_{0m}
%+\frac{1}{2}\eta_{sm}f_{0r}f_{0r}\right)\nonumber\\
%\!\!\!\!&+&\!\left(2f_{0a}f_{0(s|m)}{_{|a}}
%\!-\!f{_{0s|a}}f_{0m|a}\!+\!f{_{0a|s}}f{_{0a|m}}
%\!+\!\frac{1}{2}\eta_{sm}(f_{0r}f_{0r})_{|a|a}\right)\!\Lambda_b^{\!-\!1}.
\end{eqnarray}
Note that ${_2}\varphi$ from (\ref{singularity2}) obeys Gauss's
law,
\begin{eqnarray}
\label{Gauss} \varphi_{|a|a}=0.
\end{eqnarray}
Substituting (\ref{singularity}) into (\ref{Ttemp2}) and using
(\ref{Gauss}) gives
\begin{eqnarray}
\!\!\!\!\!\!\!\!8\pi\!\!\!\!&&\!\!\!{_4}\breve T_{sm} =-2\left(\varphi_{|s} \varphi_{|m}
+\frac{1}{2}\eta_{sm}\varphi_{|r}\varphi_{|r}\right)\nonumber\\
\!\!\!&+&\left(2\varphi_{|a}\varphi_{|s|m|a}
\!-\!\varphi_{|s|a}\varphi_{|m|a}\!+\!\varphi{_{|a|s}}\varphi{_{|a|m}}
\!+\!\frac{1}{2}\eta_{sm}(\varphi_{|r}\varphi_{|r})_{|a|a}\right)\!\Lambda_b^{\!-\!1}\\
\ifnum\ExpandDerivations=1
\!\!\!&=&-2\left(\varphi_{|s} \varphi_{|m}
+\frac{1}{2}\eta_{sm}\varphi_{|r}\varphi_{|r}\right)\nonumber\\
&&-(\varphi_{|s}\varphi_{|a|m}+\varphi_{|r}\varphi_{|r|s}\eta_{am})_{|a}\Lambda_b^{\!-\!1}
\!+\!(\varphi_{|a}\varphi_{|s|m}+\varphi_{|r}\varphi_{|r|a}\eta_{sm})_{|a}\Lambda_b^{\!-\!1}\nonumber\\
\fi
\label{T}
\!\!\!&=&-2\left(\varphi_{|s} \varphi_{|m}
+\frac{1}{2}\eta_{sm}\varphi_{|r}\varphi_{|r}\right)
-2(\varphi_{|[s}\varphi_{|a]|m}+\varphi_{|r}\varphi_{|r|[s}\eta_{a]m})_{|a}\Lambda_b^{\!-\!1}.
%&=&\!(\varphi_{|[s}\varphi_{|a]|m})_{|a}\!-\!(\varphi_{|r}\varphi_{|r|[a}\eta_{s]m})_{|a}.
\end{eqnarray}
From (\ref{Lemma}), the second group of terms in
(\ref{T}) integrates to zero in (\ref{integral}), so it
can have no effect on the equations of motion. The first group of
terms in (\ref{T}) is what one gets with
Einstein-Maxwell theory\cite{Wallace,WallaceThesis,Gorbatenko}, so
at this stage we have effectively proven that the theory predicts
a Lorentz force.

For completeness we will finish the derivation. First, we see from
(\ref{T},\ref{Gauss}) that ${_4}\breve T_{sm|s}\!=\!0$. This is to
be expected because of
(\ref{zerodivergenceEIH}), and it means that the
$8\pi\,{_4}\breve T_{sm}$ contribution to the surface integral
(\ref{integral}) will be independent of surface size and shape.
This also means that only $1/{\rm distance}^2$
terms such as $\eta_{sm}/r^2$ or $x_sx_m/r^4$ can contribute to
(\ref{integral}). The integral over a term with any other
distance-dependence would depend on the surface
radius, and therefore we know beforehand that it must vanish or
cancel with other similar terms\cite{EinsteinInfeld}. Now,
$\varphi_{|i}\!=\!\psi^1_{|i}\!+\psi^2_{|i}$ from
(\ref{singularity2}). Because $\psi^1_{|i}$ and $\psi^2_{|i}$ both
go as $1/{\rm distance}^2$, but are in different locations, it is
clear from (\ref{T}) that contributions can only come from cross
terms between the two. Including only these terms gives,
\begin{eqnarray}
\label{EIHcrossterms} 8\pi\,{_4}\breve T^c_{sm}
=-2(\psi^1_{|s}\psi^2_{|m} \!+\!\psi^2_{|s}\psi^1_{|m} \!+\!
\,\eta_{sm}\psi^1_{|r}\psi^2_{|r}).
\end{eqnarray}
Some integrals we will need can be found in \cite{EinsteinInfeld}.
With $\psi=1/\sqrt{x^s x^s}$ we have,
\begin{eqnarray}
\label{EIHintegrals} &&\frac{1}{4\pi}\!\int^0
\psi_{|m}n_mdS=-1~~~,~~~ \frac{1}{4\pi}\!\int^0
\psi_{|a}n_mdS=-\frac{1}{3}\delta_{am}.
\end{eqnarray}
Using (\ref{EIHcrossterms},\ref{EIHintegrals},\ref{singularity2})
and integrating over the first particle we get,
\begin{eqnarray}
\!\!\!\!\!\!\!\frac{1}{2\pi}\int^1\left[-8\pi\breve T_{sm}\,\right]n_mdS
&=&\frac{1}{2\pi}\!\int^1\!\!2(\psi^1_{|s}\psi^2_{|m}
\!+\!\psi^2_{|s}\psi^1_{|m}\!+\eta_{sm}\psi^1_{|r}\psi^2_{|r})n_mdS~~~~\\
\label{EIHLorentz} \!&=&\!4Q_1\psi^2_{|s}(\xi_1)
\!\left(\!-\frac{\lower2pt\hbox{1}}{3}\!-\!1\!+\!\frac{\lower2pt\hbox{1}}{3}\right)
=-4Q_1\psi^2_{|s}(\xi_1).
\end{eqnarray}
Using
(\ref{C},\ref{integral},\ref{EIHLorentz},\ref{neutral},\ref{singularity2})
we get
\begin{eqnarray}
0={_4}C_i&=&-4\left\{m_1\ddot{\xi}_1^i
-m_1m_2\frac{\partial}{\partial\xi_1^i}\left(\frac{1}{r}\right)\right\}
-4Q_1\psi^2_{|i}(\xi_1)\\
\ifnum\ExpandDerivations=1
&=&-4\left\{m_1\ddot{\xi}_1^i
-m_1m_2\frac{\partial}{\partial\xi_1^i}\left(\frac{1}{r}\right)\right\}
-4Q_1\frac{\partial}{\partial\xi_1^i}\left(\frac{Q_2}{r}\right)\nonumber\\
\fi
&=&-4\left\{m_1\ddot{\xi}_1^i
-m_1m_2\frac{\partial}{\partial\xi_1^i}\left(\frac{1}{r}\right)
+Q_1Q_2\frac{\partial}{\partial\xi_1^i}\left(\frac{1}{r}\right)\right\},
\end{eqnarray}
where
\begin{eqnarray}
r&=&\sqrt{\lower1pt\hbox{$(\xi_1^s-\xi_2^s)(\xi_1^s-\xi_2^s)$}}.
\end{eqnarray}
These are the EIH equations of motion for this theory to
$\ord(\lambda^4)$, or Newtonian/ Coulombian order.
These equations clearly exhibit the Lorentz force, and in fact they
match the $\ord(\lambda^4)$ equations of motion of
Einstein-Maxwell theory.

\section{\label{Monopole}An exact electric monopole solution}
Here we give an exact charged solution for this theory
which closely approximates the Reissner-Nordstr\"{o}m solution\cite{Reissner,Nordstrom}
of Einstein-Maxwell theory.
A MAPLE program\cite{LRESMAPLE} which checks the solution
and the derivation\cite{cShifflett} are available.
\ifnum\ExpandDerivations=1
It can be shown\cite{Papapetrou} that the assumption of spherical symmetry allows the
fundamental tensor to be written in the following form
\begin{eqnarray}
N_{\nu\mu}=
  \pmatrix{
 \gamma &-w&0&0\cr
w&-\alpha &0&0\cr
0&0&-\beta &r^2v\,{\rm sin}\,\theta\cr
0&0&-r^2v\,{\rm sin}\,\theta & -\beta\,{\rm sin}^2\theta}.
\end{eqnarray}
Both \cite{Papapetrou} and \cite{Takeno} assume this form with
$\beta=r^2,v=0$ to derive a solution to the original
Einstein-Schr\"{o}dinger field equations which looks similar to a
charged mass, but with some problems. Here we will derive a solution to the
modified field equations (\ref{JSsymmetric}-\ref{JScontractionsymmetric})
which is much closer to the Reissner-Nordstr\"{o}m
solution\cite{Reissner,Nordstrom} of electro-vac Einstein-Maxwell theory.
We will follow a similar procedure to
\cite{Papapetrou,Takeno} but will use
%the opposite sign convention for the Ricci tensor (\ref{Ricci}), and
coordinates $x_0,x_1,x_2,x_3\!=\!ct,r,\theta,\phi$ instead of
$x_1,x_2,x_3,x_4\!=\!r,\theta,\phi,ct$. We also use
the variables $a\nobreak=\nobreak1/\alpha,~b=\gamma\alpha,~\sinht=-w$,
which allow a simpler solution than the variables $\alpha,\gamma,w$. This gives
\begin{eqnarray}
\label{Nmatrix}
N_{\nu\mu}&=&
 \pmatrix{
  ab&\sinht&0&0\cr
-\sinht&- 1/a&0&0\cr
0&0&- r^2 &0 \cr
0&0&0&- r^2{\rm sin}^2\theta
},\\
N^{\dashv\mu\nu}&=&
\label{Nmatrixcontra}
 \pmatrix{
  1/ad&\sinht/d&0&0\cr
-\sinht/d & \!- ab/d&0&0\cr
0&0&\!\!\!- 1/r^2 &0 \cr
0&0&0&\!\!\!\!- 1/r^2{\rm sin}^2\theta
},\\
\label{rmN}
\rmN&=&\sqrt{d}\,r^2{\rm sin}\,\theta,
\end{eqnarray}
where
\begin{eqnarray}
\label{ddef}
d=b-\sinht^2.
\end{eqnarray}
From (\ref{Nmatrixcontra},\ref{rmN})
and the definitions (\ref{gdef},\ref{fdef}) of $g_{\nu\mu}$ and $f_{\nu\mu}$ we get
\begin{eqnarray}
\!\!\!g^{\nu\mu}\!&=&\!
\frac{1}{\cosht}\!\pmatrix{
1/ad&0&0&0\cr
0&\!\!\!\!\!\!-ab/d&0&0\cr
0&0&\!\!\!\!\!-1/r^2&0\cr
0&0&0&\!\!\!\!\!\!\!-1/r^2{\rm sin}^2\theta
}\!,~
f^{\nu\mu}\!=\!
\frac{\Lambda_b^{\!1/2}}{\rmt\,\cosht }\!\pmatrix{
\!0&\!\!\!\!\!\!\!-\sinht/d&\!\!\!0&\!\!\!0\cr
\!\sinht/d&\!\!0&\!\!\!0&\!\!\!0\cr
0&\!\!0&\!\!\!0&\!\!\!0\cr
0&\!\!0&\!\!\!0&\!\!\!0
}\!,~~~~~\\
\label{gmatrix}
\!\!\!g_{\nu\mu}\!&=&\!
 \,\cosht \pmatrix{
  ad &0&0&0\cr
0&- d/ab&0&0\cr
0&0&- r^2&0\cr
0&0&0&- r^2{\rm sin}^2\theta
}\!,~
\label{fmatrix}
f_{\nu\mu}\!=\!
\frac{\Lambda_b^{\!1/2}}{\rmt\,\cosht }\!\pmatrix{
\,0&\sinht&\!0&\!0\cr
-\sinht&0&\!0&\!0\cr
0&0&\!0&\!0\cr
0&0&\!0&\!0\,
}\!,\\
\!\!\!\rmg&=&\sqrt{b}\,r^2\sin\,\theta,
\end{eqnarray}
where
\begin{eqnarray}
\label{cdef}
\cosht =\sqrt{b/d}=\rmg/{\lower1pt\hbox{$\rmN$}}\,.
\end{eqnarray}
Using prime ($'$) to represent $\partial/\partial r$, Ampere's law (\ref{Ampere})
and (\ref{Nmatrixcontra},\ref{rmN}) require that
\begin{eqnarray}
\label{Amperesolution}
&&0=(\rmN N^{\dashv\,[01]})_{,1}
=\left(\frac{\sinht r^2 sin\,\theta}{\sqrt{d}}\right)'.
\end{eqnarray}
From (\ref{Amperesolution},\ref{ddef}), this means that for some constant $Q$ we have
\begin{eqnarray}
\label{Qintro}
&&\frac{\sinht r^2}{\sqrt{d}}=\frac{\sinht r^2}{\sqrt{b-\sinht^2}}=\frac{Q\rmt}{\Lambda_b^{1/2}}.
\end{eqnarray}
Solving this for $\sinht^2$ gives
\begin{eqnarray}
\label{wsquared}
\sinht^2=\frac{2bQ^2}{2Q^2\!-\Lambda_br^4}\,.
\end{eqnarray}
From (\ref{Qintro},\ref{wsquared}) we can derive the useful relationship
\begin{eqnarray}
\sinht'&\!=\!&\frac{(\sinht^2)'}{2\sinht}
=\frac{1}{2\sinht}\!\left(\frac{2b'Q^2}{2Q^2\!-\Lambda_br^4}+\frac{8b\Lambda_br^3Q^2}{2Q^2\!-\Lambda_br^4}\left(\frac{\sinht^2}{2bQ^2}\right)\right)
%=\frac{1}{2\sinht}\!\left(b'\frac{\sinht^2}{b}
%-\frac{4b}{r}\frac{d}{\sinht^2}\frac{\sinht^4}{b^2}\right)
\label{bitch}
=\frac{\sinht}{b}\!\left(\frac{b'}{2}-\!\frac{2d}{r}\right)\!.
\end{eqnarray}
The connection equations (\ref{JSconnection})
are solved in \cite{Papapetrou,Takeno}.
In terms of our variables, the non-zero connections are
\begin{eqnarray}
&&\tGam^1_{00}=\frac{a}{2}(ab)'+\frac{4a^2\sinht^2}{r}~~,~
\tGam^0_{10}=\tGam^0_{01}
=\frac{(ab)'}{2ab}+\frac{2\sinht^2}{br}
~~,~\tGam^1_{11}=\frac{-a'}{2a},\nonumber\\
&&\tGam^2_{12}=\tGam^2_{21}
\label{finalconnections}
=\tGam^3_{13}=\tGam^3_{31}=\frac{\lower1pt\hbox{$1$}}{r},\\
&&\tGam^1_{22}=-ar~~,~\tGam^1_{33}
=-ar\,{\rm sin}^2\theta~~,~
\tGam^3_{23}=\tGam^3_{32}={\rm cot}\,\theta~~,~
\tGam^2_{33}=-{\rm sin}\,\theta {\rm cos}\,\theta,\nonumber\\
&&\tGam^2_{02}=-\tGam^2_{20}=\tGam^3_{03}
=-\tGam^3_{30}=-\frac{\lower1pt\hbox{$a\sinht$}}{r}~~,~
\tGam^1_{10}=-\tGam^1_{01}=-\frac{\lower1pt\hbox{$2a\sinht$}}{r},\nonumber\\
\label{contractedtGam}
&&\tGam^\alpha_{\alpha0}=0,~~
\tGam^\alpha_{\alpha1}=\frac{\lower1pt\hbox{$b'$}}{2b}+\frac{\lower1pt\hbox{$2\sinht^2$}}{br}+\frac{\lower1pt\hbox{$2$}}{r},~~
\tGam^\alpha_{\alpha2}={\rm cot}\,\theta,~~\tGam^\alpha_{\alpha3}=0.
\end{eqnarray}

The Ricci tensor is also calculated in \cite{Papapetrou,Takeno}.
From (\ref{contractedtGam}) we have $\tGam^\alpha_{\alpha[\nu,\mu]}\!=0$
as expected from (\ref{funnytensor}), and this means that $\tR_{\nu\mu}\!=\!\tilde R_{\nu\mu}$.
In terms of our variables, and using our own sign convention,
the non-zero components of the Ricci tensor are
\begin{eqnarray}
\label{R00}
-\tR_{00}&=&-\frac{aba''}{2}
-\frac{a^2b''}{2}-\frac{3aa'b'}{4}+\frac{a^2b'b'}{4b}
-\frac{a}{r}(ab'+a'b)-\frac{8a^2\sinht\sinht'}{r}\nonumber\\
&&+\frac{a^2\sinht^2}{r}\left(\frac{3b'}{b}
\!-\!\frac{3a'}{a}\!-\!\frac{10}{r}\!+\!\frac{8\sinht^2}{br}\right),\\
\label{R11}
-\tR_{11}&=&\frac{a''}{2a}
+\frac{b''}{2b}-\frac{b'b'}{4b^2}+\frac{3a'b'}{4ab}+\frac{a'}{ar}
+\frac{4\sinht\sinht'}{br}+\frac{\sinht^2}{br}\left(\frac{3a'}{a}
\!+\!\frac{4\sinht^2}{br}\!-\!\frac{2}{r}\right),\\
\label{R22}
-\tR_{22}
&=&\frac{ar}{2}\left(\frac{2a'}{a}+\frac{b'}{b}\right)
+a-1+\frac{2a\sinht^2}{b},\\
\label{R33}
-\tR_{33}
\!&=&\!-\tR_{22}\,{\rm sin}^2\theta,\\
\label{R10}
-\tR_{[10]}
&=&2\left(\frac{a\sinht}{r}\right)'+\frac{6a\sinht}{r^2}.
~~\left\{{\cite{Papapetrou}~has~an~error~here}\right\}
\end{eqnarray}
From (\ref{Nmatrix},\ref{gmatrix},\ref{cdef},\ref{R33}),
the symmetric part of the field equations (\ref{JSsymmetric}) is
\begin{eqnarray}
\label{Rb00}
0&=&\tR_{00}+ \Lambda_bN_{00}+ \Lambda_zg_{00}
=\tR_{00}+\Lambda_bab+\Lambda_z\frac{\lower2pt\hbox{$ab$}}{\cosht }~,\\
\label{Rb11}
0&=&\tR_{11}+ \Lambda_bN_{11}+ \Lambda_zg_{11}
=\tR_{11}-\Lambda_b\frac{\lower2pt\hbox{$1$}}{a}-\Lambda_z\frac{\lower2pt\hbox{$1$}}{a\cosht}~,\\
\label{Rb22}
0&=&\tR_{22}+ \Lambda_bN_{22}+ \Lambda_zg_{22}
=\tR_{22}-\Lambda_br^2-\Lambda_z\cosht r^2,\\
\label{Rb33}
0&=&\tR_{33}+ \Lambda_bN_{33}+ \Lambda_zg_{33}
=(\tR_{22}+ \Lambda_bN_{22}+ \Lambda_zg_{22})\,{\rm sin}^2\theta.\frac{\vphantom{|}}{\vphantom{|}}
\end{eqnarray}
Forming a linear combination of (\ref{Rb11},\ref{Rb00}) and using
(\ref{R11},\ref{R00},\ref{bitch},\ref{ddef}), we find that many of
the terms cancel initially and we get,
\begin{eqnarray}
0&=&b\left(-\tR_{11}+\Lambda_b\frac{1}{a}
+\Lambda_z\frac{1}{a\cosht}\right)
+\frac{1}{a^2}\left(-\tR_{00}
-\Lambda_bab-\Lambda_z\frac{ab}{\cosht }\right)\\
\ifnum\ExpandDerivations=1
&=&\frac{4\sinht\sinht'}{r}
+\frac{\sinht^2}{r}\left(\frac{4\sinht^2}{br}-\frac{2}{r}\right)
-\frac{b'}{r}
-\frac{8\sinht\sinht'}{r}
+\frac{\sinht^2}{r}\left(\frac{3b'}{b}-\frac{10}{r}
+\frac{8\sinht^2}{br}\right)\nonumber\\
&=&-\frac{4\sinht}{r}\left[\frac{\sinht}{b}\left(\frac{b'}{2}
-\frac{2d}{r}\right)\right]
+\frac{12\sinht^2}{r}\left(\frac{\sinht^2}{br}-\frac{1}{r}\right)
-\frac{b'}{r}+\frac{3\sinht^2b'}{br}\nonumber\\
\fi
\label{1100}
&=&-\frac{d}{br^2}\left(4\sinht^2+rb'\right).
\end{eqnarray}
From (\ref{wsquared}) this requires
\begin{eqnarray}
\label{thing}
0=\frac{8bQ^2}{2Q^2\!-\Lambda_br^4}+rb'.
\end{eqnarray}
Solving (\ref{thing}) and using (\ref{wsquared},\ref{ddef},\ref{cdef})
gives identical results to \cite{Papapetrou,Takeno},
\begin{eqnarray}
\label{b}
b&=&1-\frac{2Q^2}{\Lambda_br^4},\\
\label{w}
\sinht&=&\sqrt{\frac{2bQ^2}{2Q^2\!-\Lambda_br^4}}=\frac{\rmt Q}{\sqrt{\Lambda_b}\,r^2},\\
\label{d}
d&=&b-\sinht^2=1,\\
\label{c}
\cosht &=&\sqrt{b/d}=\sqrt{1-\frac{2Q^2}{\Lambda_br^4}}\,.
\end{eqnarray}

To find the variable ``$a$'', the 22 component of the field equations will
be used. The solution is guessed to be that of \cite{Papapetrou,Takeno}
plus an extra term $-\Lambda_z\extra/r$,
\begin{eqnarray}
\label{adef}
a=1-\frac{2M}{r}-\frac{\Lambda_br^2}{3}-\frac{\Lambda_z\extra}{r}.
\end{eqnarray}
Because ``$b$'' and ``$\sinht$'' are the same as \cite{Papapetrou,Takeno},
we just need to look at the extra terms that result from $\Lambda_z$.
Using (\ref{Rb22},\ref{R22},\ref{adef},\ref{1100},\ref{c}) gives,
\begin{eqnarray}
\!\!\!\!\!\!0&=&-\tR_{22}+\Lambda_br^2+\Lambda_z\cosht r^2
=\frac{ar}{2}\!\left(\!\frac{2a'}{a}+\!\frac{b'}{b}\right)
\!+a\!-\!1\!+\!\frac{2a\sinht^2}{b}\!+\Lambda_br^2\!+\Lambda_z\cosht r^2~~~~\\
&=&-\Lambda_z\!\left[r\left(\frac{\extra}{r}\right)'
+\frac{\extra b'}{2b}
+\frac{\extra}{r}+\frac{2\extra \sinht^2}{rb}-\cosht r^2\right]
%&=&-\Lambda_z\!\left[\extra'+\frac{\extra}{2br}(b'r+4\sinht^2)
%-\cosht r^2\right]
\label{Veq}
=-\Lambda_z\left[\,\extra'-r^2\cosht\,\right].
\end{eqnarray}
This same equation is also obtained if the 11 or 00 components of the field
equations are used.
The solution for $V(r)$ can be written in terms of an elliptic integral
but we will not need to calculate it.
With (\ref{Veq}) and the definition
\begin{eqnarray}
\label{Vhatdef}
\hat V&=&\frac{r\Lambda_b}{Q^2}\left(V-\frac{r^3}{3}\right)
\end{eqnarray}
we get the following results which will be used shortly,
\begin{eqnarray}
\label{Vhatidentity}
\hat V'&=&\frac{\hat V}{r}+\frac{r^3\Lambda_b(\cosht-1)}{Q^2},~~~~~~~~~~~~~
\frac{Q^2}{\Lambda_br}\left(\frac{{\lower1pt\hbox{$\hat V$}}}{r^2}\right)'
=\cosht-1-\frac{Q^2\hat V}{\Lambda_br^4}.
\end{eqnarray}
\iffalse
%\ifnum\ExpandDerivations=1
The 11 component is,
\begin{eqnarray}
0&=&a\left(-\tR_{11}+\Lambda_b\frac{1}{a}+\Lambda_z\frac{1}{a\cosht}\right)\\
&=&\frac{a''}{2}+\frac{ab''}{2b}-\frac{ab'b'}{4b^2}+\frac{3a'b'}{4b}+\frac{a'}{r}
+\frac{4a\sinht\sinht'}{br}+\frac{\sinht^2}{br}\left(3a'+\frac{4a\sinht^2}{br}
-\frac{2a}{r}\right)+\Lambda_b+\frac{\Lambda_z}{\cosht}\\
&=&-\frac{\Lambda_z}{2}\left(\frac{V}{r}\right)''-\frac{\Lambda_zV b''}{2rb}
+\frac{\Lambda_zV b'b'}{4rb^2}-\frac{3\Lambda_zb'}{4b}\left(\frac{V}{r}\right)'
-\frac{\Lambda_z}{r}\left(\frac{V}{r}\right)'-\frac{4\Lambda_zV\sinht\sinht'}{br^2}\nonumber\\
&&+\frac{\sinht^2}{br}\left(-3\Lambda_z\left(\frac{V}{r}\right)'
-\frac{4\Lambda_zV\sinht^2}{br^2}+\frac{2\Lambda_zV}{r^2}\right)+\frac{\Lambda_z}{\cosht}\\
&=&-\Lambda_z\left[\frac{1}{2}\left(\frac{V'}{r}-\frac{V}{r^2}\right)'
+\frac{1}{r}\left(\frac{V'}{r}-\frac{V}{r^2}\right)-\frac{1}{\cosht}\right]\\
&=&-\Lambda_z\left[\frac{1}{2r}(r^2\cosht)'-\frac{1}{\cosht}\right]\\
&=&-\Lambda_z\left[\frac{1}{2r}\left(2r\cosht+\frac{r^2}{2\cosht}\left(\frac{8Q^2}{\Lambda_br^5}\right)\right)
-\frac{1}{\cosht}\right]\\
&=&-\Lambda_z\left[\frac{1}{\cosht}(1+\sinht^2-\sinht^2)-\frac{1}{\cosht}\right]\\
&=& 0.
\end{eqnarray}
\fi
Next we consider the antisymmetric part of the field equations (\ref{JSantisymmetric}),
where only the 10 component is non-vanishing. Using
(\ref{R10},\ref{Nmatrix},\ref{w},\ref{adef}) gives
\begin{eqnarray}
F_{01}&=&\frac{\Lambda_b^{\!-1/2}}{\rmt}(\tR_{[01]}+\Lambda_b N_{[01]})
=\frac{\Lambda_b^{\!-1/2}}{\rmt}\left[2\left(\frac{a\sinht}{r}\right)'
+\frac{6a\sinht}{r^2}+\Lambda_b\sinht\right]\\
&=&2\left(\frac{aQ}{\Lambda_br^3}\right)'+\frac{6aQ}{\Lambda_br^4}
+\frac{Q}{r^2}
\label{phipreliminary}
=\frac{Q}{r^2}\left(1+\frac{2a'}{\Lambda_br}\right)
\end{eqnarray}
Using
(\ref{gmatrix},\ref{b},\ref{w},\ref{d},\ref{c},\ref{adef},\ref{Veq},\ref{phipreliminary},\ref{Vhatdef},\ref{Vhatidentity})
we can put the solution in its final form.
\fi
The solution is
\begin{eqnarray}
\label{finalg}
ds^2&=&\cosht a dt^2-\frac{1}{\cosht a}\,dr^2-\cosht r^2 d\theta^2-\cosht r^2 sin^2\theta d\phi^2,\\
\label{finalf}
f^{10}&=&\frac{Q}{\cosht r^2},~~~~\rmN=r^2 sin\,\theta,~~~~\rmg=\cosht r^2 sin\,\theta,\\
\label{finalF}
F_{01}&=&-A_0'
=\frac{Q}{r^2}\!\left[1+\frac{4M}{\Lambda_br^3}-\frac{4\Lambda}{3\Lambda_b}
+2\left(\cosht-1-\frac{Q^2{\lower1pt\hbox{$\hat V$}}}{\Lambda_br^4}\right)\!\left(1-\frac{\Lambda}{\Lambda_b}\right)\right],\\
\label{a}
a&=&1-\frac{2M}{r}-\frac{\Lambda r^2}{3}
+\frac{Q^2\hat V}{r^2}\left(1-\frac{\Lambda}{\Lambda_b}\right),
\end{eqnarray}
where ($'$) means $\partial/\partial r$, and $\cosht$ and $\hat V$ are very close to one for ordinary radii,
\begin{eqnarray}
\label{finalc}
\!\!\!\!\!\cosht &=&\sqrt{1-\frac{2Q^2}{\Lambda_br^4}}
=1-\frac{Q^2}{\Lambda_br^4}\cdots-\frac{(2i)!}{[i!]^2 4^i(2i\!-\!1)}\!\left(\!\frac{2Q^2}{\Lambda_br^4}\!\right)^i,\,~~~\\
\label{Vhat}
\!\!\!\!\!\!\!\hat V&=&\frac{r\Lambda_b}{Q^2}\!\left(\int r^2\cosht\,dr -\frac{r^3}{3}\right)
=1+\frac{Q^2}{10\Lambda_br^4}\cdots +\frac{(2i)!}{i!(i\!+\!1)!\,4^i(4i\!+\!1)}\!\left(\!\frac{2Q^2}{\Lambda_br^4}\!\right)^i\!\!,~~~
\end{eqnarray}
and the nonzero connections are
\begin{eqnarray}
&&\tGam^1_{00}=\frac{aa'\cosht^2}{2}-\frac{4a^2Q^2}{\Lambda_b\,r^5}~~,~
\tGam^0_{10}=\tGam^0_{01}
=\frac{a'}{2a}
~~,~\tGam^1_{11}=\frac{-a'}{2a},\nonumber\\
&&\tGam^2_{12}=\tGam^2_{21}
=\tGam^3_{13}=\tGam^3_{31}=\frac{\lower1pt\hbox{$1$}}{r},\\
&&\tGam^1_{22}=-ar~~,~\tGam^1_{33}
=-ar\,{\rm sin}^2\theta~~,~
\tGam^3_{23}=\tGam^3_{32}={\rm cot}\,\theta~~,~
\tGam^2_{33}=-{\rm sin}\,\theta {\rm cos}\,\theta,\nonumber\\
&&\tGam^2_{02}=-\tGam^2_{20}=\tGam^3_{03}
=-\tGam^3_{30}=-\frac{a\rmt Q}{\sqrt{\Lambda_b}\,r^3}~~,~
\tGam^1_{10}=-\tGam^1_{01}=-\frac{2a\rmt Q}{\sqrt{\Lambda_b}\,r^3}\nonumber.
\end{eqnarray}

The solution matches the Reissner-Nordstr\"{o}m solution except for terms which
are negligible for ordinary radii.
To see this, first recall that $\Lambda/\Lambda_b\!\sim\! 10^{-122}$ from (\ref{Lambdadef},\ref{Lambdab}),
so the $\Lambda$ terms are all extremely tiny.
Ignoring the $\Lambda$ terms and keeping only the ${\mathcal O}(\Lambda_b^{-1})$ terms in
(\ref{finalF},\ref{a},\ref{finalc},\ref{Vhat}) gives
\begin{eqnarray}
\label{Fapprox}
F_{01}&=&
\frac{Q}{r^2}\!\left[1+\frac{4M}{\Lambda_br^3}-\frac{4Q^2}{\Lambda_br^4}\right]+\ord(\Lambda_b^{\!-2}),\\
\label{A0approx}
A_{0}
%&=&-\int\!F_{01}dr
&=&\frac{Q}{r}\!\left[1+\frac{M}{\Lambda_br^3}-\frac{4Q^2}{5\Lambda_br^4}\right]+\ord(\Lambda_b^{\!-2}),\\
\label{aapprox}
~~a&=&1-\frac{2M}{r}+\frac{Q^2}{r^2}\!\left[1+\frac{Q^2}{10\Lambda_br^4}\right]+\ord(\Lambda_b^{\!-2}),\\
\label{capprox}
~~\cosht &=&1-\frac{Q^2}{\Lambda_br^4}+\ord(\Lambda_b^{\!-2}).
\end{eqnarray}
%Also, near the event horizon of a solar mass extremal charged black hole we have from (\ref{BHskew})
%\begin{eqnarray}
%\label{ReisnerNordstromratio1}
%\frac{Q^2}{\Lambda_br^4}\sim 10^{-76},
%\end{eqnarray}
For the smallest radii probed by high-energy particle physics we get from (\ref{highenergyskew}),
\begin{eqnarray}
\label{ReisnerNordstromratio2}
\frac{Q^2}{\Lambda_br^4}\sim 10^{-66}.
\end{eqnarray}
The worst-case value of $M/\Lambda_b r^3$ might be near the Schwarzschild radius $r_s$
of black holes where $r\!=\!r_s\!=\!2M$ and $M/\Lambda_b r^3\!=\!1/2\Lambda_b r_s^2$.
This value will be largest for the lightest black holes,
and the lightest black hole that we can expect to
observe would be of about one solar mass, where we have
\begin{eqnarray}
\label{Qmratio1}
\frac{M}{\Lambda_b r^3}
&\sim&\frac{1}{2\Lambda_b r_s^2}
=\frac{1}{2\Lambda_b}\left(\frac{c^2}{2Gm_\odot}\right)^2\!\sim\!10^{-77}.
\end{eqnarray}

From (\ref{ReisnerNordstromratio2},\ref{Qmratio1},\ref{Lambdadef},\ref{Lambdab})
we see that our electric monopole solution (\ref{finalg}-\ref{a})
has a fractional difference from the Reissner-Nordstr\"{o}m solution\cite{Reissner,Nordstrom}
of at most $10^{-66}$ for worst-case radii accessible to measurement.
Clearly our solution does not have the deficiencies of the Papapetrou
solution\cite{Papapetrou,Takeno} in the original theory,
and it is almost certainly indistinguishable
from the Reissner-Nordstr\"{o}m solution experimentally.
Also, when this solution is expressed in Newman-Penrose
tetrad form, it can be shown to be of Petrov Type-D\cite{jShifflett}.
And of course the solution reduces to the Schwarzschild solution for $Q\!=0$.
And from (\ref{Fapprox}-\ref{capprox}) we see that
the solution goes to the Reissner-Nordstr\"{o}m solution exactly in the limit
as $\Lambda_b\!\rightarrow\!\infty$.

The only significant difference between our electric monopole solution and the
Reissner-Nordstr\"{o}m solution occurs on the Planck scale.
From (\ref{finalg},\ref{finalc}), the surface area of the solution is\cite{Chandrasekhar},
\begin{eqnarray}
\left({\rm surface}\atop{\rm area}\right)=\int_0^\pi\!d\theta\int_0^{2\pi}\!d\phi\sqrt{g_{\theta\theta}g_{\phi\phi}}
=4\pi r^2\cosht
=4\pi r^2\sqrt{1-\frac{2Q^2}{\Lambda_br^4}}.
\end{eqnarray}
The origin of the solution is where the surface area vanishes, so in our coordinates the origin
is not at $r\!=0$ but rather at
\begin{eqnarray}
\label{re}
r_0=\sqrt{Q}(2/\Lambda_b)^{1/4}.
%=\left(\frac{2\alpha l_P^2}{C_z\omega_c^4l_P^2}\right){\!\raise6pt\hbox{$^{1/4}$}}
%=\frac{1}{\omega_c}\left(\frac{2\alpha}{C_z}\right){\!\raise6pt\hbox{$^{1/4}$}}
%\sim 3.16\times 10^{-34}{\rm cm}.
\end{eqnarray}
From (\ref{redef},\ref{Lambdab}) we have $r_0\!\sim\!l_P\!\sim\!10^{-33}cm$ for an elementary charge,
and $r_0\!\ll\!2M$ for any realistic astrophysical black hole.
For $Q/M\!<\!1$ the behavior at the origin is hidden behind
an event horizon nearly identical to that of the Reissner-Nordstr\"{o}m solution.
For $Q/M\!>\!1$ where there is no event horizon, the behavior at the origin differs markedly
from the simple naked singularity of the Reissner-Nordstr\"{o}m solution.
For the Reissner-Nordstr\"{o}m solution all of the relevant fields have singularities at the origin,
with $g_{00}\!\sim\!Q^2/r^2$, $A_0\!=\!Q/r$, $F_{01}\!=\!Q/r^2$,
$R_{00}\!\sim\!2Q^4/r^6$ and $R_{11}\!\sim\!2/r^2$.
For our solution the metric has a less severe singularity at the origin,
with $g_{11}\!\sim\!-\sqrt{r}/\sqrt{r-r_0}$.
Also, the fields $N_{\mu\nu}$, $N^{\dashv\nu\mu}$, $\rmN$, $A_\nu$, $\rmg f^{\nu\mu}$,
$\rmg f_{\nu\mu}$, $\rmg g^{\nu\mu}$, $\rmg g_{\nu\mu}$, and the functions ``a'' and $\hat V$
all have finite nonzero values and derivatives at the origin, because it can be shown that
$\hat V(r_0)\!=\!\sqrt{2}\left[\Gamma(1/4)\right]^2\!/6\sqrt{\pi}\!-\!2/3=1.08137$.
The fields $F_{\nu\mu}$, $\tGam^\alpha_{\mu\nu}$ and $\rmg\,\tR_{\nu\mu}$ are also finite
and nonzero at the origin, so if we use the tensor density form of the field equations
(\ref{Einstein},\ref{Ampere}), there is no ambiguity as to whether the field equations
are satisfied at this location.

\section{\label{ppwaves}An exact electromagnetic plane-wave solution}
Here we give an exact electromagnetic plane-wave solution for this theory which
is identical to the electromagnetic plane-wave solution in Einstein-Maxwell theory,
usually called the Baldwin-Jeffery solution\cite{Baldwin,Misner,Stephani,Griffiths}.
We will not do a full derivation, but a MAPLE program\cite{LRESMAPLE} which checks the solution
is available.
We present the solution in the form of a pp-wave solution\cite{Stephani},
and a gravitational wave component is included for generality.
The solution is expressed in terms of null coordinates $x,~y,~u=(t-z)/\sqrt{2},~v=(t+z)/\sqrt{2}$,
\begin{eqnarray}
\label{gform}
\!\!\!\!\!g_{\mu\nu}&=&
\pmatrix{
-1\!&\!0\!&\!0\!&\!0\cr
0\!&\!-1\!&\!0\!&\!0\cr
0\!&\!0\!&\!H\!&\!1\cr
0\!&\!0\!&\!1\!&\!0},~~~~
\label{rmgfpp}
\rmg f^{\mu\nu}=\sqrt{2}\pmatrix{
0\!&\!0\!&\!0\!&\!\check f_x\cr
0\!&\!0\!&\!0\!&\!\check f_y\cr
0\!&\!0\!&\!0\!&\!0\cr
-\check f_x\!&\!-\check f_y\!&\!0\!&\!0\cr},~~~\\
\label{fpp}
\!\!\!\!\!f_{\mu\nu}&=&2A_{[\nu,\mu]}=2A_{,[\nu}k_{\mu]}
=\sqrt{2}\pmatrix{
0\!&\!0\!&\!-\check f_x\!&\!0\cr
0\!&\!0\!&\!-\check f_y\!&\!0\cr
\check f_x\!&\!\check f_y\!&\!0\!&\!0\cr
0\!&\!0\!&\!0\!&\!0},
~~\rmg=\rmN=1~~~~
\end{eqnarray}
where
\begin{eqnarray}
\label{Aform}
k_\mu&=&(0,0,-1,0),~~
A_\mu=(0,0,A,0),~~
A=-\sqrt{2}(x\check f_x+y\check f_y),~~\\
%\check f_x&=& \check f_x(u),~~~\check f_y= \check f_y(u),
\label{H}
H&=&2\hat H+A^2\\
\label{Hsolution}
&=&2(h_{+}x^2+h_{\times}xy-h_{+}y^2)+2(\check f_x^2+\check f_y^2)(x^2+y^2),\\
\label{Hhatsolution}
\hat H&=&h_{+}x^2+h_{\times}xy-h_{+}y^2+(y\check f_x-x\check f_y)^2.
\end{eqnarray}
and the nonzero connections are
\begin{eqnarray}
\tGam^1_{33}&=&\frac{1}{2}\frac{\partial H}{\partial x},~~~
\tGam^2_{33}=\frac{1}{2}\frac{\partial H}{\partial y},~~~
\tGam^4_{33}=\frac{1}{2}\frac{\partial H}{\partial u}
-\frac{2}{\Lambda_b}\frac{\partial(\check f_x^2+\check f_y^2)}{\partial u},\nonumber\\
\tGam^4_{13}&=&\frac{1}{2}\frac{\partial H}{\partial x}
-\frac{2i}{\sqrt{\Lambda_b}}\frac{\partial\check f_x}{\partial u},~~~
\tGam^4_{31}=\frac{1}{2}\frac{\partial H}{\partial x}
+\frac{2i}{\sqrt{\Lambda_b}}\frac{\partial\check f_x}{\partial u},\\
\tGam^4_{23}&=&\frac{1}{2}\frac{\partial H}{\partial y}
-\frac{2i}{\sqrt{\Lambda_b}}\frac{\partial\check f_y}{\partial u},~~~
\tGam^4_{32}=\frac{1}{2}\frac{\partial H}{\partial y}
+\frac{2i}{\sqrt{\Lambda_b}}\frac{\partial\check f_y}{\partial u}\,.\nonumber
\end{eqnarray}
Here $h_{+}(u),h_{\times}(u)$ characterize the
gravitational wave component, $\check f_x(u),\check f_y(u)$ characterize the electromagnetic wave
component, and all of these are arbitrary functions of the coordinate $u=(t-z)/\sqrt{2}$.

%For the parameterization (\ref{gform}-\ref{H}), it happens that $\tR_{\mu\nu}\!=\!R_{\mu\nu}$,
%and the electromagnetic field is a null field\cite{Stephani,Misner} with
%$f^\sigma{_\mu}f^\mu{_\sigma}\!=det(f^{\mu}{_{\nu}})\!=0$.
%For this case it can be shown from Appendix\nobreak~D
%of \cite{jShifflett} that all of the higher order terms in
%(\ref{approximateNbar},\ref{approximateNhat},\ref{Einstein2}) vanish so that
%$F_{\mu\nu}\!=\!f_{\mu\nu}\!=\!N_{[\mu\nu]}\Lambda_b^{\!1/2}/\rmt$
%and our Einstein and Maxwell equations are identical to those of
%Einstein-Maxwell theory. Maxwell's equations (\ref{Ampere},\ref{Faraday})
%are satisfied automatically from (\ref{fpp},\ref{gform}),
%and the Einstein equations reduce to,
%\begin{eqnarray}
%0&=&\tilde R_{33}+\Lambda_b(N_{33}-g_{33})
%=\frac{\partial^2\!\hat H}{\partial x^2}+\frac{\partial^2\!\hat H}{\partial y^2}
%-2(\check f_x^2+\check f_y^2).
%\end{eqnarray}
%This has the solution (\ref{Hhatsolution},\ref{Hsolution}).
The solution above has been discussed extensively in the literature on
Einstein-Maxwell theory\cite{Baldwin,Misner,Stephani,Griffiths} so we will
not interpret it further.
It is the same solution which forms the incoming waves for
the Bell-Szekeres colliding plane-wave solution\cite{Griffiths},
although the full Bell-Szekeres solution does not satisfy our theory
because the electromagnetic field is not null after the collision.

\ifnum\ExpandDerivations=1
\section{\label{EquationsOfMotion}Equations of motion of the electric monopole solution}
Here we assume that one body is much heavier than the other
so that we can make the approximation that this body will remain stationary,
and we will also ignore radiation reaction effects.
The Lorentz-force equation (\ref{Euler}) for this theory is the same as in Einstein-Maxwell theory,
and we have an exact electric monopole solution analogous to the Reissner-Nordstr\"{o}m solution,
so we can use the same methods as with Einstein-Maxwell theory.
The Lorentz-force equation for the classical hydrodynamics case is
\begin{eqnarray}
\label{LorentzForceEquation}
Q_2 F^\alpha{_\mu}u^\mu=\frac{du^\alpha}{d\lambda}+\Gamma^\alpha_{\mu\nu}u^\mu u^\nu,~~~u^\nu=\frac{dx^\nu}{d\lambda}.
\end{eqnarray}
The stationary and moving bodies have masses $M$, $M_2$ and charges $Q$, $Q_2$.
We are using $d\lambda=ds/M_2$ instead of $ds$ because the unitless parameter $\lambda$ is
still meaningful for the null geodesics of photons where $ds\rightarrow 0$ and $M_2\rightarrow 0$.
Using the metric (\ref{finalg})
%\begin{eqnarray}
%\label{g}
%g_{00}=\cosht a,~~~g_{11}=-1/\cosht a,~~~g_{22}=-\cosht r^2,~~~g_{33}=-\cosht r^2 sin^2\theta,
%\end{eqnarray}
and the relation $(r^2\cosht)'\!=\!2r/\cosht$ from (\ref{finalc}),
the non-zero Christoffel connections (\ref{Christoffel}) are
\begin{eqnarray}
\label{finalChristoffel}
\!\!\!\!&&\Gamma^1_{00}=\frac{a\cosht}{2}(a\cosht)'~~,~
\Gamma^0_{10}=\frac{(a\cosht)'}{2a\cosht}~~,~
\Gamma^1_{11}=-\frac{(a\cosht)'}{2a\cosht}~~,~
\Gamma^2_{12}=\Gamma^3_{13}=\frac{1}{\cosht^2 r},~~~\\
\!\!\!\!\!&&\Gamma^1_{22}=-ar~~,~
\Gamma^1_{33}=-ar{\rm sin}^2\theta~~,~
\Gamma^3_{23}={\rm cot}\,\theta~~,~
\Gamma^2_{33}=-{\rm sin}\,\theta {\rm cos}\,\theta.~~~\nonumber
\end{eqnarray}
The equations of motion are then
\begin{eqnarray}
\label{eofmr}
a\cosht Q_2 F_{01}u^t&=&\frac{du^r}{d\lambda}
-\frac{(a\cosht)'}{2a\cosht}u^r{^2}
-aru^\theta{^2}
-ar sin^2\theta u^\phi{^2}
+\!\frac{a\cosht(a\cosht)'}{2}u^t{^2},\\
\label{eofmtheta}
0&=&\frac{du^\theta}{d\lambda}
+\frac{2}{r\cosht^2}u^ru^\theta
-\sin\theta cos\theta\,u^\phi{^2},\\
\label{eofmphi}
0&=&\frac{du^\phi}{d\lambda}
+\frac{(r^2\cosht)'}{r^2\cosht}\,u^r u^\phi
+2 cot\theta\,u^\theta u^\phi,\\
\label{eofmct}
~\frac{Q_2 F_{01}}{a\cosht}\,u^r&=&\frac{du^t}{d\lambda}
+\frac{(a\cosht)'}{a\cosht}u^ru^t.
\end{eqnarray}
For motion in the equatorial plane we may put $u^\theta\!=\!0,~\theta\!=\!\pi/2$,
and (\ref{eofmtheta}) is identically satisfied.
Then from (\ref{eofmphi}) we get
\begin{eqnarray}
&&0=\frac{1}{r^2\cosht}\frac{d(u^\phi r^2\cosht)}{d\lambda},\\
\label{Leq}
&&u^\phi r^2\cosht
=(constant)=L=(angular~momentum).
\end{eqnarray}
From (\ref{eofmct},\ref{finalF}) we get
\begin{eqnarray}
&&0=\frac{1}{a\cosht}\left(\frac{d(u^ta\cosht)}{d\lambda}+Q_2A_0'u^r\right)
=\frac{1}{a\cosht}\frac{d}{d\lambda}\!\left(u^ta\cosht+Q_2A_0\right),\\
\label{Eeq}
&&u^ta\cosht+Q_2A_0
=(constant)=E=(total~energy).
\end{eqnarray}
Recalling that $d\lambda=ds/M_2$ and $u^\theta\!=\!0,~\theta\!=\!\pi/2$ we also have
\begin{eqnarray}
M_2^2&=&u^\alpha u_\alpha
\label{uu}
=a\cosht u^t{^2}\!-\!\frac{1}{a\cosht}u^r{^2}\!-\!\cosht r^2 u^\phi{^2}.
\end{eqnarray}
Eliminating $t$ and $\lambda$ from (\ref{uu}) using (\ref{Leq},\ref{Eeq}) gives
\begin{eqnarray}
\!\!\!\!\!M_2^2a\cosht \!&=&\!(a\cosht)^2 u^t{^2}
\!-\!\left(\!\frac{dr}{d\phi}u^\phi\!\!\right)^2\!\!\!-a\cosht^2 r^2 u^\phi{^2}
\label{ueq}
\!=\!(E-Q_2A_0)^2
\!-\!\left(\!\frac{dr}{d\phi}\frac{L}{r^2\cosht}\!\right)^2
\!\!-\!\frac{aL^2}{r^2}.~~~~
\end{eqnarray}
This can be rewritten as an integral
\begin{eqnarray}
\label{phisolution}
\phi=\!\int\!\!\frac{L dr/r^2}{\cosht\sqrt{(E-Q_2A_0)^2
\!-\!\!aL^2/r^2-M_2^2a\cosht}}.
%=\!\int\!\!\frac{-Ldw}{\cosht\sqrt{(E-Q_2A_0)^2
%\!-\!\!aL^2w^2-M_2^2a\cosht}}~~~
\end{eqnarray}
%where $w\!=\!1/r$.
For $\Lambda_b\!\rightarrow\!\infty,~a\!=\!1$ we have flat-space electrodynamics,
and the integral can be done analytically.
For $\Lambda_b\!\rightarrow\!\infty$ we have Einstein-Maxwell theory,
and the integral becomes an elliptic integral.
For a finite $\Lambda_b$ the integral is more complicated, but using
(\ref{A0approx},\ref{aapprox},\ref{capprox}) for $A_0,~a,~\cosht$ and neglecting
powers higher than $1/r^4$ also leads to an elliptic integral. The time dependence
can be obtained using (\ref{Leq},\ref{Eeq}) to get
\begin{eqnarray}
dt/d\phi&=&u^t/u^\phi
=(r^2\cosht/L)(E-Q_2A_0)/a\cosht
=(E -Q_2A_0)r^2/aL
\end{eqnarray}
so that from (\ref{phisolution}),
\begin{eqnarray}
\label{tsolution}
t&=&\int\frac{(E-Q_2A_0)dr}{a\cosht\sqrt{(E-Q_2A_0)^2\!-\!aL^2/r^2-M_2^2a\cosht}}\,.
\end{eqnarray}

\ifnum\ExpandDerivations=1
We can also obtain the results (\ref{phisolution},\ref{tsolution}) using
the Hamilton-Jacobi approach as in \cite{Landau},
p. 94-95 and 306-308. From (\ref{uu},\ref{finalg}), the Hamilton-Jacobi equation is
\begin{eqnarray}
M_2^2&=&g^{\mu\nu}\left(\frac{\partial S}{\partial x^\mu}+Q_2A_\mu\right)\left(\frac{\partial S}{\partial x^\nu}+Q_2A_\nu\right)\\
&=&\frac{1}{a\cosht}\left(\frac{\partial S}{\partial t}+Q_2A_0\right)^2
-a\cosht\left(\frac{\partial S}{\partial r}\right)^2
-\frac{1}{\cosht r^2}\left(\frac{\partial S}{\partial\phi}\right)^2.
\end{eqnarray}
The solution is
\begin{eqnarray}
S=-Et+L\phi+S_r(r),
~~~~~S_r=\int\frac{dr}{a\cosht}\sqrt{(E-Q_2A_0)^2\!-\!aL^2/r^2-M_2^2a\cosht}\,.
\end{eqnarray}
Then (\ref{phisolution}) is obtained from the equation $\partial S/\partial L=(constant)$,
and (\ref{tsolution}) is obtained from the equation $\partial S/\partial E =(constant)$.

Let us analyze the special case $L\!=\!Q_2\!=\!0$ using the effective potential
method of \cite{Wald,Misner}. From (\ref{ueq},\ref{Leq}) and the definitions
$\tilde E=E/M_2,~ds=M_2d\lambda$ we have
\begin{eqnarray}
a\cosht=\tilde E^2\!-\!\left(\frac{dr}{ds}\right)^2.
\end{eqnarray}
This equation can be expressed as a non-relativistic potential problem,
\begin{eqnarray}
\frac{1}{2}\left(\frac{dr}{ds}\right)^2=\frac{\tilde E^2-1}{2}-\tilde V,
\end{eqnarray}
where $(dr/ds)^2/2$ corresponds to the kinetic energy per mass, and $\tilde V$ is the
so-called ``effective potential'',
\begin{eqnarray}
\tilde V=\frac{a\cosht-1}{2}.
\end{eqnarray}
Since $\hat V(r_e)\!=\!\sqrt{2}\left[\Gamma(1/4)\right]^2\!/6\sqrt{\pi}\!-\!2/3=1.08137$,
we can assume that $\hat V\approx 1$ for present purposes, and the effective potential becomes,
\begin{eqnarray}
\label{radialeffectiveV}
\tilde V&\approx&
\frac{1}{2}\!\left(1-\frac{2M}{r}+\!\frac{Q^2}{r^2}\right)\sqrt{1-\!\frac{2Q^2}{\Lambda_br^4}}-\frac{1}{2}.
\end{eqnarray}

Let us consider the case for elementary particles where $Q\!\gg\!M$. This case is
more interesting than astronomical objects because there is no event horizon to hide
the behavior close to $r_e\!=\!(2Q^2/\Lambda_b)^{1/4}\!=\!3.16\times 10^{-34}cm$
where $\cosht\!=\!0$. Assuming an electron charge and mass we
have $Q\!=\!Q_e\!=\!e\sqrt{G/c^4}\!=\!\sqrt{\alpha}\,l_P\!=\!1.38\times 10^{-34}cm$ and
$M\!=\!Gm_e/c^2\!=\!7\!\times\!10^{-56}cm$. In this case the mass term in ``a'' is
smaller that the charge term for $r\!<\!Q^2/2M\!=\!1.36\times 10^{-13}cm$,
which is close to the classical electron radius.
The following table shows the rough behavior of $\tilde V$,
\bigskip

\noindent
$\tilde V$ vs. $r$ for three $Q/Q_e$ values\\
$\phantom{|}$~~~~~~~~~~~~~~~~~~~our charged solution
~~~~~~~~~~~~~~~~~~~Reissner-Nordstr\"{o}m solution
\begin{eqnarray}
\matrix{
r/Q_e\backslash Q/Q_e & .265 & 1.06    & 1.86~~~~~~~~~& .265 & 1.06 & 1.86\cr
  1.68 &  -.05602 &     -    &     -   ~~~~~~~~~& ~~.01250 &  .20000 & .61250\cr
  2.66 &  -.00519 &  -.15032 &     -   ~~~~~~~~~& ~~.00500 &  .08000 & .24500\cr
  3.76 &  -.00002 &  -.00478 &  -.05315~~~~~~~~~& ~~.00250 &  .04000 & .12250\cr
  4.60 & ~~.00055 & ~~.00769 & ~~.01525~~~~~~~~~& ~~.00166 &  .02666 & .08166\cr
  5.32 & ~~.00062 & ~~.00953 & ~~.02611~~~~~~~~~& ~~.00125 &  .02000 & .06125\cr
  5.94 & ~~.00060 & ~~.00937 & ~~.02722~~~~~~~~~& ~~.00100 &  .01600 & .04900\cr
10^{21}&  -.00001 & ~~.00000 & ~~.00001~~~~~~~~~&  -.00001 &  .00000 & .00001\cr
10^{25}& <0&<0&<0~~~~~~~~~&<0&<0&<0\cr
\infty & ~~.00000 & ~~.00000 & ~~.00000~~~~~~~~~& ~~.00000 &  .00000 & ~.00000
}
\end{eqnarray}
For $Q\!=\!Q_e$ we find that $\tilde V$ has a zero at
$r_e/Q_e\!=\!(2/\Lambda_bQ_e^2)^{1/4}=2.3$, then it rises quickly to a maximum
of $\sim.00953$ at $r/Q_e\!=\!5.3$, then it falls slowly to $0$ near the classical
electron radius calculated above, and it remains slightly below $0$ as
$r\!\rightarrow\!\infty$ where it goes to $0$. Radii where
$\tilde V=(\tilde E\!-\!1)/2$ are turning points where the
radial motion reverses, so a falling body would bounce back only if
$(\tilde E\!-\!1)/2\!<\!.00953$. It is the $\cosht$ term that causes $\tilde V$
to have a zero at $r_e/Q_e=2.3$, which limits the potential for small radii.
Particles falling into the Reissner-Nordstr\"{o}m solution with
$L\!=\!0,~Q\!\gg\!M,~\cosht\!=\!1$ would always bounce back near $r\!\sim\!Q_e\!\sim\!l_P$
regardless of their energy, because the potential goes to infinity as $r\!\rightarrow\!0$.
This is a clear difference between our charged solution
and the Reissner-Nordstr\"{o}m solution, although it is unclear whether it has
any significance from an experimental viewpoint.

Now let us consider a massless particle with $M_2\!=\!L\!=\!Q_2\!=\!0$ falling into a body
with $Q\!\gg\!M$. Setting $M_2\!=\!L\!=\!Q_2\!=\!0$ in (\ref{tsolution}) gives
\begin{eqnarray}
t=\int\frac{dr}{a\cosht}
=\int\frac{dr}{(1-2M/r+\!Q^2/r^2)\sqrt{1-\!2Q^2/\Lambda_br^4}}
,~~~~~~~~dr=a\cosht dt.
\end{eqnarray}
Here the $\cosht$ term causes a pole at
$r_e\!=\!(2Q_e^2/\Lambda_b)^{1/4}\!=\!3.16\times10^{-34}cm$.
Ignoring the ``a'' term, one gets an elliptic integral which evaluates to some
finite value when the lower limit is set to $r_e$. This means that a particle would
take a finite time to reach the radius $r_e$, at which point it presumably disappears.
For the Reissner-Nordstr\"{o}m solution with $\cosht\!=\!1$, the integrand becomes
$\sim r^2/Q^2$ near $r\!=\!0$, so a particle would take a finite time to reach $r\!=\!0$.
Therefore, in contrast to the massive case, a massless neutral particle of any energy
will fall into the singularity in a finite time for either our charged solution
or the Reissner-Nordstr\"{o}m solution.
\fi
\fi

\ifnum\ExpandDerivations=1
\section{\label{PeriastronAdvance}Periastron advance}
Here we calculate the periastron advance for a body with mass and charge $M_2,Q_2$
rotating around a more massive body with mass and charge $M,Q$.
We will use the effective potential method of \cite{Wald,Misner},
together with the Lorentz force equation (\ref{Euler},\ref{LorentzForceEquation})
and the resulting equations of motion calculated in \S\ref{EquationsOfMotion}.
Using (\ref{ueq},\ref{Leq}) and the definitions
\begin{eqnarray}
\label{M2scaling}
\tilde Q_2=Q_2/M_2,~~~\tilde L=L/M_2,~~~\tilde E=E/M_2,~~~ds=M_2d\lambda,
\end{eqnarray}
we have
\begin{eqnarray}
a\cosht=(\tilde E -\tilde Q_2 A_0)^2
\!-\!\left(\frac{dr}{ds}\right)^2\!-\!\frac{a{\tilde L}^2}{r^2}.
\end{eqnarray}
This equation can be expressed in the form of a non-relativistic potential problem,
\begin{eqnarray}
\frac{1}{2}\left(\frac{dr}{ds}\right)^2=\frac{\tilde E ^2-1}{2}-\tilde V,
\end{eqnarray}
where $(dr/ds)^2/2$ corresponds to the kinetic energy per mass, and $\tilde V$ is the
so-called ``effective potential'',
\begin{eqnarray}
\tilde V=-\frac{1}{2}(\tilde E -\tilde Q_2 A_0)^2
%\frac{\tilde E \tilde Q_2 Q}{r}
%-\frac{\tilde Q_2^2 Q^2}{2r^2}
+\!\frac{a{\tilde L}^2}{2r^2}
+\frac{a\cosht}{2}+\frac{(\tilde E^2-1)}{2}.
\end{eqnarray}
Using the expressions (\ref{A0approx},\ref{aapprox},\ref{capprox}) for $A_0,~a,~\cosht$ and keeping
only terms which fall off as $1/r^4$ or slower we get
\ifnum\ExpandDerivations=1
\begin{eqnarray}
\!\!\tilde V\!\approx\!\tilde E\tilde Q_2\!\left(\frac{Q}{r}
\!+\!\frac{QM}{\Lambda_br^4}\right)
\!-\!\frac{\tilde Q_2^2 Q^2}{2r^2}
\!+\!\frac{{\tilde L}^2}{2r^2}\!\left(1\!-\!\frac{2M}{r}+\!\frac{Q^2}{r^2}\right)
\!+\!\frac{1}{2}\!\left(-\frac{2M}{r}+\!\frac{Q^2}{r^2}-\!\frac{Q^2}{\Lambda_br^4}\right).~~~
\end{eqnarray}
Combining the powers of $1/r$ gives
\fi
\begin{eqnarray}
&&~~~~~~~~~~~~~~~~
\tilde V_{SR}~~~\tilde V_{RN1}~~~~\tilde V_{GR}~~~~~~\tilde V_{RN2}~~~~~~~~~\tilde V_{ES}\nonumber\\
\!\!\!\!\tilde V&=&-\frac{\mu}{r}
+\frac{({\tilde L}^2\!-\!\tilde Q_2^2 Q^2\!+\!Q^2)}{2r^2}
-\frac{M{\tilde L}^2}{r^3}
+\!\frac{(Q^2{\tilde L}^2\!+\!Q(2\tilde E\tilde Q_2 M\!-Q)/\Lambda_b)}{2r^4},~~~
\end{eqnarray}
where
\begin{eqnarray}
\label{alphadef}
\mu=M\!-\!\tilde E \tilde Q_2 Q.
\end{eqnarray}
Here $-\mu/r$ is the combined Newtonian/Coulombian potential,
and the term ${\tilde L}^2/2r^2$ is sometimes called the ``centrifugal potential energy''.
These terms characterize the nonrelativistic
Newtonian/Coulombian central force problem, and the associated orbits will be ellipses with no precession.
The additional terms $\tilde V_{SR},\tilde V_{GR},\tilde V_{RN1},\tilde V_{RN2},\tilde V_{ES}$ are due
respectively to special relativity, general relativity, the $Q^2/r^2$ term of the
Reissner-N\"{o}rdstrom solution, and finally our $\Lambda$-renormalized Einstein-Sch\"{o}dinger theory.
All of these additional terms are small relative to the first terms for ordinary radii,
and they can be treated as perturbations of the Newtonian/Coulombian case.
Setting $d\tilde V/dr\!=\nobreak\!0$
gives the radius ``$r_0$'' of a stable circular orbit.
If a body is displaced slightly from $r_0$
it will oscillate in radius about $r_0$, executing
simple harmonic motion with proper time radial frequency $\omega_r$ given by
\begin{eqnarray}
\omega_r=\sqrt{[d^2\tilde V/dr^2]_{r=r_0}}\,.
\end{eqnarray}
Subtracting this from the proper time angular frequency from (\ref{Leq})
\begin{eqnarray}
\label{omegaphifull}
\omega_\phi={\tilde L}/{\cosht r^2}
\end{eqnarray}
gives the periastron advance,
\begin{eqnarray}
\omega_p=\omega_\phi-\omega_r.
\end{eqnarray}
\ifnum\ExpandDerivations=1
The derivatives of the effective potential are
\begin{eqnarray}
\label{Vp}
\!\!\!\!\frac{\partial\tilde V}{\partial r}&=&\frac{\mu}{r^2}
-\frac{({\tilde L}^2\!-\!\tilde Q_2^2 Q^2\!+\!Q^2)}{r^3}
+\frac{3M{\tilde L}^2}{r^4}
-\!\frac{2(Q^2{\tilde L}^2\!+\!Q(2\tilde E\tilde Q_2 M\!-Q)/\Lambda_b)}{r^5},\\
\label{Vpp}
\!\!\!\!\!\frac{\partial^2\tilde V}{\partial r^2}&=&-\frac{2\mu}{r^3}
+\frac{3({\tilde L}^2\!-\!\tilde Q_2^2 Q^2\!+\!Q^2)}{r^4}
-\frac{12M{\tilde L}^2}{r^5}
+\!\frac{10(Q^2{\tilde L}^2\!+\!Q(2\tilde E\tilde Q_2 M\!-Q)/\Lambda_b)}{r^6}.~~~
\end{eqnarray}

For the Newtonian/Coulombian case we have
\begin{eqnarray}
0=\frac{\partial\tilde V}{\partial r}&=&\frac{\mu}{r^2}-\frac{{\tilde L}^2}{r^3}
=\frac{1}{r^3}(\mu r-{\tilde L}^2)
\end{eqnarray}
which has the solution
\begin{eqnarray}
\label{r0NewtonianCoulombian}
r_0&=&{\tilde L}^2/\mu\\
\omega_r&=&\sqrt{\frac{d^2\tilde V}{dr^2}}
=\sqrt{-\frac{2\mu}{r_0^3}+\frac{3{\tilde L}^2}{r_0^4}}=\frac{\tilde L}{r_0^2}.
\end{eqnarray}
From (\ref{Leq}) the orbital frequency using proper time is
\begin{eqnarray}
\label{omegaphi}
\omega_\phi=u^\phi=\frac{\tilde L}{r_0^2}.
\end{eqnarray}
So for the Newtonian/Coulombian case there is no periastron advance
\begin{eqnarray}
\omega_p=\omega_\phi-\omega_r=0.
\end{eqnarray}
Each of the additional potential terms will have the uninteresting effect of
changing the dependence of the orbital parameters on the constants $\tilde L$ and $\tilde E$.
However these terms will also have the more fundamental effect of introducing periastron advance.
Here we will calculate the periastron advance for each of the additional potential terms,
and compare these to the results to our theory.

Including the special relativistic term $V_{SR}$ gives
\begin{eqnarray}
\label{SpecialRelativistic}
0=\frac{\partial\tilde V}{\partial r}
=\frac{\mu}{r^2}-\frac{{\tilde L}^2}{r^3}+\frac{\tilde Q_2^2Q^2}{r^3}
=\frac{1}{r^3}(\mu r-{\tilde L}^2+\tilde Q_2^2Q^2),
\end{eqnarray}
which has the solution
\begin{eqnarray}
r_0&=&({\tilde L}^2\!-\!\tilde Q_2^2 Q^2)/\mu,\\
\omega_r&=&\sqrt{\frac{d^2\tilde V}{dr^2}}
=\sqrt{-\frac{2\mu}{r_0^3}+\frac{3({\tilde L}^2-\tilde Q_2^2Q^2)}{r_0^4}}
%=\frac{1}{r_0^2}\sqrt{-2\mu r_0+3({\tilde L}^2-\tilde Q_2^2Q^2)}\\
=\frac{\tilde L}{r_0^2}\sqrt{(1-\tilde Q_2^2Q^2/{\tilde L}^2)}\\
&=&\omega_\phi\left(1-\frac{\tilde Q_2^2Q^2}{2{\tilde L}^2}\right).
\end{eqnarray}
So the periastron advance caused by the special relativistic term $V_{SR}$ is
\begin{eqnarray}
\omega_{pSR}=\omega_\phi-\omega_r=\frac{\tilde Q_2^2Q^2\omega_\phi}{2{\tilde L}^2}.
\end{eqnarray}

Including the general relativistic term $V_{GR}$ gives
\begin{eqnarray}
\label{GeneralRelativistic}
0=\frac{\partial\tilde V}{\partial r}
=\frac{\mu}{r^2}-\frac{{\tilde L}^2}{r^3}+\frac{3M{\tilde L}^2}{r^4}
=\frac{1}{r^4}(\mu r^2-{\tilde L}^2r+3M{\tilde L}^2).
\end{eqnarray}
This equation could potentially be solved by using the solution of a quadratic equation.
However, we will instead find an approximation based on a perturbation of the
Newtonian case. We assume that
\begin{eqnarray}
r_{GR}={\tilde L}^2/\mu+\Delta r_0.
\end{eqnarray}
Assuming that $\Delta r_0$ is small compared to ${\tilde L}^2/\mu$ we can make the approximation
$({\tilde L}^2/\mu+\Delta r_0)^n\approx({\tilde L}^2/\mu+n\Delta r_0)({\tilde L}^2/\mu)^{n-1}$.
Substituting into (\ref{GeneralRelativistic}) gives
\begin{eqnarray}
0=\mu({\tilde L}^2/\mu+2\Delta r_0){\tilde L}^2/\mu-{\tilde L}^2({\tilde L}^2/\mu+\Delta r_0)\!+\!3M{\tilde L}^2
={\tilde L}^2\Delta r_0\!+\!3M{\tilde L}^2,
\end{eqnarray}
which has the solution
\begin{eqnarray}
\!\!\!\Delta r_0&=&-3M,\\
\!\!\!\omega_r&=&\sqrt{\frac{d^2\tilde V}{dr^2}}
=\sqrt{-\frac{2\mu}{r_0^3}+\frac{3{\tilde L}^2}{r_0^4}-\frac{12M{\tilde L}^2}{r_0^5}}
=\frac{1}{r_0^2}\sqrt{-2\mu r_0+3{\tilde L}^2-12M{\tilde L}^2/r_0}~~~\\
&=&\frac{1}{r_0^2}\sqrt{-2\mu({\tilde L}^2/\mu-3M)+3{\tilde L}^2-12M{\tilde L}^2({\tilde L}^2/\mu+3M)({\tilde L}^2/\mu)^{-2}}\\
&=&\frac{\tilde L}{r_0^2}\sqrt{1-6\mu M/{\tilde L}^2 -36M^2\mu^2/{\tilde L}^4}\\
&=&\omega_\phi\left(1-\frac{3\mu M}{{\tilde L}^2}\right).
\end{eqnarray}
So the periastron advance caused by the general relativistic term $V_{GR}$ is
\begin{eqnarray}
\omega_{pGR}=\omega_\phi-\omega_r=\frac{3\mu M\omega_\phi}{{\tilde L}^2}.
\end{eqnarray}

The periastron advance caused by the 1st Reissner-Nordstr\"{o}m term $V_{RN1}$ can be derived from the
calculations for the $V_{SR}$ term by letting $\tilde Q_2^2Q^2\rightarrow -Q^2$,
\begin{eqnarray}
\omega_{pRN1}=\omega_\phi-\omega_r=-\frac{Q^2\omega_\phi}{2{\tilde L}^2}.
\end{eqnarray}

Including the 2nd Reissner-Nordstr\"{o}m term $V_{RN2}$ gives
\begin{eqnarray}
\label{LRN2}
0=\frac{\partial\tilde V}{\partial r}
&=&\frac{\mu}{r^2}-\frac{{\tilde L}^2}{r^3}-\frac{2Q^2{\tilde L}^2}{r^5}
=\frac{1}{r^5}\left(\mu r^3-{\tilde L}^2r^2-2Q^2{\tilde L}^2\right).
\end{eqnarray}
Again we assume that
\begin{eqnarray}
r_{ES}={\tilde L}^2/\mu+\Delta r_0.
\end{eqnarray}
Substituting into (\ref{LRN2}) and using
$({\tilde L}^2/\mu+\Delta r_0)^n\approx({\tilde L}^2/\mu+n\Delta r_0)({\tilde L}^2/\mu)^{n-1}$
gives
\begin{eqnarray}
0&=&\mu({\tilde L}^2/\mu+3\Delta r_0)({\tilde L}^2/\mu)^2
-{\tilde L}^2({\tilde L}^2/\mu+2\Delta r_0){\tilde L}^2/\mu-2Q^2{\tilde L}^2\\
&=&\Delta r_0{\tilde L}^4/\mu-2Q^2{\tilde L}^2,
\end{eqnarray}
which has the solution
\begin{eqnarray}
\!\!\!\Delta r_0&=&2\mu Q^2/{\tilde L}^2,\\
\!\!\!\omega_r&=&\sqrt{\frac{d^2\tilde V}{dr^2}}
=\sqrt{-\frac{2\mu}{r_0^3}+\frac{3{\tilde L}^2}{r_0^4}+\frac{10Q^2{\tilde L}^2}{r_0^6}}
=\frac{1}{r_0^2}\sqrt{-2\mu r_0+3{\tilde L}^2+\frac{10Q^2{\tilde L}^2}{r_0^2}}~~~\\
&=&\frac{1}{r_0^2}\sqrt{-2\mu({\tilde L}^2/\mu+\Delta r_0)+3{\tilde L}^2+10Q^2{\tilde L}^2({\tilde L}^2/\mu-2\Delta r_0)({\tilde L}^2/\mu)^{-3}}\\
&=&\frac{1}{r_0^2}\sqrt{-4\mu^2Q^2/{\tilde L}^2+{\tilde L}^2+10Q^2\mu^2/{\tilde L}^2-40Q^4\mu^4/{\tilde L}^6}\\
&=&\frac{{\tilde L}}{r_0^2}\sqrt{1+6\mu^2Q^2/{\tilde L}^4-40Q^4\mu^4/{\tilde L}^8}
\label{tempRN2}
\approx\frac{\tilde L}{r_0^2}\left(1+\frac{3Q^2\mu^2}{{\tilde L}^4}\right)\\
&=&\omega_\phi\left(1+\frac{3Q^2\mu^2}{{\tilde L}^4}\right).
\end{eqnarray}
So the periastron advance caused by the 2nd Reissner-Nordstr\"{o}m term $V_{RN2}$ is
\begin{eqnarray}
\omega_{pRN2}=\omega_\phi-\omega_r=-\frac{3Q^2\mu^2\omega_\phi}{{\tilde L}^4}.
\end{eqnarray}

The periastron advance caused by the Einstein-Schr\"{o}dinger term $V_{ES}$ can be derived from the
calculations for the 2nd Reissner-Nordstr\"{o}m term $V_{RN2}$ by
letting $Q^2{\tilde L}^2\rightarrow Q(2\tilde E\tilde Q_2 M\!-Q)/\Lambda_b$, except that
$\omega_\phi=\tilde L/\cosht r^2$ from (\ref{omegaphifull}) instead of
$\omega_\phi=\tilde L/r^2$ from (\ref{omegaphi}).
Ignoring the correction to $r_0=\tilde L^2/\mu$ from (\ref{r0NewtonianCoulombian}) we have
\begin{eqnarray}
\omega_\phi&=&\frac{\tilde L}{r_0^2\sqrt{1-2Q^2/\Lambda_br_0^4}}
=\frac{\tilde L}{r_0^2\sqrt{1-2Q^2\mu^4/\Lambda_b{\tilde L}^8}}
\approx\frac{\tilde L}{r_0^2}\left(1+\frac{Q^2\mu^4}{\Lambda_b{\tilde L}^8}\right),
\end{eqnarray}
and from (\ref{tempRN2}),
\begin{eqnarray}
\omega_r&=&\frac{\tilde L}{r_0^2}\!\left(1+\frac{3Q(2\tilde E\tilde Q_2 M\!-Q)\mu^2}{\Lambda_b{\tilde L}^6}\right)
%=\omega_\phi\left(1-\frac{Q^2\mu^4}{\Lambda_b{\tilde L}^8}\right)\!\left(1-\frac{3Q(Q+2\tilde E\tilde Q_2 M)\mu^2}{\Lambda_b{\tilde L}^6}\right)\\
%&\approx&\omega_\phi\left(1-\frac{3Q(Q+2\tilde E\tilde Q_2 M)\mu^2}{\Lambda_b{\tilde L}^6}-\frac{Q^2\mu^4}{\Lambda_b{\tilde L}^8}\right)\\
=\frac{\tilde L}{r_0^2}\left[1-\left(3-\frac{6\tilde E \tilde Q_2 M}{Q}\right)\frac{Q^2\mu^2}{\Lambda_b{\tilde L}^6}\right].
\end{eqnarray}
So the total periastron advance caused by the Einstein-Schr\"{o}dinger term $V_{ES}$ is
\begin{eqnarray}
\omega_{pES}&=&\omega_\phi-\omega_r
=\left(3-\frac{6\tilde E \tilde Q_2 M}{Q}+\frac{\mu^2}{{\tilde L}^2}\right)\frac{Q^2\mu^2\omega_\phi}{\Lambda_b{\tilde L}^6}.
\end{eqnarray}
\fi

Combining all of the calculations, the total periastron advance comes to
\begin{eqnarray}
\label{PeriastronShift}
&&
\omega_{pSR}~~~~\omega_{pGR}~~~\omega_{pRN1}~~~\omega_{pRN2}~~~~~~~~~~~~~\omega_{pES}\nonumber\\
\!\!\!\!\frac{\omega_p}{\omega_\phi}&=&\frac{\tilde Q_2^2Q^2}{2{\tilde L}^2}
+\frac{3\mu M}{{\tilde L}^2}
-\frac{Q^2}{2{\tilde L}^2}
-\frac{3Q^2\mu^2}{{\tilde L}^4}
+\!\left(3-\frac{6\tilde E \tilde Q_2 M}{Q}+\frac{\mu^2}{{\tilde L}^2}\right)\!\frac{Q^2\mu^2}{{\tilde L}^6\Lambda_b}.~~~
\end{eqnarray}
Here the special relativity term $\omega_{pSR}$ agrees with \cite{Landau,Wallace,Barker},
the general relativity term $\omega_{pGR}$ agrees with \cite{Wald,Adler},
and the first Reissner-N\"{o}rdstrom term $\omega_{pRN1}$ agrees with \cite{Balakin,Rathod}.
The $\omega_{pES}$ term is due to our $\Lambda$-renormalized Einstein-Sch\"{o}dinger theory.
%where
%\begin{eqnarray}
%\mu&=&M-\frac{QQ_2\tilde E}{M_2}
%\end{eqnarray}
\ifnum\ExpandDerivations=1
All of these calculations were done by assuming nearly circular orbits. However the
$\omega_{pGR}$ result can be shown to be correct for arbitrary eccentricity\cite{Wald} if
we replace $\omega_\phi=\tilde L/r^2$ from (\ref{omegaphi}) with the more general expression
from Newtonian mechanics
\begin{eqnarray}
\label{Newtonian}
\omega_\phi&=&\frac{\lower2pt\hbox{${\tilde L}^2$}}{\sqrt{\mu}\,(1-e_s^2)a_s^{5/2}}\\
e_s&=&(eccentricity)\\
a_s&=&(semimajor~axis)
\end{eqnarray}
This also true of $\omega_{pSR}$, as can be seen from p.94 of \cite{Landau},
and it is probably true for the other $\omega_p$ results as well.
Also, using $\omega_\phi=\tilde L/r^2$ from (\ref{omegaphi}) and (\ref{Newtonian})
we can reproduce the result (\ref{r0NewtonianCoulombian}) for nearly circular orbits
\begin{eqnarray}
\tilde L\approx\sqrt{\mu r}
\end{eqnarray}
%Note that the $\omega_{pRN1}$ and $\omega_{pRN2}$ results have certainly been derived before,
%but we are unaware of a reference to verify them.
\fi

For a first test case we choose the Bohr atom with $M\!=\!M_P$, $M_2\!=\!M_e$, $Q\!=\!-Q_2\!=\!|Q_e|$
because of its approximate physical relevance, and because the Einstein-Schr\"{o}dinger term will have the greatest
effect at small radii.
\ifnum\ExpandDerivations=1
Using $M^{(geom)}\!=\!M^{(cgs)}G/c^2$, $Q^{(geom)}\!=\!Q^{(cgs)}\sqrt{G/c^4}$
and $esu\!=\!cm^{3/2}g^{1/2}\!/s$ we have
\begin{eqnarray}
M&=&M_P=1.67\!\times 10^{-24}g\!\left(\frac{6.67\!\times\! 10^{-8}cm^3\!/g\!\cdot\!s^2}{(3\!\times\! 10^{10} cm/s)^2}\right)
=1.24\!\times\!10^{-52} cm,\\
\!\!M_2&=&M_e=9.11\times 10^{-28}g\!\left(\frac{6.67\!\times\! 10^{-8}cm^3\!/g\!\cdot\!s^2}{(3\!\times\! 10^{10} cm/s)^2}\right)
=6.75\times 10^{-56}cm,\\
Q&=&-Q_2=|Q_e|=4.8\!\times\! 10^{-10}esu\sqrt{\frac{6.67\!\times\! 10^{-8}cm^3\!/g\!\cdot\!s^2}{(3\!\times\! 10^{10} cm/s)^4}}
=1.38\!\times\!10^{-34}cm,~~\\
r&=&a_0=.529\!\times\! 10^{-8} cm,\\
\Lambda_b&=&10^{66} cm^{-2},\\
\!\!\tilde E&=&(total~energy)/M_2\approx 1,\\
\mu&=&-\frac{QQ_2\tilde E}{M_2}=\frac{(1.38\times 10^{-34}cm)^2\times 1}{6.75\times 10^{-56}cm}
=2.81\times 10^{-13} cm,\\
\tilde L&=&\sqrt{\mu a_0}=\sqrt{2.81\times 10^{-13}cm\times .529\times 10^{-8}cm}
=3.86\times 10^{-11} cm,\\
\label{omegaphival1}
\!\!\omega_\phi&=&\frac{c\tilde L}{r^2}=\frac{3\times 10^{10}cm/s\times 3.86\times 10^{-11} cm}{(.529\!\times\! 10^{-8} cm)^2}
=4.14\times 10^{16} rad/s.
\end{eqnarray}
and
\begin{eqnarray}
\frac{\omega_{pSR}}{\omega_\phi}
&=&\frac{(1.38\times 10^{-34}cm)^4}{2(3.86\times 10^{-11}cm)^2(6.75\times 10^{-56}cm)^2}
=2.65\times 10^{-5},\\
\frac{\omega_{pGR}}{\omega_\phi}
&=&\frac{3\times 2.81\times 10^{-13}cm\times 1.24\times 10^{-52}cm}{(3.86\times 10^{-11}cm)^2}
=7.00\times 10^{-44},\\
\!\!\frac{\omega_{pRN1}}{\omega_\phi}
&=&-\frac{(1.38\times 10^{-34}cm)^2}{2(3.86\times 10^{-11}cm)^2}
=-6.36\times 10^{-48},\\
\!\!\frac{\omega_{pRN2}}{\omega_\phi}
&=&-\frac{3(1.38\!\times\! 10^{-34}cm)^2(2.81\!\times\! 10^{-13}cm)^2}{(3.86\times 10^{-11}cm)^4}
=-2.02\!\times\! 10^{-51},\\
\frac{\omega_{pES}}{\omega_\phi}
&=&\frac{6(1.38\!\times\! 10^{-34}cm)^2(2.81\!\times\! 10^{-13}cm)^2\!\times\!1.24\!\times\!10^{-52} cm}{(3.86\times 10^{-11}cm)^6\times 6.75\times 10^{-56}cm\!\times\! 10^{66}cm^{-2}}
=5.0\!\times\! 10^{-93}.~~
\end{eqnarray}
\fi
From (\ref{PeriastronShift},\ref{omegaphifull}), the periastron advances are
\begin{eqnarray}
\omega_{pSR}&=&2.65\times 10^{-5}\omega_\phi~=1.10\times 10^{12}rad/s,\\
\omega_{pGR}&=&7.00\times 10^{-44}\omega_\phi=2.90\times 10^{-27}rad/s,\\
\omega_{pRN1}&=&-6.36\times 10^{-48}\omega_\phi=-2.63\times 10^{-31}rad/s,\\
\omega_{pRN2}&=&-2.02\times 10^{-51}\omega_\phi=-8.36\times 10^{-35}rad/s,\\
\omega_{pES}&=&5.0\times 10^{-93}\omega_\phi=2.07\times 10^{-76}rad/s.
\end{eqnarray}

For a second test case we choose $M\!=\!M_\odot$ because this is the smallest
black hole we can expect to observe, and the smallest black hole will create the
worst-case observable spatial curvature. We choose an extremal black hole with
$Q\!=\!M$ because this is the worst-case charge which avoids a naked singularity,
and we choose the orbital radius to be $r\!=\!4M$ because this is close to the
smallest stable orbit. For the second body we choose $Q_2\!=\!0$ and $M_2\!<<\!M$,
so that $M_2$ does not enter into the equations.
\ifnum\ExpandDerivations=1
Using $M^{(geom)}\!=\!M^{(cgs)}G/c^2$ we have
\begin{eqnarray}
\mu&=&M=Q=M_\odot=1.99\!\times\!10^{33}g\!\left(\frac{6.67\!\times\! 10^{-8}cm^3\!/g\!\cdot\!s^2}{(3\!\times\! 10^{10} cm/s)^2}\right)
=1.47\times 10^5 cm,~~~\\
r&=&4M=5.90\!\times\!10^{5} cm,\\
Q_2&=&0cm,\\
\Lambda_b&=&10^{66} cm^{-2},\\
\tilde E&=&(total~energy)/M_2\approx 1,\\
\tilde L&=&\sqrt{\mu r}=\sqrt{1.47\times 10^{5}cm\times 5.90\times 10^5cm}
=2.95\times 10^{5} cm,\\
\label{omegaphival2}
\omega_\phi&=&\frac{c{\tilde L}}{r^2}
=\frac{3\times 10^{10}cm/s\times 2.95\times 10^{5}cm}{(5.90\times 10^{5}cm)^2}
=2.55\times 10^{4} rad/s,
\end{eqnarray}
and
\begin{eqnarray}
\frac{\omega_{pSR}}{\omega_\phi}
&=&0,\\
\frac{\omega_{pGR}}{\omega_\phi}
&=&\frac{3\times 1.47\times 10^{5}cm\times 1.47\times 10^{5}cm}{(2.95\times 10^{5}cm)^2}
=.747,\\
\frac{\omega_{pRN1}}{\omega_\phi}
&=&-\frac{(1.47\times 10^{5}cm)^2}{2(2.95\times 10^{5}cm)^2}
=-.125,\\
\frac{\omega_{pRN2}}{\omega_\phi}
&=&-\frac{3(1.47\times 10^{5}cm)^2(1.47\times 10^{5}cm)^2}{(2.95\times 10^{5}cm)^4}
=-.186,\\
\frac{\omega_{pES}}{\omega_\phi}
&=&\frac{3(1.47\times 10^{5}cm)^2(1.47\times 10^{5}cm)^2}{(2.95\times 10^{5}cm)^6\times 10^{66}cm^{-2}}
=2.14\times 10^{-78}.
\end{eqnarray}
\fi
From (\ref{PeriastronShift},\ref{omegaphifull}), the periastron advances are
\begin{eqnarray}
\omega_{pSR}&=&0\omega_\phi~=0rad/s,\\
\omega_{pGR}&=&.747\omega_\phi=1.90\times 10^{4}rad/s,\\
\omega_{pRN1}&=&-.125\omega_\phi=-3.19\times 10^{3}rad/s,\\
\omega_{pRN2}&=&-.186\omega_\phi=-4.74\times 10^{3}rad/s,\\
\omega_{pES}&=&2.14\times 10^{-78}\omega_\phi=5.46\times 10^{-74}rad/s.
\end{eqnarray}

Obviously $\omega_{pES}$ is too small to measure for either of the test cases,
with a fractional difference from the Einstein-Maxwell result of $<\!10^{-78}$.
However, this result is
one more indication that the $\Lambda$-renormalized Einstein-Schr\"{o}dinger
theory closely approximates Einstein-Maxwell theory.
%It is also a demonstration
%that experimental results can be derived from this theory, and that the
%calculations are not much more complicated than with Einstein-Maxwell theory.
\fi

\ifnum\ExpandDerivations=1
\section{\label{NullGeodesics}Deflection and time delay of light}

Null geodesics can be found by letting $M_2\!=\!Q_2\!=\!0$ in the
Lorentz force equation (\ref{Euler},\ref{LorentzForceEquation}) from \S\ref{EquationsOfMotion}.
Since both the Lorentz force equation and our electromagnetic plane-wave solution
are exactly the same as in Einstein-Maxwell theory,
we may expect that light rays will follow null geodesics.
We will find it convenient to rewrite the equations of
\S\ref{EquationsOfMotion} in terms of the impact parameter ``$b$'' instead of ``$L$'' and ``$E$''.
The impact parameter is the distance of closest approach of a line drawn from
the initial $r\rightarrow\infty$ asymptote of the path, and it is where $d\phi/dt=1/b$
would occur if the path was not bent. From (\ref{Leq},\ref{Eeq}) the impact parameter is given by
\begin{eqnarray}
b=L/E.
\end{eqnarray}
Using (\ref{ueq}) with $b=L/E$ and $Q_2\!=\!M_2\!=\!0$ gives
\begin{eqnarray}
\label{ub}
0 &=&\frac{1}{b^2}
-\left(\frac{dr}{d\phi}\frac{1}{r^2\cosht}\right)^2
-\frac{a}{r^2}.
\end{eqnarray}
For these calculations the factor $\cosht$ from (\ref{capprox}) dominates over the
extra $1/r^6$ term in ``a'' from (\ref{aapprox}), so for present purposes we will assume that
\begin{eqnarray}
a=1-\frac{2M}{r}+\frac{Q^2}{r^2}=1-2Mw+Q^2w^2,~~~~~w=1/r.
\end{eqnarray}
From (\ref{ub}) with $dr/d\phi\!=\!0$, the distance of closest approach $R_0$
is related to the impact parameter $b$ by
\begin{eqnarray}
\label{bvsR0}
\frac{1}{b^2}=\frac{1}{R_0^2}\left(1-\frac{2M}{R_0}+\frac{Q^2}{R_0^2}\right).
\end{eqnarray}

To find the angular deflection of light we will use the method of \cite{Wald}.
Integrating (\ref{ub}) or using (\ref{phisolution}) with $b=L/E$ and $Q_2\!=\!M_2\!=\!0$ gives
\begin{eqnarray}
\label{phisolutionnull}
\phi&=&\int\frac{dr/r^2}{\cosht\sqrt{1/b^2\!-\!a/r^2}}
=\int\frac{-dw}{\cosht\sqrt{1/b^2\!-\!aw^2}}\,,~~~~~w=1/r.
\end{eqnarray}
We assume that
$\delta\phi\!\approx\!M[\partial\phi/\partial M]_{M=Q^2=0}\!+\!Q^2[\partial\phi/\partial(Q^2)]_{M=Q^2=0}$,
where the differentiation of (\ref{phisolutionnull}) and the substitution $M\!=\!Q^2\!=\!0$
is done before the integration to simplify the calculations. Following \cite{Wald},
the differentiation is done with a fixed $R_0$ from (\ref{bvsR0}) instead of a fixed ``b''.
\ifnum\ExpandDerivations=1
Using (\ref{bvsR0},\ref{phisolutionnull}), and changing sign by convention gives
\begin{eqnarray}
\delta\phi&=&M\left[\frac{\partial}{\partial M}\int_0^{1/R_0}\frac{2dw}{\cosht\sqrt{1/b^2\!-\!aw^2}}\right]_{M=Q=0}\nonumber\\
&&+Q^2\left[\frac{\partial}{\partial (Q^2)}\int_0^{1/R_0}\frac{2dw}{\cosht\sqrt{1/b^2\!-\!aw^2}}\right]_{M=Q=0}\\
&=&M\int_0^{1/R_0} dw\left[\frac{2(1/R_0^3-w^3)}{\cosht(1/b^2\!-\!aw^2)^{3/2}}\right]_{M=Q=0}\nonumber\\
&&+Q^2\int_0^{1/R_0} dw\left[\frac{(-1/R_0^4+w^4)}{\cosht(1/b^2\!-\!aw^2)^{3/2}}
+\frac{2w^4}{\cosht^3\Lambda_b\sqrt{1/b^2\!-\!aw^2}}\right]_{M=Q=0}\\
&=&M\int_0^{1/b} dw\left[\frac{2(1/b^3-w^3)}{(1/b^2\!-\!w^2)^{3/2}}\right]\nonumber\\
&&+Q^2\int_0^{1/b} dw\left[\frac{(-1/b^4+w^4)}{(1/b^2\!-\!w^2)^{3/2}}
+\frac{2w^4}{\Lambda_b\sqrt{1/b^2\!-\!w^2}}\right]\\
&=&2M\left[\frac{w/b}{\sqrt{1/b^2-w^2}}
-\frac{1/b^2}{\sqrt{1/b^2-w^2}}
-\sqrt{1/b^2-w^2}\right]_0^{1/b}\nonumber\\
&&+Q^2\left[\frac{w}{2}\sqrt{1/b^2-w^2}
-\frac{3\,asin(wb)}{2b^2}\right]_0^{1/b}\nonumber\\
&&+\frac{Q^2}{\Lambda_b}\left[
-\frac{w^3}{2}\sqrt{1/b^2-w^2}
-\frac{3w}{4b^2}\sqrt{1/b^2-w^2}
+\frac{3\,asin(wb)}{4b^4}\right]_0^{1/b}.
\end{eqnarray}
Therefore the total angular deflection is
\begin{eqnarray}
\delta\phi=\frac{4M}{b}-\frac{3\pi Q^2}{4b^2}+\frac{3\pi Q^2}{8\Lambda_bb^4}.
\end{eqnarray}

This same result can also be obtained by another method\cite{Adler}.
Using $w=1/r$ and $w'=\partial w/\partial\phi$ we can write (\ref{ub}) as
\begin{eqnarray}
0 &=&\frac{1}{b^2}
-\frac{w'^2}{\cosht^2}
-aw^2.
\end{eqnarray}
Using (\ref{aapprox},\ref{capprox}) for $a,~\cosht$
and keeping only terms which fall off as $1/r^4$ or slower we get
\begin{eqnarray}
0 &=&\frac{\cosht^2}{b^2}
-w'^2
-\cosht^2(1-2Mw+Q^2w^2)w^2\\
&\approx&\frac{1}{b^2}
-w'^2
-w^2+2Mw^3-Q^2w^4-\frac{2Q^2w^4}{\Lambda_b}\left(\frac{1}{b^2}-w^2\right).
\end{eqnarray}
Here the last term is the modification due to the $\Lambda$-renormalized
Einstein-Schr\"{o}dinger theory.
Taking the derivative of this equation gives
\begin{eqnarray}
0 &=&
2w'\!\left[-w''
\!-\!w+3Mw^2-2Q^2w^3-\frac{2Q^2w^3}{\Lambda_b}\left(\frac{2}{b^2}-3w^2\right)\right].
\end{eqnarray}
Removing the $2w'$ gives
\begin{eqnarray}
w''+w &=&3Mw^2-2Q^2w^3-\frac{2Q^2w^3}{\Lambda_b}\left(\frac{2}{b^2}-3w^2\right).
\end{eqnarray}
If the terms on the right-hand side were absent, this equation
would have the solution $1/r=w=sin(\phi)/b$, which is a
straight line along the equatorial plane in spherical coordinates.
The terms on the right-hand side can be regarded as perturbations due to the presence of
the stationary body, so we seek a solution of the form
\begin{eqnarray}
w=sin(\phi)/b+f(\phi).
\end{eqnarray}
Substituting this and keeping only the first order terms gives
\begin{eqnarray}
f''+f&=&3M \frac{sin^2(\phi)}{b^2}
-2Q^2\frac{sin^3(\phi)}{b^3}
-\frac{2Q^2}{\Lambda_b}\frac{sin^3(\phi)}{b^3}\left(\frac{2}{b^2}-\frac{3sin^2(\phi)}{b^2}\right).
\end{eqnarray}
Using the identities $sin^2(\phi)=(1-cos(2\phi))/2$,~$sin^3(\phi)=(3sin(\phi)-sin(3\phi))/4$
and $sin^5(\phi)=(10sin(\phi)-5sin(3\phi)+sin(5\phi))/16$ gives
\begin{eqnarray}
f''+f&=&\frac{3M}{2b^2}(1-cos(2\phi))
-\frac{Q^2}{2b^3}\left(1+\frac{2}{\Lambda_bb^2}\right)(3sin(\phi)-sin(3\phi))\\
&&+\frac{3Q^2}{8\Lambda_bb^5}(10sin(\phi)-5sin(3\phi)+sin(5\phi)).
\end{eqnarray}
Using $(\phi cos(\phi))''=(cos\phi-\phi sin\phi)'=-2sin(\phi)-\phi cos(\phi)$,
it is easy to see that this has the solution
\begin{eqnarray}
f(\phi)&=&\frac{3M}{2b^2}+\frac{M}{2b^2}cos(2\phi)
-\frac{Q^2}{8b^3}\left(-6+\frac{3}{\Lambda_bb^2}\right)\left(\phi-\frac{\pi}{2}\right) cos(\phi)\\
&&-\frac{Q^2}{8b^3}\left(\frac{1}{2}-\frac{7}{8\Lambda_bb^2}\right)sin(3\phi)
-\frac{Q^2}{64\Lambda_bb^5}sin(5\phi).
\end{eqnarray}
So the full perturbed solution is
\begin{eqnarray}
\!\!w&=&\frac{sin(\phi)}{b}+\!\frac{3M}{2b^2}+\frac{M}{2b^2}cos(2\phi)
-\!\frac{3Q^2}{8b^3}\left(2-\frac{1}{\Lambda_bb^2}\right)\left(\frac{\pi}{2}-\!\phi\right)cos(\phi)\\
&&-\frac{Q^2}{16b^3}\left(1-\frac{7}{4\Lambda_bb^2}\right)sin(3\phi)
-\frac{Q^2}{64\Lambda_bb^5}sin(5\phi).
\end{eqnarray}
At $r=\infty,~w=0$ the unperturbed solution requires $\phi=0$. So if we set $w=0$ for
the perturbed solution, the resulting $\phi$ will be half of the deflection, and we
change sign by convention $\phi\!=\!-\delta\phi/2$.
Doing this and neglecting higher order terms gives,
\begin{eqnarray}
\!\!\!0&=&\frac{sin(-\delta\phi/2)}{b}+\!\frac{3M}{2b^2}+\frac{M}{2b^2}cos(\delta\phi)
-\!\frac{3Q^2}{8b^3}\left(2-\frac{1}{\Lambda_bb^2}\right)\left(\frac{\pi}{2}+\!\frac{\delta\phi}{2}\right)cos(\delta\phi/2)~~~\\
&&-\frac{Q^2}{16b^3}\left(1-\frac{7}{4\Lambda_bb^2}\right)sin(-3\delta\phi/2)
-\frac{Q^2}{64\Lambda_bb^5}sin(-5\delta\phi/2)\\
&\approx&-\frac{\delta\phi}{2b}+\frac{3M}{2b^2}+\frac{M}{2b^2}
-\frac{3Q^2}{8b^3}\left(2-\frac{1}{\Lambda_bb^2}\right)\left(\frac{\pi}{2}+\frac{\delta\phi}{2}\right)\\
&&+\frac{Q^2}{16b^3}\left(1-\frac{7}{4\Lambda_bb^2}\right)\frac{3\delta\phi}{2}
+\frac{Q^2}{64\Lambda_bb^5}\frac{5\delta\phi}{2}\\
&\approx&-\frac{\delta\phi}{2b}+\frac{2M}{b^2}
-\frac{3\pi Q^2}{16b^3}\left(2-\frac{1}{\Lambda_bb^2}\right).
\end{eqnarray}
\fi
The resulting angular deflection is
\begin{eqnarray}
~~~~~~~\delta\phi_{GR}~~~\delta\phi_{RN}~~~~\delta\phi_{ES}\nonumber\\
\delta\phi=\frac{4M}{b}
-\frac{3\pi Q^2}{4b^2}
+\frac{3\pi Q^2}{8\Lambda_bb^4}.
\end{eqnarray}
Here the first term $\delta\phi_{GR}$ is the ordinary general relativistic deflection of light,
and this result agrees with \cite{Wald}.
The second term $\delta\phi_{RN}$ is from the Reissner-Nordstr\"{o}m $Q^2/r^2$ term,
and this result agrees with \cite{Jeffery,Bhadra,Amore}. The last term $\delta\phi_{ES}$ is
from the $\Lambda$-renormalized Einstein-Schr\"{o}dinger theory.

To find the time delay of light
we will use a method much like the one used for computing angular deflection,
again from \cite{Wald}.
Using (\ref{tsolution}) with $b=L/E$ and $Q_2\!=\!M_2\!=\!0$ gives the time dependence
\begin{eqnarray}
\label{tsolutionnull}
t&=&\int\frac{dr}{a\cosht\sqrt{1\!-\!ab^2/r^2}}
=\int\frac{rdr}{a\cosht\sqrt{r^2\!-\!ab^2}}.
\end{eqnarray}
We assume that
$\delta t\!\approx\!M[\partial t/\partial M]_{M=Q^2=0}\!+\!Q^2[\partial t/\partial(Q^2)]_{M=Q^2=0}$,
where the differentiation of (\ref{tsolutionnull}) and the substitution $M\!=\!Q^2\!=\!0$
is done before the integration to simplify the calculations. Following \cite{Wald},
the differentiation is done with a fixed $R_0$ from (\ref{bvsR0}) instead of a fixed ``b''.
The integration is done from an initial radius $R_i$ to the distance of closest approach $R_0$,
and then from $R_0$ to the final radius $R_f$.
\ifnum\ExpandDerivations=1
Using (\ref{bvsR0},\ref{tsolutionnull}) gives the time delay from $R_0$ to $R_f$,
\begin{eqnarray}
\delta t&=&M\left[\frac{\partial}{\partial M}\int_{R_0}^{R_f}\frac{rdr}{a\cosht\sqrt{r^2\!-\!ab^2}}\right]_{M=Q=0}\nonumber\\
&&+Q^2\left[\frac{\partial}{\partial (Q^2)}\int_{R_0}^{R_f}\frac{rdr}{a\cosht\sqrt{r^2\!-\!ab^2}}\right]_{M=Q=0}\\
&=&M\int_{R_0}^{R_f} dr\left[\frac{2}{a^2\cosht \sqrt{r^2\!-\!ab^2}}+\frac{(-b^2+arb^4/R_0^3)}{a\cosht(r^2\!-\!ab^2)^{3/2}}\right]_{M=Q=0}\nonumber\\
&&+Q^2\int_{R_0}^{R_f} dr\left[-\frac{1/r}{a^2\cosht \sqrt{r^2\!-\!ab^2}}+\frac{(b^2/r-arb^4/R_0^4)}{2a\cosht(r^2\!-\!ab^2)^{3/2}}
+\frac{1/r^3}{a\Lambda_b\cosht^3\sqrt{r^2\!-\!ab^2}}\right]_{M=Q=0}\\
&=&M\int_{R_0}^{R_f} dr\left[\frac{2}{\sqrt{r^2\!-\!R_0^2}}+\frac{(-R_0^2+rR_0)}{(r^2\!-\!R_0^2)^{3/2}}\right]\nonumber\\
&&+Q^2\int_{R_0}^{R_f} dr\left[-\frac{3/r}{2\sqrt{r^2\!-\!R_0^2}}
+\frac{1/r^3}{\Lambda_b\sqrt{r^2\!-\!R_0^2}}\right]\\
&=&M\left[2 ln(r+\sqrt{r^2\!-\!R_0^2}\,)+\frac{r}{\sqrt{r^2\!-\!R_0^2}}-\frac{R_0}{\sqrt{r^2\!-\!R_0^2}}\right]_{R_0}^{R_f}\nonumber\\
&&+Q^2\left[-\frac{3\,acos(R_0/r)}{2R_0}\right]_{R_0}^{R_f}\nonumber\\
&&+\frac{Q^2}{2\Lambda_b}\left[\frac{\sqrt{r^2\!-\!R_0^2}}{r^2R_0^2}+\frac{acos(R_0/r)}{R_0^3}\right]_{R_0}^{R_f}\\
&=&M\left[2 ln(r\!+\!\sqrt{r^2\!-\!R_0^2}\,)+\sqrt{\frac{r\!-\!R_0}{r\!+\!R_0}}\right]_{R_0}^{R_f}\nonumber\\
&&-\frac{3Q^2}{2}\left[\frac{acos(R_0/r)}{R_0}\right]_{R_0}^{R_f}\nonumber\\
&&+\frac{Q^2}{2\Lambda_b}\left[\frac{\sqrt{r^2\!-\!R_0^2}}{r^2R_0^2}+\frac{acos(R_0/r)}{R_0^3}\right]_{R_0}^{R_f}.
\end{eqnarray}
The variable $r$ is always positive, so to get the time delay from $R_i$ to $R_0$ we can use this same
expression but with $R_f\!\rightarrow\! R_i$.
\fi
The resulting time delay in geometrized units is
\begin{eqnarray}
\delta t
&=&\delta t_{GR}+\delta t_{RN}+\delta t_{ES}\\
&=&M\left[2 ln\left(\frac{(R_i\!+\!\sqrt{R_i^2\!-\!R_0^2}\,)(R_f\!+\!\sqrt{\smash{R_f^2}\!-\!R_0^2}\,)}{R_0^2}\right)
\!+\!\sqrt{\frac{R_i\!-\!R_0}{R_i\!+\!R_0}}\!+\!\sqrt{\frac{\smash{R_f}\!-\!R_0}{\smash{R_f}\!+\!R_0}}\right]\nonumber\\
&&-\frac{3Q^2}{2}\left[\frac{acos(R_0/R_i)}{R_0}+\frac{acos(R_0/R_f)}{R_0}\right]\nonumber\\
&&+\frac{Q^2}{2\Lambda_b}\left[\frac{\sqrt{R_i^2\!-\!R_0^2}}{R_i^2R_0^2}+\frac{\sqrt{\smash{R_f^2}\!-\!R_0^2}}{R_f^2R_0^2}+\frac{acos(R_0/R_i)}{R_0^3}+\frac{acos(R_0/R_f)}{R_0^3}\right].
\end{eqnarray}
Here the first line, $\delta t_{GR}$ is the ordinary general relativistic time delay,
and this result agrees with \cite{Wald}.
The last line $\delta t_{ES}$ is from the $\Lambda$-renormalized Einstein-Schr\"{o}dinger theory.

For a first test case we choose $b\!=\!R_i/2\!=\!R_f/2\!=\!R_0\!=\!a_0$, the Bohr radius,
and $M\!=\!M_P,~Q\!=\!Q_P$, for a proton because this case has some approximate physical relevance,
and because the Einstein-Schr\"{o}dinger term will have the strongest effect for small radii.
\ifnum\ExpandDerivations=1
Using $M^{(geom)}\!=\!M^{(cgs)}G/c^2$, $Q^{(geom)}\!=\!Q^{(cgs)}\sqrt{G/c^4}$
and $esu\!=\!cm^{3/2}g^{1/2}\!/s$ we have
\begin{eqnarray}
M&=&M_P=1.67\!\times 10^{-24}g\!\left(\frac{6.67\!\times\! 10^{-8}cm^3\!/g\!\cdot\!s^2}{(3\!\times\! 10^{10} cm/s)^2}\right)
=1.24\!\times\!10^{-52} cm,\\
Q&=&|Q_e|=4.8\!\times\! 10^{-10}esu\sqrt{\frac{6.67\!\times\! 10^{-8}cm^3\!/g\!\cdot\!s^2}{(3\!\times\! 10^{10} cm/s)^4}}
=1.38\!\times\!10^{-34}cm,\\
b&=&R_i/2=R_f/2=R_0=a_0=.529\!\times\! 10^{-8} cm,\\
\Lambda_b&=&10^{66} cm^{-2}.
\end{eqnarray}
\fi
For the angular deflections we get
\begin{eqnarray}
&&\delta\phi_{GR}
\ifnum\ExpandDerivations=1
=\frac{4\times1.24\times 10^{-52}cm}{.529\times 10^{-8} cm}
\fi
=9.36\times 10^{-44} rad,\\
&&\delta\phi_{RN}
\ifnum\ExpandDerivations=1
=-\frac{3\pi(1.38\times 10^{-34}cm)^2}{4(.529\times 10^{-8} cm)^2}
\fi
=-1.60\times 10^{-51} rad,\\
&&\delta\phi_{ES}
\ifnum\ExpandDerivations=1
=\frac{3\pi(1.38\times 10^{-34}cm)^2}{8(.529\times 10^{-8} cm)^4\times 10^{66}cm^{-2}}
\fi
=2.85\times 10^{-101} rad.
\end{eqnarray}
For the times delays we get
\begin{eqnarray}
\!\!\!&&\delta t_{GR}
\ifnum\ExpandDerivations=1
=\frac{1.24\times 10^{-52}cm}{3\times 10^{10} cm/s}\left[4 ln(2+\sqrt{3})+\frac{2}{\sqrt{3}}\right]
\fi
=2.65\times 10^{-62}s,\\
\!\!\!&&\delta t_{RN}
\ifnum\ExpandDerivations=1
=-\frac{3(1.38\times 10^{-34}cm)^2}{2\times 3\times 10^{10} cm/s}\left[\frac{2\pi/3}{.529\times 10^{-8} cm}\right]
\fi
=-3.75\times 10^{-70}s,\\
\!\!\!&&\delta t_{ES}
\ifnum\ExpandDerivations=1
=\frac{(1.38\times 10^{-34}cm)^2}{3\times 10^{10}cm/s\times 2\times 10^{66} cm^{-2}}\left[\frac{\sqrt{3}/2+2\pi/3}{(.529\times 10^{-8} cm)^3}\right]
\fi
=6.31\times 10^{-120}s.~~~~
\end{eqnarray}
For reference purposes, these results may be compared to the travel time of light across a Bohr radius, $a_0/c\!=\!1.76\!\times\! 10^{-19}s$.
%or to the period of an $n\!=\!1$ Bohr orbit, $P\!=\!h/\alpha^2 m_e c^2\!=\!1.5\!\times\! 10^{-16}s$.

For a second test case we choose $M\!=\!M_\odot$ because this is the smallest
black hole we can expect to observe, and the smallest black hole will create the
worst-case observable spatial curvature. We choose an extremal black hole with
$Q\!=\!M$ because this is the worst-case charge which avoids a naked singularity,
and we choose $b\!=\!R_i\!=\!R_f\!=\!2R_0\!=\!4M$ because this is close to the gravitational radius.
\ifnum\ExpandDerivations=1
Using $M^{(geom)}\!=\!M^{(cgs)}G/c^2$ we have
\begin{eqnarray}
M&=&Q=M_\odot=1.99\!\times\!10^{33}g\!\left(\frac{6.67\!\times\! 10^{-8}cm^3\!/g\!\cdot\!s^2}{(3\!\times\! 10^{10} cm/s)^2}\right)
=1.47\times 10^5 cm,\\
b&=&R_i=R_f=2R_0=4M=5.90\!\times\! 10^{5} cm,\\
\Lambda_b&=&10^{66} cm^{-2}.
\end{eqnarray}
\fi
For the angular deflections we get
\begin{eqnarray}
&&\delta\phi_{GR}
\ifnum\ExpandDerivations=1
=\frac{4\times 1.47\times 10^5cm}{5.90\times 10^{5} cm}
\fi
=1.0 rad,\\
&&\delta\phi_{RN}
\ifnum\ExpandDerivations=1
=-\frac{3\pi(1.47\times 10^5 cm)^2}{4(5.90\times 10^{5} cm)^2}
\fi
=-.147 rad,\\
&&\delta\phi_{ES}
\ifnum\ExpandDerivations=1
=\frac{3\pi(1.47\times 10^5 cm)^2}{8(5.90\times 10^{5} cm)^4\times 10^{66}cm^{-2}}
\fi
=2.11\times 10^{-79} rad.
\end{eqnarray}
For the time delays we get
\begin{eqnarray}
\!\!\!&&\delta t_{GR}
\ifnum\ExpandDerivations=1
=\frac{1.47\times 10^5}{3\times 10^{10} cm/s}\left[4 ln(2+\sqrt{3})+\frac{2}{\sqrt{3}}\right]
\fi
=3.16\times 10^{-5}s,\\
\!\!\!&&\delta t_{RN}
\ifnum\ExpandDerivations=1
=-\frac{3(1.47\times 10^5 cm)^2}{2\times 3\times 10^{10} cm/s}\left[\frac{4\pi/3}{5.90\times 10^{5} cm}\right]
\fi
=-7.73\times 10^{-6}s,\\
\!\!\!&&\delta t_{ES}
\ifnum\ExpandDerivations=1
=\frac{(1.47\times 10^5 cm)^2}{3\times 10^{10} cm/s\times 2\times 10^{66} cm^{-2}}\left[\frac{4\sqrt{3}+16\pi/3}{(5.90\times 10^{5} cm)^3}\right]
\fi
=4.18\times 10^{-83}s.~~~~
\end{eqnarray}
For reference purposes, these results may be compared to the travel time of light across $R_i/c\!=\!1.97\!\times\! 10^{-5}s$.

The contributions $\delta\phi_{ES}$ and $\delta t_{ES}$ from the
$\Lambda$-renormalized Einstein-Schr\"{o}dinger theory are too tiny to measure,
with a fractional difference from the Einstein-Maxwell result of $<\!10^{-57}$.
Again this shows how closely the theory matches Einstein-Maxwell theory.
\fi

\ifnum\ExpandDerivations=1
\section{\label{test}Shift in Hydrogen atom energy levels}

Here we estimate the energy shift of a Hydrogen atom that would result in our
theory as compared to Einstein-Maxwell theory.
This is an important case to consider because these energy levels can be measured so accurately.
It is also significant because it demonstrates that predictions can be done when additional
fields are included in the theory.
When a spin-1/2 field is added onto our Lagrangian, the theory predicts the ordinary
Dirac equation in curved space. We will only consider the effect of the difference between our
electric monopole potential (\ref{A0approx}) and the Reissner-Nordstr\"{o}m $Q/r$ potential.
We will neglect the difference of the metrics, and in fact we will neglect the difference
of the metric from that of flat space. Because of this, we do not expect the calculated energy shift
to be accurate in an absolute sense. We are only attempting to get an order of magnitude
estimate of the energy shift of our charge solution vs. the Reissner-Nordstr\"{o}m solution.
Using (\ref{A0approx}) the potential energy difference between the two solutions is
\begin{eqnarray}
\Delta V=Q_e\Delta A_0
=\frac{Q_e^2}{\Lambda_b}\left(\frac{M_e}{r^4}-\frac{4Q_e^2}{5r^5}\right).
\end{eqnarray}
Using this result, an estimate of the shift in the energy levels can be calculated using
perturbation theory. It is sufficient to treat the problem non-relativistically.
The lowest energy level of a Hydrogen atom is spherically symmetric with
\begin{eqnarray}
\label{psi0}
\psi_0=\sqrt{1/\pi a_0^3}\,e^{-a_0/r}.
\end{eqnarray}
Unlike the Reissner-Nordstr\"{o}m solution, the vector potential of our charged solution
is finite at the origin. However, the origin is at $r=r_0$ from (\ref{re}) instead of at $r=0$.
The energy shift will then be roughly
\begin{eqnarray}
\!\!\!\!\!\!\Delta E_0&\approx& <\!\psi_0|\Delta V|\psi_0\!>
=\frac{Q_e^2}{\Lambda_b}\left(\frac{1}{\pi a_0^3}\right)\int_{r_0}^\infty\!\!e^{-2r/a_0}\left(\frac{M_e}{r^4}-\frac{4Q_e^2}{5r^5}\right)4\pi r^2dr~~~\\
&\approx&\frac{4Q_e^2}{a_0^3\Lambda_b}\int_{r_0}^\infty\left(\frac{M_e}{r^2}-\frac{4Q_e^2}{5r^3}\right)dr
=\frac{4Q_e^2}{a_0^3\Lambda_b}\left(\frac{M_e}{r_0}-\frac{2Q_e^2}{5r_0^2}\right).
\end{eqnarray}
Using $r_0=\sqrt{Q}(2/\Lambda_b)^{1/4}$ from (\ref{re}) and $Q_e=\sqrt{\alpha}\,l_P$
from (\ref{redef}), the second term dominates and we get
\begin{eqnarray}
\Delta E_0\approx-\frac{4\sqrt{2}\,Q_e^3}{5a_0^3\sqrt{\Lambda_b}}
\approx-\left(\frac{Q_e^2}{2a_0}\right)\frac{8\sqrt{2\alpha}\,l_P}{5a_0^2\sqrt{\Lambda_b}}.
\end{eqnarray}
The term in the parenthesis is the ground state energy of a Hydrogen atom.
With $E_0=e^2/2a_0\sim 13.6 eV$, $l_P=1.6\times 10^{-33}cm$, $\Lambda_b\sim 10^{66}cm^{-2}$,
$h\sim 4\times 10^{-15} eV\cdot s$, and $a_0=\hbar^2/m_ee^2\sim 5\times 10^{-9}cm$ we get
\begin{eqnarray}
\!\!\!\!\!\frac{8\sqrt{2\alpha}\,l_P}{5a_0^2\sqrt{\Lambda_b}}\!\sim\! 10^{-50},~~
\Delta E_0\!\sim\! \frac{e^2}{2a_0}10^{-50}\!\sim\! 10^{-49}eV,~~
\Delta f_0\!\sim\! \frac{\Delta E_0}{h}\!\sim\! 10^{-34}Hz.~~~
\end{eqnarray}
This is clearly unmeasurable.

\ifnum\ExpandDerivations=1
There is another method of calculating an energy shift.
We showed in \S\ref{WeakField} that the weak field Lagrangian (\ref{Lmshifted})
can be written more simply in terms of a shifted vector potential (\ref{Aredefinition2}).
Now, the Dirac equation contains $A_\alpha$
whereas our shifted Maxwell equations derived from (\ref{Lmshifted}) will contain
$\breve A_\alpha$ from (\ref{Aredefinition2}), and no longer contain $j_\alpha$ except
for the normal source term. The shifted Maxwell equations might maintain
$\breve A_0\approx Q/r$, so we might expect that charge currents will cause a
perceived shift in the $A_0$ present in the Dirac equation. On the other hand,
this would imply that an electron is feeling its own charge current, and this
is not correct. It would be much like assuming that an electron is screened from
the nucleus by its own current density. Screening is to be expected for multiple
electron atoms, where one electron can screen the nucleus from other electrons,
but an electron is never assumed to be screened by its own current density.
Nevertheless, if there was such a potential shift as discussed above
we should calculate the associated energy shift just in case.
We can treat the problem by calculating
$j_\nu$ with the normal Dirac equation and then treating the shift in $A_\nu$
using perturbation theory. From (\ref{Aredefinition2}) the potential shift in cgs units becomes
\begin{eqnarray}
\Delta V=e\Delta A_0=\frac{4\pi 2e^2}{3\Lambda_b}\psi_0^*\psi_0
\end{eqnarray}
where $\psi_0$ comes from (\ref{psi0}).
The energy shift will then be,
\begin{eqnarray}
\Delta E_0&\approx& <\psi_0|\Delta V|\psi_0>
=\frac{4\pi 2e^2}{3\Lambda_b}\left(\frac{1}{\pi a_0^3}\right)^2\int_0^\infty e^{-4r/a_0} 4\pi r^2dr\\
&=&\frac{8\pi e^2}{3\Lambda_b}\left(\frac{1}{\pi a_0^3}\right)^2 2\left(\frac{a_0}{4}\right)^3 4\pi
=\left(\frac{e^2}{2a_0}\right)\frac{2}{3a_0^2\Lambda_b}
\end{eqnarray}
The term in parenthesis is the ground state energy of a Hydrogen atom.
With $E_0=e^2/2a_0\sim 13.6 eV$, $l_P=1.6\times 10^{-33}cm$, $\Lambda_b\sim 10^{66}cm^{-2}$,
$h\sim 4\times 10^{-15} eV\cdot s$, and $a_0=\hbar^2/m_ee^2\sim 5\times 10^{-9}cm$ we get
\begin{eqnarray}
\frac{2}{3a_0^2\Lambda_b}\sim 10^{-50},~~
\Delta E_0\sim \frac{e^2}{2a_0}10^{-50}\sim 10^{-49}eV,~~
\Delta f_0\sim \frac{\Delta E_0}{h}\sim 10^{-34}Hz.
\end{eqnarray}
So this second method gives the same result as the first method.
If we assume that energy levels can be measured and predicted down to a fractional error
of perhaps $10^{-10}$, the energy shift would be $10^{-40}$ too small to detect.
Alternatively we may interpret this as a lower bound of $\Lambda_b>10^{26}cm^{-2}$.
This is significant because it would mean that the theory would not work using
$\Lambda_b\!\sim\!10^{-5}\!-\!10^{-1}cm^{-2}$ resulting from the Standard Model Higgs field.
\fi
\fi

\ifnum\ExpandDerivations=1
\section{\label{Discussion}Discussion}
%There are other ways of generalizing the symmetric field Lagrangian
%density (\ref{GR}) to nonsymmetric fields which are just as natural as (\ref{Palatini}).
%The literature on the original Einstein-Schr\"{o}dinger theory can be confusing because
%the theory results from many different Lagrangian densities.
The original Einstein-Schr\"{o}dinger theory results from many different Lagrangian densities.
In fact it results from any Lagrangian density of the form,
\begin{eqnarray}
\label{JSlag2_d}
{\mathcal L}(\nGam^{\lambda}_{\!\rho\tau},N_{\rho\tau})
&=&\!-\frac{\lower2pt\hbox{$1$}}{16\pi}\rmN\left[N^{\dashv\mu\nu}(R_{\nu\mu}(\tGam)
\!+\ca\tGam^\alpha_{\alpha[\nu,\mu]}
\!+2\Aphi_{[\nu,\mu]}\rmt\Lambda_b^{\!1/2})\right.\nonumber\\
&&~~~~~~~~~~~~~~~~~~~~~~~~~~~~~~~~~~~~~~~~~~~~\left.\phantom{0_0^{|}}+(n\!-\!2)\Lambda_b\,\right]\!,
\end{eqnarray}
where $\ca,\cb,\cc$ are arbitrary constants and
\begin{eqnarray}
R_{\nu\mu}(\tGam)
&=&\tGam^\alpha_{\nu\mu,\alpha}
-\tGam^\alpha_{\nu\alpha,\mu}
+\tGam^\sigma_{\nu\mu}\tGam^\alpha_{\sigma\alpha}
-\tGam^\sigma_{\nu\alpha}\tGam^\alpha_{\sigma\mu},\\
\label{gamma_tilde_d}
~~~~~\tGam{^\alpha_{\nu\mu}}
&=&{\nGam}{^\alpha_{\nu\mu}}
\!+\frac{\lower2pt\hbox{$2$}}{(n\!-\!1)}\left[\,\cb\delta^\alpha_\mu{\nGam}{^\sigma_{\![\sigma\nu]}}
\!+\!(\cb-1)\,\delta^\alpha_\nu {\nGam}{^\sigma_{\![\sigma\mu]}}\right],\\
\label{A_d}
~~~~~~\Aphi_\nu&=&{\nGam}{^\sigma_{\![\sigma\nu]}}/\cc.
%\cc,\cb,zeta_3&=&({\rm arbitrary~constants).
\end{eqnarray}
Contracting (\ref{gamma_tilde_d}) on the right and left gives
\begin{eqnarray}
%\label{JScontractionsymmetric2}
\tGam^\alpha_{\beta\alpha}
=\frac{\lower2pt\hbox{$1$}}{(n\!-\!1)}\left[(\cb n+\cb-1)\nGam^\alpha_{\alpha\beta}
-(\cb n+\cb-n)\nGam^\alpha_{\beta\alpha}\right]=\tGam^\alpha_{\alpha\beta},
\end{eqnarray}
so $\tGam{^\alpha_{\nu\mu}}$
has only $n^3\!-n$ independent components.
%Also note that the term $\tGam^\alpha_{\alpha[\nu,\mu]}$ in (\ref{JSlag2}) is a tensor
%because $\tGam^\alpha_{\alpha[\nu,\mu]}\!=\!\tR^\alpha{_{\!\alpha\mu\nu}}/2$.
Also, from (\ref{gamma_tilde_d},\ref{A_d}) we have
\begin{eqnarray}
\label{gamma_natural_d}
{\nGam}{^\alpha_{\nu\mu}}\!&=&\!\tGam{^\alpha_{\nu\mu}}
\!-\frac{\lower2pt\hbox{$2\cc$}}{(n\!-\!1)}\left[\,\cb\delta^\alpha_\mu \Aphi_\nu
+(\cb-1)\delta^\alpha_\nu \Aphi_\mu\right],
\end{eqnarray}
so $\tGam^\alpha_{\nu\mu}$ and $A_\nu$ fully parameterize
$\nGam^\alpha_{\nu\mu}$ and can be treated as independent variables.
Therefore setting $\delta{\mathcal L}/\delta\tGam^\alpha_{\nu\mu}\!=0$
and $\delta{\mathcal L}/\delta A_\nu\!=0$
must give the same field equations as
$\delta{\mathcal L}/\delta\nGam^\alpha_{\nu\mu}\!=0$.
%Therefore when we set to zero the variational derivative of (\ref{JSlag2_d}) with respect to
%$\tGam^\alpha_{\nu\mu}$~and $A_\nu$, the same field equations must result as when
%$\nGam^\alpha_{\nu\mu}$ is used.
%In this theory $\nGam^\alpha_{\nu\mu}$ is really the
%``fundamental'' field, and $\tGam^\alpha_{\nu\mu}$ and $A_\nu$ are derived from it.
Because the field equations can be derived in this way,
the constants $\cb$ and $\cc$ are clearly arbitrary,
and because of (\ref{funnytensor}) with $j^\sigma\!=0$, the constant $\ca$ is also arbitrary.

For $\ca\!=\!1,\,\cb\!=\!1/2,\,\cc\!=\!-(n\!-\!1)\rmt\Lambda_b^{\!1/2}$,
(\ref{JSlag2_d}) reduces to (\ref{Palatini})
formed from the Hermitianized Ricci tensor (\ref{HermitianizedRicci}), where
we have the invariance properties from (\ref{transposition},\ref{gauge}),
\begin{eqnarray}
\label{transposition_d}
\!\!\!\!\!\!A_\nu\!\rightarrow\!-A_\nu
,\tGam^\alpha_{\nu\mu}\!\rightarrow\!\tGam^\alpha_{\mu\nu}
,\nGam^\alpha_{\nu\mu}\!\rightarrow\!\nGam^\alpha_{\mu\nu}
,N_{\nu\mu}\!\rightarrow\!N_{\mu\nu}
,N^{\dashv\mu\nu}\!\!\rightarrow\!N^{\dashv\nu\mu}
~\Rightarrow~{\mathcal L}\!\rightarrow\!{\mathcal L},~~\\
\label{gauge_d}
\!\!\!\!\!\!A_\alpha\!\rightarrow\! A_\alpha\!-\!\frac{\hbar}{Q}\phi_{,\alpha},
~\tGam^\alpha_{\rho\tau}\!\rightarrow\!\tGam^\alpha_{\rho\tau}
,~\nGam^\alpha_{\rho\tau}\!\rightarrow\!\nGam^\alpha_{\rho\tau}\!+\frac{2\hbar}{Q}\delta^\alpha_{[\rho}\phi_{,\tau]}\rmt\Lambda_b^{\!1/2}
\Rightarrow~{\mathcal L}\!\rightarrow\!{\mathcal L}.~~~
\end{eqnarray}
For this case we have
$\tGam^\alpha_{\sigma\alpha}\!=\!{\tGam}{^\alpha_{\alpha\sigma}}\!=\!{\nGam}{^\alpha_{\!(\alpha\sigma)}}$,
and from (\ref{der0},\ref{para},\ref{JSlag2_d}) the field equations require
a generalization of the result ${\mathcal L}_{,\sigma}\!-\!{\Gamma}{^\alpha_{\alpha\sigma}}{\mathcal L}\!=\!0$
that occurs with the Lagrangian density (\ref{GR}) of ordinary vacuum general relativity,
that is
\begin{eqnarray}
\label{selftransplantation0}
{\mathcal L}_{,\sigma}\!-\!{\nGam}{^\alpha_{(\alpha\sigma)}}{\mathcal L}=0
~~~{\rm or}~~~
{\mathcal L}_{,\sigma}\!-\!Re({\nGam}{^\alpha_{\alpha\sigma}}){\mathcal L}=0.
\end{eqnarray}
For the alternative choice, $\ca\!=\!0,\,\cb\!=\!n/(n\!+\!1),\,\cc\!=\!-(n\!-\!1)\rmt\Lambda_b^{\!1/2}/2$,
we have $\tGam^\alpha_{\sigma\alpha}\!=\!{\tGam}{^\alpha_{\alpha\sigma}}\!=\!{\nGam}{^\alpha_{\alpha\sigma}}$ and
from (\ref{der0},\ref{para},\ref{JSlag2_d}) the field equations require
\begin{eqnarray}
\label{selftransplantation}
{\mathcal L}_{,\sigma}\!-\!{\nGam}{^\alpha_{\alpha\sigma}}{\mathcal L}=0.
\end{eqnarray}
%In \cite{Shifflett} it is shown that the
%theory appears to be unique in that it can be derived from a
%Lagrangian density with the property (\ref{selftransplantation}).
%at least when the Lagrangian density is
%assumed to depend only on a nonsymmetric ${^\natural\Gamma}{^\alpha_{\nu\mu}}$.
%This property is destroyed by including an extrinsic cosmological constant
%from zero-point fluctuations, and also by modelling matter with spin-0
%or spin-1/2 wave-functions rather than as singular solutions of the field equations.
%This suggests the hypothesis that nature requires a purely classical Lagrangian density
%to have the property (\ref{selftransplantation}), but that quantization modifies this requirement.
%A vague idea is presented in \cite{Shifflett} as to why the property (\ref{selftransplantation})
%might be expected.
%However, it is legitimate to propose a simple principle without knowing why it should apply.
%As an example which makes this obvious, consider that when Newton proposed
%that mass attracts itself with a $1/r^2$ force, he certainly could not explain why.
%After all, no one knows why the laws of physics are what they are,
%so it would be unreasonable to expect an exception in this case.
\ifnum\ExpandDerivations=1
For the alternative choice $\ca\!=\!1,\,\cb\!=\!0,\,\cc\!=\!-(n\!-\!1)\rmt\Lambda_b^{\!1/2}$,
(\ref{JSlag2_d}) reduces to
\begin{eqnarray}
\label{Palatini_d}
{\mathcal L}(\nGam^{\lambda}_{\!\rho\tau},N_{\rho\tau})
=-\frac{\lower2pt\hbox{$1$}}{16\pi}\rmN\left[N^{\dashv\mu\nu}
\Re_{\nu\mu}({\nGam})+(n\!-\!2)\Lambda_b\,\right],
\end{eqnarray}
where $\Re_{\nu\mu}({\nGam})$ is a fairly simple generalization of the ordinary Ricci tensor
\begin{eqnarray}
\label{HermitianizedRicciB}
\Re_{\nu\mu}(\nGam)
&=&\nGam^\alpha_{\nu\mu,\alpha}
-\nGam^\alpha_{\!(\nu|\alpha,|\mu)}
+\nGam^\sigma_{\nu\mu}\nGam^\alpha_{\sigma\alpha}
-\nGam^\sigma_{\nu\alpha}\nGam^\alpha_{\sigma\mu}.
\end{eqnarray}
\fi
For another alternative choice $\ca\!=\!0,\,\cb\!=\!0,\,\cc\!=\!-(n\!-\!1)\rmt\Lambda_b^{\!1/2}/2$,
(\ref{JSlag2_d}) reduces to
\begin{eqnarray}
\label{ordinary}
{\mathcal L}(\nGam^{\lambda}_{\!\rho\tau},N_{\rho\tau})
=-\frac{\lower2pt\hbox{$1$}}{16\pi}\rmN\left[N^{\dashv\mu\nu}
R_{\nu\mu}({\nGam})+(n\!-\!2)\Lambda_b\,\right],
\end{eqnarray}
where $R_{\nu\mu}(\nGam)$ is the ordinary Ricci tensor
\begin{eqnarray}
R_{\nu\mu}(\nGam)
&=&\nGam^\alpha_{\nu\mu,\alpha}
-\nGam^\alpha_{\!\nu\alpha,\mu}
+\nGam^\sigma_{\nu\mu}\nGam^\alpha_{\sigma\alpha}
-\nGam^\sigma_{\nu\alpha}\nGam^\alpha_{\sigma\mu}.
\end{eqnarray}
The original Einstein-Schr\"{o}dinger theory (including the cosmological constant)
can even be derived from purely affine versions
of the Lagrangian densities described above, such as the Lagrangian density
used by Schr\"{o}dinger\cite{SchrodingerI},
\begin{eqnarray}
\label{Schrodingerslag}
{\mathcal L}(\nGam)=\sqrt{-det(\tR_{\nu\mu}(\nGam))}\,.
\end{eqnarray}

Whether one prefers the Lagrangian density (\ref{Palatini},\ref{HermitianizedRicci}) with the properties
(\ref{transposition_d},\ref{gauge_d},\ref{selftransplantation0})
or one of the alternatives, it is clear that the original
Einstein-Schr\"{o}dinger theory can be derived from rather simple principles.
The theory proposed in this paper is a natural extension of the original Einstein-Schr\"{o}dinger theory
to account for zero-point fluctuations.
One might perhaps also regard a spin-1/2 ${\mathcal L}_m$ term as a quantization effect,
that is as the ``first quantization'' of our charged solution, in which case all of
one-particle quantum electrodynamics results by including quantization effects in the
original Einstein-Schr\"{o}dinger theory.
The search for simple principles has led to many advances in physics,
and is what led Einstein to general relativity and also
to the Einstein-Schr\"{o}dinger theory\cite{Schilpp,EinsteinBianchi}.
%He says ``I have learned something else from the theory of gravitation:
%No ever so inclusive collection of empirical facts can ever lead to the setting up of such
%complicated equations. A theory can be tested by experience, but there is no way from experience
%to the setting up of a theory. Equations of such complexity as are the equations of the
%gravitational field can be found only through the discovery of a logically simple mathematical
%condition which determines them completely, or [at least] almost completely. Once one
%has those sufficiently strong formal conditions, one requires only little knowledge of
%facts for the setting up of a theory; in the case of the equations of gravitation it is
%the four-dimensionality and the symmetric tensor as expression for the structure of
%space which, together with the invariance concerning the continuous transformation-group,
%determine the equations almost completely.''
Einstein disliked the term
$\rmg F^{\nu\mu}\!F_{\mu\nu}/16\pi$ in the Einstein-Maxwell Lagrangian density. Referring to the equation
$G_{\nu\mu}\!\nobreak=\nobreak\!8\pi T_{\nu\mu}$ he states\cite{Schilpp} ``The right side is a
formal condensation of all things whose comprehension in the sense of a field-theory is still
problematic. Not for a moment, of course, did I doubt that this formulation was merely a
makeshift in order to give the general principle of relativity a preliminary closed
expression. For it was essentially not anything more than a theory of the gravitational field,
which was somewhat artificially isolated from a total field of as yet unknown structure.''
In modern times the term $\rmg F^{\nu\mu}\!F_{\mu\nu}/16\pi$ has become standard and
is rarely questioned.
The theory presented here suggests that this term should be questioned, and offers an
alternative which is based on simple principles
and which genuinely unifies gravitation and electromagnetism.
%and suggests an alternative and more natural explanation for electromagnetism.
%The author believes that this term should be questioned, and that the theory presented here
%offers a more natural explanation for electromagnetism.

%But why should it be there? Also, the Lagrangian density of classical general relativity
%and electromagnetism is evidently rather important, yet what does it's value actually represent?
%It is not (kinetic energy)-(potential energy) as in classical mechanics, so what is it?
%Why should its integral over all space be extremized? Of what significance is the covariant
%derivative of a Lagrangian density, and what happens if it is required to vanish?
%Such questions can guide the search for the fundamental physical laws,
%and as such they are legitimate questions to investigate.
\fi

\section{\label{Conclusions}Conclusions}
The Einstein-Schr\"{o}dinger theory is modified to include a cosmological constant $\Lambda_z$
which multiplies the symmetric metric.
This is assumed to be nearly cancelled by
Schr\"{o}dinger's ``bare'' cosmological constant $\Lambda_b$
which multiplies the nonsymmetric fundamental tensor,
such that the total cosmological constant $\Lambda\!=\!\Lambda_b\!+\!\Lambda_z$ matches measurement.
%Other fields could be included in a similar manner.
The resulting theory closely
approximates Einstein-Maxwell theory for $|\Lambda_z|\!\sim\!1/({\rm Planck~length})^2$,
and it becomes exactly Einstein-Maxwell theory
in the limit as $|\Lambda_z|\!\rightarrow\!\infty$.

\section*{Acknowledgements}
I am grateful to Clifford Will for discussions
and for helpful comments on drafts of this manuscript.
%Thanks also to Claude Bernard for his help.
This work was supported in part by the National Science Foundation under grant PHY~03-53180.
%\bigskip\\

%\newpage
\begin{appendix}

\ifnum\ExpandDerivations=1
\section{\label{UsefulIdentity}A divergence identity}
Here we derive (\ref{usefulidentity})
using only the definitions (\ref{gdef},\ref{fdef}) of $g_{\nu\mu}$
and $f_{\nu\mu}$, and the identity (\ref{sqrtdetcomma}),
\begin{eqnarray}
&&\left(N^{(\mu}{_{\nu)}}-\frac{1}{2}\delta^\mu_\nu
N^\rho_\rho\right)\!{_{;\,\mu}}
-\frac{3}{2}f^{\sigma\rho}N_{[\sigma\rho,\nu]}\rmt\Lambda_b^{\!-1/2}\\
&&=\frac{1}{2}g^{\sigma\rho}
(N_{(\rho\nu);\sigma}+N_{(\nu\sigma);\rho}-N_{(\rho\sigma);\nu})
%-\frac{3}{2}f^{\sigma\rho}N_{[\sigma\rho,\nu]}\\
-\frac{3}{2}f^{\sigma\rho}N_{[\sigma\rho;\nu]}\rmt\Lambda_b^{\!-1/2}\\
&&=\frac{1}{2}\frac{\rmN}{\rmg}\left[N^{\dashv(\sigma\rho)}
(N_{(\rho\nu);\sigma}+N_{(\nu\sigma);\rho}
-N_{(\rho\sigma);\nu})-3N^{\dashv[\rho\sigma]}
N_{[\sigma\rho;\nu]}\right]\\
&&=\frac{1}{2}\frac{\rmN}{\rmg}\left[\,N^{\dashv\sigma\rho}
(N_{(\rho\nu);\sigma}+N_{(\nu\sigma);\rho}
-N_{(\rho\sigma);\nu})+3N^{\dashv\sigma\rho}
N_{[\rho\nu;\sigma]}\right]\\
&&=\frac{1}{2}\frac{\rmN}{\rmg}N^{\dashv\sigma\rho}
(N_{\rho\nu;\sigma}+N_{\nu\sigma;\rho}-N_{\rho\sigma;\nu})\\
&&=\frac{1}{2}\frac{\rmN}{\rmg}\left[N^{\dashv\sigma\rho}
(N_{\rho\nu;\sigma}+N_{\nu\sigma;\rho})
-N^{\dashv\sigma\rho}(N_{\rho\sigma,\nu}
-\Gamma^\alpha_{\rho\nu}N_{\alpha\sigma}
-\Gamma^\alpha_{\sigma\nu}N_{\rho\alpha})\right]~~~~~\\
&&=-\frac{1}{2}\frac{\rmN}{\rmg}
(N^{\dashv\sigma\rho}{_{;\sigma}}N_{\rho\nu}
+N^{\dashv\sigma\rho}{_{;\rho}}N_{\nu\sigma})
-\frac{1}{\rmg}(\rmN\,)_{;\nu}\\
&&=-\frac{1}{2}\left[
\left(\frac{\rmN}{\rmg}N^{\dashv\sigma\rho}\right)
{_{\!\!;\sigma}}N_{\rho\nu}
+\left(\frac{\rmN}{\rmg}N^{\dashv\sigma\rho}\right)
{_{\!\!;\rho}}N_{\nu\sigma}\right]\\
&&=-\frac{1}{2}\left[
(g^{\rho\sigma}+f^{\rho\sigma}\rmt\Lambda_b^{\!-1/2}){_{;\sigma}}N_{\rho\nu}
+(g^{\rho\sigma}+f^{\rho\sigma}\rmt\Lambda_b^{\!-1/2}){_{;\rho}}N_{\nu\sigma}\right]\\
&&=f^{\sigma\rho}{_{;\sigma}}N_{[\rho\nu]}\rmt\Lambda_b^{\!-1/2}.
\end{eqnarray}
\fi

\ifnum\ExpandDerivations=1
\section{\label{ExtractionofConnectionAddition}Extraction of a connection
addition from the Hermitianized Ricci tensor}
Substituting $\tGam^\alpha_{\nu\mu}\!=\!\Gamma^\alpha_{\nu\mu}
\!+\!\Upsilon^\alpha_{\nu\mu}$ from (\ref{gammadecomposition},\ref{Christoffel})
into (\ref{HermitianizedRiccit})
%and using the notation $\bUps^\alpha_{\nu\mu}
%=\Upsilon^\alpha_{(\nu\mu)}$,
%$\cUps^\alpha_{\nu\mu}=\Upsilon^\alpha_{[\nu\mu]}$
gives
\begin{eqnarray}
%\label{Ricciaddition}
\hR_{\nu\mu}(\tGam)
\!&=&2[(\Gamma^\alpha_{\nu[\mu}
\!+\!\Upsilon^\alpha_{\nu[\mu}){_{,\alpha]}}
+(\Gamma^\sigma_{\nu[\mu}
\!+\!\Upsilon^\sigma_{\nu[\mu})(\Gamma^\alpha_{\sigma|\alpha]}
\!+\!\Upsilon^\alpha_{\sigma|\alpha]})]
+\!(\Gamma^\alpha_{\alpha[\nu}
\!+\!\Upsilon^\alpha_{\alpha[\nu})_{,\mu]}\\
\iftrue
&=&R_{\nu\mu}(\Gamma)+\Upsilon^\alpha_{\nu\mu,\alpha}
-\Gamma^\sigma_{\nu\alpha}\Upsilon^\alpha_{\sigma\mu}
+\Gamma^\alpha_{\sigma\alpha}\Upsilon^\sigma_{\nu\mu}
-\Gamma^\alpha_{\sigma\mu}\Upsilon^\sigma_{\nu\alpha}\nonumber\\
\nopagebreak
&&~~~~~~~~~-\Upsilon^\alpha_{\alpha(\nu,\mu)}
+\Gamma^\sigma_{\nu\mu}\Upsilon^\alpha_{\sigma\alpha}
-\Upsilon^\sigma_{\nu\alpha}\Upsilon^\alpha_{\sigma\mu}
+\Upsilon^\sigma_{\nu\mu}\Upsilon^\alpha_{\sigma\alpha}\\
\fi
%\label{Ricciaddition2}
\!&=&R_{\nu\mu}(\Gamma)+\Upsilon^\alpha_{\nu\mu;\alpha}
-\Upsilon^\alpha_{\alpha(\nu;\mu)}
-\Upsilon^\sigma_{\nu\alpha}\Upsilon^\alpha_{\sigma\mu}
+\Upsilon^\sigma_{\nu\mu}\Upsilon^\alpha_{\sigma\alpha},\\
\!R_{(\nu\mu)}(\tGam)
\label{Ricciadditionsymmetric}
\!&=& R_{\nu\mu}(\Gamma)+{\Upsilon}^\alpha_{\!(\nu\mu);\alpha}
\!-\!\Upsilon^\alpha_{\alpha(\nu;\mu)}
\!-\!{\Upsilon}^\sigma_{\!(\nu\alpha)}{\Upsilon}^\alpha_{\!(\sigma\mu)}
\!-\!{\Upsilon}^\sigma_{\![\nu\alpha]}{\Upsilon}^\alpha_{\![\sigma\mu]}
\!+\!{\Upsilon}^\sigma_{\!(\nu\mu)}\Upsilon^\alpha_{\sigma\alpha},\\
\!R_{[\nu\mu]}(\tGam)
\label{Ricciadditionantisymmetric}
\!&=&{\Upsilon}^\alpha_{\![\nu\mu];\alpha}
%\!-\!\Upsilon^\alpha_{\alpha[\nu;\mu]}
\!-\!{\Upsilon}^\sigma_{\!(\nu\alpha)}{\Upsilon}^\alpha_{\![\sigma\mu]}
\!-\!{\Upsilon}^\sigma_{\![\nu\alpha]}{\Upsilon}^\alpha_{\!(\sigma\mu)}
\!+\!{\Upsilon}^\sigma_{\![\nu\mu]}\Upsilon^\alpha_{\sigma\alpha}.
\end{eqnarray}
Also, substituting $\nGam^\alpha_{\!\nu\mu}\!=\!\tGam^\alpha_{\!\nu\mu}
\!+[\,\delta^\alpha_\mu \Aphi_\nu\!-\delta^\alpha_\nu \Aphi_\mu]\!\rmt\Lambda_b^{\!1/2}$
from (\ref{gamma_natural}) into (\ref{HermitianizedRicci}) and using
$\tGam^\alpha_{\nu\alpha}\!=\!\nGam^\alpha_{\!(\nu\alpha)}\!=\!\tGam^\alpha_{\alpha\nu}$ gives
\begin{eqnarray}
\hR_{\nu\mu}(\nGam)
\!&=&\tGam^\alpha_{\!\nu\mu,\alpha}
\!+[\,\delta^\alpha_\mu \Aphi_\nu\!-\delta^\alpha_\nu \Aphi_\mu]_{,\alpha}\!\rmt\Lambda_b^{\!1/2}
\!-\tGam^\alpha_{\!\alpha(\nu,\mu)}\nonumber\\
&&+\left(\tGam^\sigma_{\!\nu\mu}+[\,\delta^\sigma_\mu \Aphi_\nu
\!-\delta^\sigma_\nu \Aphi_\mu]\rmt\Lambda_b^{\!1/2}\right)\tGam^\alpha_{\!\sigma\alpha}\nonumber\\
&&-\left(\tGam^\sigma_{\!\nu\alpha}\!+[\,\delta^\sigma_\alpha \Aphi_\nu
-\delta^\sigma_\nu \Aphi_\alpha]\rmt\Lambda_b^{\!1/2}\right)\!
\left(\tGam^\alpha_{\!\sigma\mu}\!+[\,\delta^\alpha_\mu \Aphi_\sigma-\delta^\alpha_\sigma \Aphi_\mu]\rmt\Lambda_b^{\!1/2}\right)\nonumber\\
&&+2(n\!-\!1)A_\nu A_\mu\Lambda_b\\
\!&=&\tGam^\alpha_{\!\nu\mu,\alpha}
\!+2\Aphi_{[\nu,\mu]}\rmt\Lambda_b^{\!1/2}
-\tGam^\alpha_{\!\alpha(\nu,\mu)}+\tGam^\sigma_{\!\nu\mu}\tGam^\alpha_{\!\sigma\alpha}
+[\,\Aphi_\nu\tGam^\alpha_{\!\mu\alpha}
-\Aphi_\mu\tGam^\alpha_{\!\nu\alpha}]\rmt\Lambda_b^{\!1/2}\nonumber\\
&&-\tGam^\sigma_{\!\nu\alpha}\tGam^\alpha_{\!\sigma\mu}
-[\,\tGam^\sigma_{\!\nu\mu}\Aphi_\sigma-\tGam^\sigma_{\!\nu\sigma}\Aphi_\mu]\rmt\Lambda_b^{\!1/2}
-[\,\Aphi_\nu\tGam^\alpha_{\!\alpha\mu}
-\Aphi_\alpha\tGam^\alpha_{\!\nu\mu}]\rmt\Lambda_b^{\!1/2}\nonumber\\
&&+2\Aphi_\nu\Aphi_\mu(1-n-1+1)\Lambda_b
+2(n\!-\!1)A_\nu A_\mu\Lambda_b\\
&=&\tGam^\alpha_{\!\nu\mu,\alpha}
-\tGam^\alpha_{\alpha(\nu,\mu)}+\tGam^\sigma_{\!\nu\mu}\tGam^\alpha_{\!\sigma\alpha}
-\tGam^\sigma_{\!\nu\alpha}\tGam^\alpha_{\!\sigma\mu}
\!+2\Aphi_{[\nu,\mu]}\rmt\Lambda_b^{\!1/2}\\
\label{RnGam}
&=&\hR_{\nu\mu}(\tGam)\!+2\Aphi_{[\nu,\mu]}\rmt\Lambda_b^{\!1/2}.
\end{eqnarray}
\fi

\ifnum\ExpandDerivations=1
\section{\label{VariationalDerivative}Variational derivatives for fields with
the symmetry $\tGam^\sigma_{\![\mu\sigma]}\!=\!0$}
The field equations associated with a field with symmetry properties must have the
same number of independent components as the field. For a field with the symmetry
$\tGam^\sigma_{\![\mu\sigma]}=0$, the field equations can be found by introducing
a Lagrange multiplier $\Omega^\mu$,
\begin{eqnarray}
0=\delta\int({\mathcal L}+\Omega^\mu\tGam^\sigma_{\![\mu\sigma]})d^n x.
\end{eqnarray}
Minimizing the integral with respect to $\Omega^\mu$ shows that the symmetry is enforced.
Using the definition,
\begin{eqnarray}
\frac{\Delta{\mathcal L}}{\Delta\tGam^\beta_{\tau\rho}}
=\frac{\partial{\mathcal L}}{\partial\tGam^\beta_{\tau\rho}}
-\left(\frac{\partial{\mathcal L}}
{\partial\tGam^\beta_{\tau\rho,\omega}}\right)\!{_{,\,\omega}}~...~,
\end{eqnarray}
and minimizing the integral with respect to $\tGam^\beta_{\!\tau\rho}$ gives
\begin{eqnarray}
\label{minimization}
0=\frac{\Delta{\mathcal L}}{\Delta\tGam^\beta_{\tau\rho}}
+\Omega^\mu\delta^\sigma_\beta\delta^\tau_{[\mu}\delta^\rho_{\sigma]}
=\frac{\Delta{\mathcal L}}{\Delta\tGam^\beta_{\tau\rho}}
+\frac{1}{2}(\Omega^\tau\delta^\rho_\beta-\delta^\tau_\beta\Omega^\rho).
\end{eqnarray}
Contracting this on the left and right gives
\begin{eqnarray}
\label{lagrangemultiplier}
\Omega^\rho=\frac{2}{(n\!-\!1)}\frac{\Delta{\mathcal L}}{\Delta\tGam^\alpha_{\alpha\rho}}
=-\frac{2}{(n\!-\!1)}\frac{\Delta{\mathcal L}}{\Delta\tGam^\alpha_{\rho\alpha}}.
\end{eqnarray}
Substituting (\ref{lagrangemultiplier}) back into (\ref{minimization}) gives
\begin{eqnarray}
\label{usefulresult}
0&=&\frac{\Delta{\mathcal L}}
{\Delta\tGam^\beta_{\tau\rho}}
-\frac{\delta^\tau_\beta}{(n\!-\!1)}
\frac{\Delta{\mathcal L}}{\Delta\tGam^\alpha_{\alpha\rho}}
-\frac{\delta^\rho_\beta}{(n\!-\!1)} \frac{\Delta{\mathcal L}}
{\Delta\tGam^\alpha_{\tau\alpha}}.
\end{eqnarray}
In (\ref{lagrangemultiplier},\ref{usefulresult}) the index contractions occur after
the derivatives. Contracting (\ref{usefulresult}) on the right and left gives the
same result, so it has the same number of independent components as
$\tGam^\alpha_{\mu\nu}$. This is a general expression for the field equations
associated with a field having the symmetry $\tGam^\sigma_{\![\mu\sigma]}=0$.
\fi

\section{\label{ApproximateFandg}Solution for $N_{\nu\mu}$ in terms
of $g_{\nu\mu}$ and $f_{\nu\mu}$}
Here we invert the definitions (\ref{gdef},\ref{fdef}) of
$g_{\nu\mu}$ and $f_{\nu\mu}$ to obtain
(\ref{approximateNbar},\ref{approximateNhat}), the approximation of
$N_{\nu\mu}$ in terms of $g_{\nu\mu}$ and $f_{\nu\mu}$.
First let us define the notation
\begin{eqnarray}
\label{hfdef}
\hf^{\nu\mu}\!=\!f^{\nu\mu}\rmt\,\Lambda_b^{\!-1/2}.
\end{eqnarray}
We assume that $|\hf^\nu{_\mu}|\!\ll\!1$ for all components of the
unitless field $\hf^\nu{_\mu}$, and find a solution
in the form of a power series expansion in $\hf^\nu{_\mu}$.
Lowering an index on (\ref{Wdef}) gives
%$(\rmN/\rmg\,)N^{\dashv\mu\nu}\!=g^{\nu\mu}\!+\hf^{\nu\mu}$
%from (\ref{gdef},\ref{fdef}) gives
\begin{eqnarray}
\label{gminusF2}
(\rmN/\rmg~)N^{\dashv\mu}{_\alpha}
=\delta^\mu_\alpha-\hf^\mu{_\alpha}.
\end{eqnarray}
Let us consider the tensor $\hf^\mu{_\alpha}\!=\!\hf^{\mu\nu}g_{\nu\alpha}$.
Because $g_{\nu\alpha}$ is symmetric and $\hf^{\mu\nu}$ is antisymmetric, it is
clear that $\hf^\alpha{_\alpha}\!=\!0$. Also because $\hf_{\nu\sigma}\hf^\sigma{_\mu}$ is
symmetric it is clear that $\hf^\nu{_\sigma}\hf^\sigma{_\mu}\hf^\mu{_\nu}=0$.
In matrix language therefore $tr(\hf)\!=\!0,~tr(\hf^3)\!=\!0$,
and in fact $tr(\hf^p)\!=\!0$ for any odd p.
Using the well known formula $det(e^M)=exp\,(tr(M))$ and
the power series $ln(1\!-\!x)=-x-x^2/2-x^3/3-x^4/4\dots$
we then get\cite{Deif},
\begin{eqnarray}
\label{lndetspecial}
ln(det(I\!-\!\hf))&=&tr(ln(I\!-\!\hf))
=-\hf^\rho{_\sigma}\hf^\sigma{_\rho}/2+(\hf^4)\dots
\end{eqnarray}
Here the notation $(\hf^4)$ refers to terms like
$\hf^\tau{_\alpha}\hf^\alpha{_\sigma}\hf^\sigma{_\rho}\hf^\rho{_\tau}$.
Taking $ln(det())$ on both sides of (\ref{gminusF2}) using (\ref{lndetspecial})
and the identities $det(sM^{})\!=s^n det(M^{})$, $det(M^{-1}_{})\!=1/det(M^{})$
gives
\begin{eqnarray}
ln\!\left(\!\frac{\lower2pt\hbox{$\rmN$}}{\rmg}\right)
\!&=&\!\frac{1}{(n\!-\!2)}\,ln\!\left(\!\frac{\lower2pt\hbox{$N^{(n/2-1)}$}}{g^{(n/2-1)}}\right)
\label{lnapproxdetN}
=-\frac{1}{2(n\!-\!2)}\,\hf^\rho{_\sigma}\hf^\sigma{_\rho}
+(\hf^4)\dots
\end{eqnarray}
Taking $e^x$ on both sides of (\ref{lnapproxdetN}) and using $e^x=1+x+x^2/2\dots$ gives
\begin{eqnarray}
\label{approxdetN}
\frac{\lower2pt\hbox{$\rmN$}}{\rmg}
=1\!-\!\frac{1}{2(n\!-\!2)}\,\hf^{\rho\sigma}\!\hf_{\sigma\rho}
+(\hf^4)\dots
\end{eqnarray}
Using the power series $(1\!-\!x)^{-1}\!=\!1+x+x^2+x^3\dots$,
or multiplying (\ref{gminusF2}) term by term,
we can calculate the inverse of (\ref{gminusF2}) to get\cite{Deif}
\begin{eqnarray}
(\rmg/\rmN\,) N^\nu{_\mu}
=\delta^\nu_\mu+\hf^\nu{_\mu}+\hf^\nu{_\sigma}\hf^\sigma{_\mu}
+\hf^\nu{_\rho}\hf^\rho{_\sigma}\hf^\sigma{_\mu}+(\hf^4)\dots\\
%\end{eqnarray}
%and therefore,
%\begin{eqnarray}
\label{approxN}
N_{\nu\mu}
=(\rmN/\rmg~)(g_{\nu\mu}+\hf_{\nu\mu}+\hf_{\nu\sigma}\hf^\sigma{_\mu}
+\hf_{\nu\rho}\hf^\rho{_\sigma}\hf^\sigma{_\mu}+(\hf^4)\dots).
\end{eqnarray}
Here the notation $(\hf^4)$ refers to terms like
$\hf_{\nu\alpha}\hf^\alpha{_\sigma}\hf^\sigma{_\rho}\hf^\rho{_\mu}$.
Since $\hf_{\nu\sigma}\hf^\sigma{_\mu}$ is symmetric and
$\hf_{\nu\rho}\hf^\rho{_\sigma}\hf^\sigma{_\mu}$ is antisymmetric, we obtain from
(\ref{approxN},\ref{approxdetN},\ref{hfdef}) the final result
(\ref{approximateNbar},\ref{approximateNhat}).
\iffalse
\begin{eqnarray}
N_{(\nu\mu)}&=&g_{\nu\mu}+\hf_{\nu\sigma}\h\hf^\sigma{_\mu}
-\frac{\lower2pt\hbox{$1$}}{2(n\!-\!2)}g_{\nu\mu}\h\hf^{\rho\sigma}\!\hf_{\sigma\rho}+(\hf^4),\\
\label{approxNhat}
N_{[\nu\mu]}&=&\hf_{\nu\mu}+(\hf^3).
\end{eqnarray}
\fi

\ifnum\ExpandDerivations=1
\section{\label{ApproximateGamma}Solution for
${\tGam^\alpha_{\nu\mu}}$ in terms of $g_{\nu\mu}$ and
$f_{\nu\mu}$, with sources}
Here we derive the approximate solution
(\ref{upsilonsymmetric},\ref{upsilonantisymmetric})
to the connection equations (\ref{JSconnection}).
First let us define the notation
\begin{eqnarray}
\label{notation}
\hf^{\nu\mu}\!=\!f^{\nu\mu}\rmt\,\Lambda_b^{\!-1/2},~~~
\hj^\sigma\!=\!j^\sigma\rmt\,\Lambda_b^{\!-1/2},~~~
\bUps^\alpha_{\nu\mu}\!=\!\Upsilon^\alpha_{(\nu\mu)},~~~
\cUps^\alpha_{\nu\mu}\!=\!\Upsilon^\alpha_{[\nu\mu]}.
\end{eqnarray}
We assume that $|\hf^\nu{_\mu}|\!\ll\!1$ for all components of the
unitless field $\hf^\nu{_\mu}$, and find a solution
in the form of a power series expansion in $\hf^\nu{_\mu}$.
%It is easier to work with the contravariant connection equations
%(\ref{JSconnection}) than the covariant connection equations (\ref{JSconnection0}).
Using (\ref{contravariant}) and
\begin{eqnarray}
\tGam^\sigma_{ \sigma\alpha}
&=&\frac{(\rmN)_{,\alpha}}{\rmN}+\frac{8\pi}{(n\!-\!2)(n\!-\!1)}\hj^\sigma N_{[\sigma\alpha]}
\end{eqnarray}
from (\ref{der0}) and
%$(\!\rmN)_{,\alpha}\!=\!\rmN\,\tGam^\sigma_{ \sigma\alpha}$,
$\tGam^\alpha_{\nu\mu}\!=\!\Gamma^\alpha_{\nu\mu}\!+\!\Upsilon^\alpha_{\nu\mu}$,
$(\rmN/\rmg\,)N^{\dashv\mu\nu}\!=\!g^{\nu\mu}\!+\!\hf^{\nu\mu}$
from (\ref{gammadecomposition},\ref{gdef},\ref{fdef}) we get
\begin{eqnarray}
\label{contravariant2}
0&=&\frac{\rmN}{\rmg}(N^{\dashv \mu\nu}{_{,\alpha}}
+\tGam^\nu_{\tau\alpha}N^{\dashv\mu\tau}
+\tGam^\mu_{\alpha\tau}N^{\dashv\tau\nu})\nonumber\\
&& -\frac{8\pi}{(n\!-\!1)}\!\left(
\hj^{[\mu}\delta^{\nu]}_\alpha
+\frac{1}{(n\!-\!2)}\hj^\tau N_{[\tau\alpha]}N^{\dashv\mu\nu}\!\right)\\
&=&\left(\!\frac{\rmN N^{\dashv \mu\nu}}{\rmg}\right)\!{_{,\alpha}}
+\frac{\rmN}{\rmg}(\tGam^\nu_{\tau\alpha}N^{\dashv\mu\tau}
+\tGam^\mu_{\alpha\tau}N^{\dashv\tau\nu}
-(\tGam^\sigma_{ \sigma\alpha}
\!-\!\Gamma^\sigma_{\sigma\alpha})N^{\dashv\mu\nu})\nonumber\\
&& -\frac{8\pi}{(n\!-\!1)}\hj^{[\mu}\delta^{\nu]}_\alpha\\
&=&(g^{\nu\mu}+\hf^{\nu\mu}){_{;\alpha}}
+\Upsilon^\nu_{\tau\alpha}(g^{\tau\mu}   +\hf^{\tau\mu})
+\Upsilon^\mu_{   \alpha\tau}(g^{\nu\tau}+\hf^{\nu\tau})
-\Upsilon^\sigma_{ \sigma\alpha}(g^{\nu\mu}  +\hf^{\nu\mu})\nonumber\\
&& -\frac{8\pi}{(n\!-\!1)}\hj^{[\mu}\delta^{\nu]}_\alpha\\
\label{contravariantreduced}
&=&\hf^{\nu\mu}{_{;\alpha}}
+\Upsilon^\nu_{\tau\alpha}g^{\tau\mu}
+\Upsilon^\nu_{\tau\alpha}\hf^{\tau\mu}
+\Upsilon^\mu_{\alpha\tau}g^{\nu\tau}
+\Upsilon^\mu_{\alpha\tau}\hf^{\nu\tau}
-\Upsilon^\sigma_{\sigma\alpha}g^{\nu\mu}
-\Upsilon^\sigma_{\sigma\alpha}\hf^{\nu\mu}\nonumber\\
&& +\frac{4\pi}{(n\!-\!1)}(\hj^\nu\delta^\mu_\alpha-\hj^\mu\delta^\nu_\alpha).
\end{eqnarray}
Contracting this with $g_{\nu\mu}$ gives
\begin{eqnarray}
\label{contractedupsilon}
0=(2-n)\Upsilon^\sigma_{\sigma\alpha}
-2{\cUps}^\sigma_{\alpha\tau}\hf^\tau{_\sigma}
~~\Rightarrow ~~\Upsilon^\sigma_{\sigma\alpha}
=\frac{2}{(n\!-\!2)}{\cUps}_{\sigma\tau\alpha}\hf^{\tau\sigma}.
\end{eqnarray}
Lowering the indices of (\ref{contravariantreduced}) and making linear
combinations of its permutations gives
\begin{eqnarray}
\Upsilon_{\alpha\nu\mu}
=\Upsilon_{\alpha\nu\mu}
&+&\frac{1}{2}\left(\hf_{\nu\mu;\alpha}
+\!\Upsilon_{\nu\mu\alpha}+\!\Upsilon_{\nu\tau\alpha}\hf^\tau{_\mu}
+\!\Upsilon_{\mu\alpha\nu}+\!\Upsilon_{\mu\alpha\tau}\hf_\nu{^\tau}
-\!\Upsilon^\sigma_{\sigma\alpha}g_{\nu\mu}
-\!\Upsilon^\sigma_{\sigma\alpha}\hf_{\nu\mu}\phantom{\frac{1}{2}}\right.\nonumber\\
&& ~~~~+\left.\frac{4\pi}{(n\!-\!1)}(\hj_\nu g_{\alpha\mu}-\hj_\mu g_{\nu\alpha})\right)\nonumber\\
&-&\frac{1}{2}\left(\hf_{\mu\alpha;\nu}
+\!\Upsilon_{\mu\alpha\nu}+\!\Upsilon_{\mu\tau\nu}\hf^\tau{_\alpha}
+\!\Upsilon_{\alpha\nu\mu}+\!\Upsilon_{\alpha\nu\tau}\hf_\mu{^\tau}
-\!\Upsilon^\sigma_{\sigma\nu}g_{\mu\alpha}
-\!\Upsilon^\sigma_{\sigma\nu}\hf_{\mu\alpha}\phantom{\frac{1}{2}}\right.\nonumber\\
&& ~~~~+\left.\frac{4\pi}{(n\!-\!1)}(\hj_\mu g_{\nu\alpha}-\hj_\alpha g_{\mu\nu})\right)\nonumber\\
&-&\frac{1}{2}\left(\hf_{\alpha\nu;\mu}
+\!\Upsilon_{\alpha\nu\mu}+\!\Upsilon_{\alpha\tau\mu}\hf^\tau{_\nu}
+\!\Upsilon_{\nu\mu\alpha}+\!\Upsilon_{\nu\mu\tau}\hf_\alpha{^\tau}
-\!\Upsilon^\sigma_{\sigma\mu}g_{\alpha\nu}
-\!\Upsilon^\sigma_{\sigma\mu}\hf_{\alpha\nu}\phantom{\frac{1}{2}}\right.\nonumber\\
&& ~~~~+\left.\frac{4\pi}{(n\!-\!1)}(\hj_\alpha g_{\mu\nu}-\hj_\nu g_{\alpha\mu})\right).
\end{eqnarray}
Cancelling out the $\Upsilon_{\alpha\nu\mu}$ terms on the right-hand side, collecting terms,
and separating out the symmetric and antisymmetric parts gives,
\begin{eqnarray}
\label{barupsilonofN}
\bUps_{\alpha\nu\mu}&=&
{\cUps}_{[\alpha\mu]\tau}\hf^\tau{_\nu}
+{\cUps}_{[\alpha\nu ]\tau}\hf^\tau{_\mu}
+{\cUps}_{(\nu\mu)\tau}\hf^\tau{_\alpha}
-\frac{1}{2}\Upsilon^\sigma_{\sigma\alpha}g_{\nu\mu}
+\Upsilon^\sigma_{\sigma(\nu}g_{\mu)\alpha}\\
\label{hatupsilonofN}
\cUps_{\alpha\nu\mu}&=&
-{\bUps}_{(\alpha\mu)\tau}\hf^\tau{_\nu}
+{\bUps}_{(\alpha\nu )\tau}\hf^\tau{_\mu}
+{\bUps}_{[\nu\mu]\tau}\hf^\tau{_\alpha}
-\frac{1}{2}\Upsilon^\sigma_{\sigma\alpha}\hf_{\nu\mu}
+\Upsilon^\sigma_{\sigma[\nu}\hf_{\mu]\alpha}\nonumber\\
&&+\frac{1}{2}
(\hf_{\nu\mu;\alpha}+\hf_{\alpha\mu;\nu}-\hf_{\alpha\nu;\mu})
+\frac{8\pi}{(n\!-\!1)}\hj_{[\nu} g_{\mu]\alpha}.
\end{eqnarray}
Substituting (\ref{barupsilonofN}) into (\ref{hatupsilonofN})
\ifnum\ExpandDerivations=1
%\pagebreak
\begin{eqnarray}
\cUps_{\alpha\nu\mu}
&=&-\frac{1}{2}\left({\cUps}_{[\alpha\tau]\sigma}\hf^\sigma{_\mu}
+{\cUps}_{[\alpha\mu ]\sigma}\hf^\sigma{_\tau}
+{\cUps}_{(\mu\tau)\sigma}\hf^\sigma{_\alpha}
-\frac{1}{2}\Upsilon^\sigma_{\sigma\alpha}g_{\mu\tau}
+\Upsilon^\sigma_{\sigma(\mu}g_{\tau)\alpha}\right)\hf^\tau{_\nu}\nonumber\\
&&-\frac{1}{2}\left({\cUps}_{[\mu\tau]\sigma}\hf^\sigma{_\alpha}
+{\cUps}_{[\mu\alpha ]\sigma}\hf^\sigma{_\tau}
+{\cUps}_{(\alpha\tau)\sigma}\hf^\sigma{_\mu}
-\frac{1}{2}\Upsilon^\sigma_{\sigma\mu}g_{\alpha\tau}
+\Upsilon^\sigma_{\sigma(\alpha}g_{\tau)\mu}\right)\hf^\tau{_\nu}\nonumber\\
&&+\frac{1}{2}\left({\cUps}_{[\alpha\tau]\sigma}\hf^\sigma{_\nu}
+{\cUps}_{[\alpha\nu ]\sigma}\hf^\sigma{_\tau}
+{\cUps}_{(\nu\tau)\sigma}\hf^\sigma{_\alpha}
-\frac{1}{2}\Upsilon^\sigma_{\sigma\alpha}g_{\nu\tau}
+\Upsilon^\sigma_{\sigma(\nu}g_{\tau)\alpha}\right)\hf^\tau{_\mu}\nonumber\\
&&+\frac{1}{2}\left({\cUps}_{[\nu\tau]\sigma}\hf^\sigma{_\alpha}
+{\cUps}_{[\nu\alpha ]\sigma}\hf^\sigma{_\tau}
+{\cUps}_{(\alpha\tau)\sigma}\hf^\sigma{_\nu}
-\frac{1}{2}\Upsilon^\sigma_{\sigma\nu}g_{\alpha\tau}
+\Upsilon^\sigma_{\sigma(\alpha}g_{\tau)\nu}\right)\hf^\tau{_\mu}\nonumber\\
&&+\frac{1}{2}\left({\cUps}_{[\nu\tau]\sigma}\hf^\sigma{_\mu}
+{\cUps}_{[\nu\mu ]\sigma}\hf^\sigma{_\tau}
+{\cUps}_{(\mu\tau)\sigma}\hf^\sigma{_\nu}
-\frac{1}{2}\Upsilon^\sigma_{\sigma\nu}g_{\mu\tau}
+\Upsilon^\sigma_{\sigma(\mu}g_{\tau)\nu}\right)\hf^\tau{_\alpha}\nonumber\\
&&-\frac{1}{2}\left({\cUps}_{[\mu\tau]\sigma}\hf^\sigma{_\nu}
+{\cUps}_{[\mu\nu ]\sigma}\hf^\sigma{_\tau}
+{\cUps}_{(\nu\tau)\sigma}\hf^\sigma{_\mu}
-\frac{1}{2}\Upsilon^\sigma_{\sigma\mu}g_{\nu\tau}
+\Upsilon^\sigma_{\sigma(\nu}g_{\tau)\mu}\right)\hf^\tau{_\alpha}\nonumber\\
&&-\frac{1}{2}\Upsilon^\sigma_{\sigma\alpha}\hf_{\nu\mu}
+\Upsilon^\sigma_{\sigma[\nu}\hf_{\mu]\alpha}\nonumber\\
&&+\frac{1}{2}
(\hf_{\nu\mu;\alpha}+\hf_{\alpha\mu;\nu}-\hf_{\alpha\nu;\mu})
+\frac{8\pi}{(n\!-\!1)}\hj_{[\nu} g_{\mu]\alpha}\nonumber\\
%\nonumber\\
%\nonumber\\
&=&-\frac{1}{2}\left({\cUps}_{\alpha\tau\sigma}\hf^\sigma{_\mu}
+{\cUps}_{\mu\tau\sigma}\hf^\sigma{_\alpha}
\right)\hf^\tau{_\nu}\nonumber\\
&&+\frac{1}{2}\left({\cUps}_{\alpha\tau\sigma}\hf^\sigma{_\nu}
+{\cUps}_{\nu\tau\sigma}\hf^\sigma{_\alpha}
\right)\hf^\tau{_\mu}\nonumber\\
&&+\frac{1}{2}\left({\cUps}_{\tau\mu\sigma}\hf^\sigma{_\nu}
+2{\cUps}_{[\nu\mu]\sigma}\hf^\sigma{_\tau}
-{\cUps}_{\tau\nu\sigma}\hf^\sigma{_\mu}
\right)\hf^\tau{_\alpha}\nonumber\\
&&+\frac{1}{2}\Upsilon^\sigma_{\sigma\alpha}\hf_{\mu\nu}
+\Upsilon^\sigma_{\sigma\tau}\hf^\tau{_{[\mu}}g_{\nu]\alpha}\nonumber\\
&&+\frac{1}{2}
(\hf_{\nu\mu;\alpha}+\hf_{\alpha\mu;\nu}-\hf_{\alpha\nu;\mu})
+\frac{8\pi}{(n\!-\!1)}\hj_{[\nu} g_{\mu]\alpha}\nonumber,
\end{eqnarray}
\fi
and using (\ref{contractedupsilon}) gives,
\begin{eqnarray}
\label{upsilonhateq}
\cUps_{\alpha\nu\mu}
&=&{\cUps}_{\alpha\sigma\tau}\hf^\sigma{_\mu}\hf^\tau{_\nu}
+{\cUps}_{(\mu\sigma)\tau}\hf^\sigma{_\alpha}\hf^\tau{_\nu}
-{\cUps}_{(\nu\sigma)\tau}\hf^\sigma{_\alpha}\hf^\tau{_\mu}
+{\cUps}_{[\nu\mu]\sigma}\hf^\sigma{_\tau}
\hf^\tau{_\alpha}\nonumber\\
&&+\frac{1}{(n\!-\!2)}{\cUps}_{\sigma\tau\alpha}\hf^{\tau\sigma}\hf_{\mu\nu}
+\frac{2}{(n\!-\!2)}{\cUps}_{\sigma\rho\tau}\hf^{\rho\sigma}
\hf^\tau{_{[\mu}}g_{\nu]\alpha}\nonumber\\
&&+\frac{1}{2}
(\hf_{\nu\mu;\alpha}+\hf_{\alpha\mu;\nu}-\hf_{\alpha\nu;\mu})
+\frac{8\pi}{(n\!-\!1)}\hj_{[\nu} g_{\mu]\alpha}.
\end{eqnarray}
Equation (\ref{upsilonhateq}) is useful for finding exact solutions
to the connection equations because it
consists of only $n^2(n-1)/2$ equations in the $n^2(n-1)/2$ unknowns
${\cUps}_{\alpha\nu\mu}$.
Also, from (\ref{upsilonhateq}) we can immediately see that
\begin{eqnarray}
\label{upsilonantisymmetriclowered}
\cUps_{\alpha\nu\mu}
=\frac{1}{2}(\hf_{\nu\mu;\alpha}+\hf_{\alpha\mu;\nu}-\hf_{\alpha\nu;\mu})
+\frac{4\pi}{(n\!-\!1)}(\hj_{\nu} g_{\mu\alpha}-\hj_{\mu} g_{\nu\alpha})+(\hf^{3\prime})\dots.
\end{eqnarray}
Here the notation $(\hf^{3\prime})$ refers to terms like
$\hf_{\alpha\tau}\hf^\tau{_\sigma}\hf^\sigma{_{[\nu;\mu]}}$.
With (\ref{upsilonantisymmetriclowered}) as a starting point, one can calculate more
accurate $\cUps_{\alpha\nu\mu}$ by recursively substituting
the current $\cUps_{\alpha\nu\mu}$ into (\ref{upsilonhateq}).
%When the desired accuracy is achieved,
Then this
$\cUps_{\alpha\nu\mu}$ can be substituted into
(\ref{contractedupsilon},\ref{barupsilonofN}) to get $\bUps_{\alpha\nu\mu}$.
For our purposes (\ref{upsilonantisymmetriclowered}) will be accurate enough.
Substituting (\ref{upsilonantisymmetriclowered}) into (\ref{contractedupsilon}) we get
\begin{eqnarray}
%\label{upsilonantisymmetriclowered}
%{\cUps}_{\alpha\nu\mu}
%&\approx&\frac{1}{2}(\hf_{\nu\mu;\alpha}
%+\hf_{\alpha\mu;\nu}-\hf_{\alpha\nu;\mu})
%+\frac{8\pi}{(n\!-\!1)}\hj_{[\nu} g_{\mu]\alpha},\\
{\cUps}_{(\alpha\nu)\mu}
&=&-\hf_{\mu(\nu;\alpha)}
+\frac{4\pi}{(n\!-\!1)}(\hj_{(\nu} g_{\alpha)\mu}-\hj_\mu g_{\nu\alpha})+(\hf^{3\prime})\dots,\\
{\cUps}_{[\alpha\nu]\mu}&=&\frac{1}{2}\hf_{\nu\alpha;\mu}
+\frac{4\pi}{(n\!-\!1)}\hj_{[\nu} g_{\alpha]\mu}+(\hf^{3\prime})\dots,\\
\Upsilon^\sigma_{\sigma\alpha}&=&
\ifnum\ExpandDerivations=1
\frac{2}{(n\!-\!2)}\left(\frac{1}{2}\hf_{\tau\sigma;\alpha}
+\frac{4\pi}{(n\!-\!1)}\hj_{[\tau} g_{\sigma]\alpha}\right)\hf^{\tau\sigma}+(\hf^{4\prime})\dots\nonumber\\
&=&
\fi
\label{upsiloncontractedappendix}
\frac{-1}{2(n\!-\!2)}(\hf^{\rho\sigma}\!\hf_{\sigma\rho})_{,\alpha}
+\frac{8\pi}{(n\!-\!1)(n\!-\!2)}\hj^\tau \hf_{\tau\alpha}+(\hf^{4\prime})\dots.
\end{eqnarray}
%where
%\begin{eqnarray}
%\ff=\hf^{\tau\sigma}\hf_{\sigma\tau}.
%\end{eqnarray}
Substituting these equations into (\ref{barupsilonofN}) gives
\begin{eqnarray}
{\bUps}_{\alpha\nu\mu}&=&
\ifnum\ExpandDerivations=1
-\left(\frac{1}{2}\hf_{\alpha\mu;\tau}
+\frac{2\pi}{(n\!-\!1)}(\hj_{\alpha}g_{\mu\tau}-\hj_{\mu}g_{\alpha\tau})\right)\hf^\tau{_\nu}\nonumber\\
&&+\left(\frac{1}{2}\hf_{\nu\alpha;\tau}
+\frac{2\pi}{(n\!-\!1)}(\hj_{\nu}g_{\alpha\tau}-\hj_{\alpha}g_{\nu\tau})\right)\hf^\tau{_\mu}\nonumber\\
&&+\left(-\hf_{\tau(\mu;\nu)}
+\frac{2\pi}{(n\!-\!1)}(\hj_\mu g_{\nu\tau}+\hj_\nu g_{\mu\tau}-2 \hj_\tau g_{\mu\nu})\right)\hf^\tau{_\alpha}\nonumber\\
&&-\frac{1}{2}\left(\frac{-1}{2(n\!-\!2)}(\hf^{\rho\sigma}\!\hf_{\sigma\rho})_{,\alpha}
+\frac{8\pi}{(n\!-\!1)(n\!-\!2)}\hj^\tau \hf_{\tau\alpha}\right)g_{\nu\mu}\nonumber\\
&&+\left(\frac{-1}{2(n\!-\!2)}(\hf^{\rho\sigma}\!\hf_{\sigma\rho})_{,(\nu}
+\frac{8\pi}{(n\!-\!1)(n\!-\!2)}\hj^\tau \hf_{\tau(\nu}\right)g_{\mu)\alpha}+(\hf^{4\prime})\dots\nonumber\\
&=&
\fi
\label{upsilonsymmetriclowered2}
\hf^\tau{_{(\nu}}\hf_{\mu)}{_{\alpha;\tau}}
+\hf_\alpha{^\tau}\hf_{\tau(\nu;\mu)}
+\frac{1}{4(n\!-\!2)}\left((\hf^{\rho\sigma}\!\hf_{\sigma\rho})_{,\alpha}g_{\nu\mu}
\!-2(\hf^{\rho\sigma}\!\hf_{\sigma\rho})_{,(\nu}g_{\mu)\alpha}\right)\nonumber\\
&&+\frac{4\pi}{(n\!-\!2)}\hj^\tau\left(\hf_{\alpha\tau}g_{\nu\mu}
+\frac{2}{(n\!-\!1)}\hf_{\tau(\nu}g_{\mu)\alpha}\right)+(\hf^{4\prime})\dots.
\end{eqnarray}
Here the notation $(\hf^{4\prime})$ refers to terms like
$\hf_{\alpha\tau}\hf^\tau{_\sigma}\hf^\sigma{_\rho}\hf^\rho{_{(\nu;\mu)}}$.
Raising the indices on
(\ref{upsilonsymmetriclowered2},\ref{upsilonantisymmetriclowered},\ref{upsiloncontractedappendix})
and using (\ref{notation}) gives the final result
(\ref{upsilonsymmetric},\ref{upsilonantisymmetric},\ref{upsiloncontracted}).
\fi

\ifnum\ExpandDerivations=1
\section{\label{Bianchi}Derivation of the generalized contracted Bianchi identity}
Here we derive the generalized contracted Bianchi identity
(\ref{contractedBianchi}) from the connection equations
%in their contravariant density form
(\ref{JSconnection}), and from the
symmetry (\ref{JScontractionsymmetric}) of $\tGam^\alpha_{\nu\mu}$.
Whereas \cite{Antoci3} derived the identity
by performing an infinitesimal coordinate transformation on an invariant integral,
%although the full derivation was not shown.
we will instead use a direct method similar to \cite{EinsteinBianchi},
but generalized to include charge currents.  First we make the
following definitions,
\begin{eqnarray}
\label{Mdef}
\mathbf{W}^{\tau\rho}&=&\rmg\,W^{\tau\rho}=\rmN N^{\dashv\rho\tau}
=\rmg\,(g^{\tau\rho}+\hf^{\tau\rho}),\\
\label{notation0}
\hf^{\nu\mu}&=&f^{\nu\mu}\rmt\Lambda_b^{\!-1/2},~~~~\hbj^\alpha=\rmg j^\alpha\rmt\Lambda_b^{\!-1/2},\\
\label{HermitianizedRiemann}
\tR^\tau{_{\nu\alpha\mu}}&=&\tGam^\tau_{\nu\mu,\alpha}-\tGam^\tau_{\nu\alpha,\mu}
+\tGam^\sigma_{\nu\mu}\tGam^\tau_{\sigma\alpha}
-\tGam^\sigma_{\nu\alpha}\tGam^\tau_{\sigma\mu}
+\delta^\tau_\nu\tGam^\sigma_{\sigma[\alpha,\mu]},\\
\tR_{\nu\mu}=\tR^\alpha{_{\nu\alpha\mu}}
\label{HermitianizedRicci2}
&=&\tGam^\alpha_{\nu\mu,\alpha}-\tGam^\alpha_{\nu\alpha,\mu}
+\tGam^\sigma_{\nu\mu}\tGam^\alpha_{\sigma\alpha}
-\tGam^\sigma_{\nu\alpha}\tGam^\alpha_{\sigma\mu}
+\tGam^\sigma_{\sigma[\nu,\mu]}.
\end{eqnarray}
Here $\tR_{\nu\mu}$ is the Hermitianized Ricci tensor
(\ref{HermitianizedRiccit}), which has the property from (\ref{transpositionsymmetric}),
\begin{eqnarray}
\label{Hermiticity}
{\mathcal R}_{\nu\mu}(\tGam^T)=\tR_{\mu\nu}.
\end{eqnarray}
The tensors $\tR_{\nu\mu}$ and $\tR^\tau{_{\nu\alpha\mu}}$ reduce to the
ordinary Ricci and Riemann tensors for symmetric fields where
$\Gamma^\sigma_{\sigma[\nu,\mu]}\!=\!R^\sigma_{~\sigma\mu\nu}/2\!=\!0$.

Rewriting the
%contravariant density
connection equations (\ref{JSconnection})
in terms of the definitions above gives,
\begin{eqnarray}
\label{Lconnections}
0&=&\mathbf{W}^{\tau\rho}{_{,\lambda}}
+\tGam^\tau_{\sigma\lambda}\mathbf{W}^{\sigma\rho}
+\tGam^\rho_{\lambda\sigma}\mathbf{W}^{\tau\sigma}
-\tGam^\sigma_{\sigma\lambda}\mathbf{W}^{\tau\rho}
-\frac{4\pi}{(n\!-\!1)}(\hbj^{\rho}\delta^{\tau}_\lambda-\hbj^{\tau}\delta^{\rho}_\lambda).
\end{eqnarray}
Differentiating (\ref{Lconnections}), antisymmetrizing, and substituting
(\ref{Lconnections}) for $\mathbf{W}^{\tau\rho}{_{,\lambda}}$ gives,
\begin{eqnarray}
0&=&\!\left(\mathbf{W}^{\tau\rho}{_{,[\lambda}}
+\tGam^\tau_{\sigma[\lambda}\mathbf{W}^{\sigma\rho}
+\tGam^\rho_{[\lambda\vert\sigma}\mathbf{W}^{\tau\sigma}
-\tGam^\sigma_{\sigma[\lambda}\mathbf{W}^{\tau\rho}
-\frac{4\pi}{(n\!-\!1)}(\hbj^\rho\delta^{\tau}_{[\lambda}
-\hbj^\tau\delta^{\rho}_{[\lambda})\right){_{\!\!,\,\nu]}}\\
&=&\tGam^\tau_{\sigma[\lambda,\nu]}\mathbf{W}^{\sigma\rho}
+\tGam^\rho_{[\lambda\vert\sigma,\vert\nu]}\mathbf{W}^{\tau\sigma}
-\tGam^\sigma_{\sigma[\lambda,\nu]}\mathbf{W}^{\tau\rho}
-\frac{4\pi}{(n\!-\!1)}(\hbj^\rho{_{,[\nu}}\delta^\tau_{\lambda]}
-\hbj^\tau{_{,[\nu}}\delta^\rho_{\lambda]})\nonumber\\
&&+\tGam^\tau_{\sigma[\lambda}\mathbf{W}^{\sigma\rho}{_{,\nu]}}
+\tGam^\rho_{[\lambda\vert\sigma}\mathbf{W}^{\tau\sigma}{_{,\nu]}}
-\tGam^\sigma_{\sigma[\lambda}\mathbf{W}^{\tau\rho}{_{,\nu]}}\\
%\end{eqnarray}
%%\bigskip
%\begin{eqnarray}
\label{mess}
&=&\tGam^\tau_{\sigma[\lambda,\nu]}\mathbf{W}^{\sigma\rho}
+\tGam^\rho_{[\lambda\vert\sigma,\vert\nu]}\mathbf{W}^{\tau\sigma}
-\tGam^\sigma_{\sigma[\lambda,\nu]}\mathbf{W}^{\tau\rho}
-\frac{4\pi}{(n\!-\!1)}(\hbj^\rho{_{,[\nu}}\delta^\tau_{\lambda]}
-\hbj^\tau{_{,[\nu}}\delta^\rho_{\lambda]})\nonumber\\
&&-\tGam^\tau_{\sigma[\lambda}\left(
\tGam^\sigma_{\alpha\vert\nu]}\mathbf{W}^{\alpha\rho}
+\tGam^\rho_{\nu]\alpha}\mathbf{W}^{\sigma\alpha}
-\tGam^\alpha_{\nu]\alpha}\mathbf{W}^{\sigma\rho}
-\frac{4\pi}{(n\!-\!1)}(\hbj^\rho\delta^\sigma_{\nu]}
-\hbj^\sigma\delta^\rho_{\nu]})\right)\nonumber\\
&&-\tGam^\rho_{[\lambda\vert\sigma}\!\left(
\tGam^\tau_{\alpha\vert\nu]}\mathbf{W}^{\alpha\sigma}
+\tGam^\sigma_{\nu]\alpha}\mathbf{W}^{\tau\alpha}
-\tGam^\alpha_{\nu]\alpha}\mathbf{W}^{\tau\sigma}
-\frac{4\pi}{(n\!-\!1)}(\hbj^\sigma\delta^\tau_{\nu]}
-\hbj^\tau\delta^\sigma_{\nu]})\right)\nonumber\\
&&+\tGam^\sigma_{\sigma[\lambda}\left(
\tGam^\tau_{\alpha\vert\nu]}\mathbf{W}^{\alpha\rho}
+\tGam^\rho_{\nu]\alpha}\mathbf{W}^{\tau\alpha}
-\tGam^\alpha_{\nu]\alpha}\mathbf{W}^{\tau\rho}
-\frac{4\pi}{(n\!-\!1)}(\hbj^\rho\delta^\tau_{\nu]}
-\hbj^\tau\delta^\rho_{\nu]})\right).
\end{eqnarray}
Cancelling the terms 2B-3A, 2C-4A, 3C-4B and using (\ref{HermitianizedRiemann}) gives,
\begin{eqnarray}
\label{result}
0&=&\frac{1}{2}\left[\mathbf{W}^{\sigma\rho}\tR^\tau{_{\sigma\nu\lambda}}
+\mathbf{W}^{\tau\sigma}{\mathcal R}^\rho{_{\sigma\nu\lambda}}(\tGam^T)\right]
+\frac{4\pi}{(n\!-\!1)}\left[\tGam{^\tau_{\![\nu\lambda]}}\hbj^\rho
-\tGam{^\rho_{\![\lambda\nu]}}\hbj^\tau\right]\nonumber\\
&&+\frac{4\pi}{(n\!-\!1)}\left[(\hbj^\tau{_{,[\nu}}
\!+\!\tGam^\tau_{\sigma[\nu}\hbj^\sigma
\!-\!\tGam^\sigma_{\sigma[\nu}\hbj^\tau)\delta^\rho_{\lambda]}
-(\hbj^\rho{_{,[\nu}}
\!+\!\tGam^\rho_{\![\nu\vert\sigma}\hbj^\sigma
\!-\!\tGam^\sigma_{\sigma[\nu}\hbj^\rho)\delta^\tau_{\lambda]}\right].
\end{eqnarray}
%\bigskip\\
Multiplying by 2, contracting over $^\rho_\nu$, and using (\ref{Hermiticity})
and $\hbj^\nu_{,\nu}\!=\!0$ from (\ref{continuity}) gives,
\begin{eqnarray}
0&=&\mathbf{W}^{\sigma\nu}\tR^\tau{_{\sigma\nu\lambda}}
+\mathbf{W}^{\tau\sigma}{\mathcal R}^\nu{_{\sigma\nu\lambda}}(\tGam^T)
+\frac{8\pi}{(n\!-\!1)}\left[\tGam{^\tau_{\![\nu\lambda]}}\hbj^\nu
-\tGam{^\nu_{\![\lambda\nu]}}\hbj^\tau\right]\nonumber\\
&&+\frac{8\pi}{(n\!-\!1)}\left[(\hbj^\tau{_{,[\nu}}
\!+\!\tGam^\tau_{\sigma[\nu}\hbj^\sigma
\!-\!\tGam^\sigma_{\sigma[\nu}\hbj^\tau)\delta^\nu_{\lambda]}
-(\hbj^\nu{_{,[\nu}}
\!+\!\tGam^\nu_{\![\nu\vert\sigma}\hbj^\sigma
\!-\!\tGam^\sigma_{\sigma[\nu}\hbj^\nu)\delta^\tau_{\lambda]}\right]\\
\label{intermediate}
&=&\mathbf{W}^{\sigma\nu}\tR^\tau{_{\sigma\nu\lambda}}
+\mathbf{W}^{\tau\sigma}\tR_{\lambda\sigma}
-\frac{4\pi(n\!-\!2)}{(n\!-\!1)}(\hbj^\tau{_{,\lambda}}
\!+\!\tGam^\tau_{\sigma\lambda}\hbj^\sigma
\!-\!\tGam^\sigma_{\sigma\lambda}\hbj^\tau).
\end{eqnarray}
This is a generalization of the symmetry $R^\tau{_\lambda}=R_\lambda{^\tau}$
%that occurs with
of the ordinary Ricci tensor.

%\newpage
Next we will use the generalized uncontracted Bianchi identity\cite{EinsteinBianchi},
which can be verified by direct computation,
\begin{eqnarray}
\label{PBianchi}
  \tR{^{\Stacksymbols{+}{\tau}{0}{1}}}{_{\Stacksymbols{\sigma}{+}{6}{1}{\Stacksymbols{\nu}{-}{6}{1}}{\Stacksymbols{\alpha}{+}{6}{1}};\lambda}}
 +\tR{^{\Stacksymbols{+}{\tau}{0}{1}}}{_{\Stacksymbols{\sigma}{+}{6}{1}{\Stacksymbols{\alpha}{+}{6}{1}}{\Stacksymbols{\lambda}{+}{6}{1}};\nu}}
 +\tR{^{\Stacksymbols{+}{\tau}{0}{1}}}{_{\Stacksymbols{\sigma}{+}{6}{1}{\Stacksymbols{\lambda}{-}{6}{1}}{\Stacksymbols{\nu}{-}{6}{1}};\alpha}}
=0.
\end{eqnarray}
The $+/-$ notation is from \cite{EinsteinBianchi} and indicates that covariant
derivative is being done with $\tGam^\alpha_{\nu\mu}$ instead of the usual
$\Gamma^\alpha_{\nu\mu}$. A plus by
an index means that the associated derivative index is to be placed on the right side of the
connection, and a minus means that it is to be placed on the left side.
Note that the identity (\ref{PBianchi}) is true for either the ordinary Riemann tensor or
for our definition (\ref{HermitianizedRiemann}).
This is because the two tensors differ by the term $\delta^\tau_\nu\tGam^\sigma_{\sigma[\alpha,\mu]}$,
so that the expression (\ref{PBianchi}) would differ by the term
$\delta^\tau_\sigma(
 \tGam^\rho_\rho{_{[{\Stacksymbols{\nu}{-}{6}{1}},{\Stacksymbols{\alpha}{+}{6}{1}]};\lambda}}
+\tGam^\rho_\rho{_{[{\Stacksymbols{\alpha}{+}{6}{1}},{\Stacksymbols{\lambda}{+}{6}{1}]};\nu}}
+\tGam^\rho_\rho{_{[{\Stacksymbols{\lambda}{-}{6}{1}},{\Stacksymbols{\nu}{-}{6}{1}]};\alpha}})$.
But this difference vanishes because for an arbitrary curl $Y_{[\alpha,\lambda]}$ we have
\begin{eqnarray}
Y{_{[{\Stacksymbols{\nu}{-}{6}{1}},{\Stacksymbols{\alpha}{+}{6}{1}]};\lambda}}
   +Y{_{[{\Stacksymbols{\alpha}{+}{6}{1}},{\Stacksymbols{\lambda}{+}{6}{1}]};\nu}}
   +Y{_{[{\Stacksymbols{\lambda}{-}{6}{1}},{\Stacksymbols{\nu}{-}{6}{1}]};\alpha}}
&=&Y_{[\nu,\alpha],\lambda}
-\tGam^\sigma_{\lambda\nu}Y_{[\sigma,\alpha]}
-\tGam^\sigma_{\alpha\lambda}Y_{[\nu,\sigma]}\nonumber\\
&+&Y_{[\alpha,\lambda],\nu}
-\tGam^\sigma_{\alpha\nu}Y_{[\sigma,\lambda]}
-\tGam^\sigma_{\lambda\nu}Y_{[\alpha,\sigma]}\nonumber\\
&+&Y_{[\lambda,\nu],\alpha}
-\tGam^\sigma_{\alpha\lambda}Y_{[\sigma,\nu]}
-\tGam^\sigma_{\alpha\nu}Y_{[\lambda,\sigma]}=0.
\end{eqnarray}

A simple form of the generalized contracted Bianchi identity results if we contract (\ref{PBianchi})
over $\mathbf{W}^{\sigma\nu}$ and $^\tau_\alpha$, then substitute (\ref{intermediate}) for
$\mathbf{W}^{\sigma\nu}\tR^\tau{_{\sigma\nu\lambda}}$ and (\ref{Lconnections}) for
$\mathbf{W}^{\Stacksymbols{+}{\sigma}{0}{1}\Stacksymbols{-}{\nu}{0}{1}}{_{;\tau}}$,
\begin{eqnarray}
0&=&\mathbf{W}^{\sigma\nu}( \tR{^{\Stacksymbols{+}{\tau}{0}{1}}}{_{\Stacksymbols{\sigma}{+}{6}{1}{\Stacksymbols{\nu}{-}{6}{1}}{\Stacksymbols{\tau}{+}{6}{1}};\lambda}}
   +\tR{^{\Stacksymbols{+}{\tau}{0}{1}}}{_{\Stacksymbols{\sigma}{+}{6}{1}{\Stacksymbols{\tau}{+}{6}{1}}{\Stacksymbols{\lambda}{+}{6}{1}};\nu}}
   +\tR{^{\Stacksymbols{+}{\tau}{0}{1}}}{_{\Stacksymbols{\sigma}{+}{6}{1}{\Stacksymbols{\lambda}{-}{6}{1}}{\Stacksymbols{\nu}{-}{6}{1}};\tau}})\\
&=&-\mathbf{W}^{\sigma\nu}\tR{_{\Stacksymbols{\sigma}{+}{6}{1}{\Stacksymbols{\nu}{-}{6}{1}};\lambda}}
   +\mathbf{W}^{\sigma\nu}\tR{_{\Stacksymbols{\sigma}{+}{6}{1}{\Stacksymbols{\lambda}{+}{6}{1}};\nu}}
   -\mathbf{W}^{\sigma\nu}\tR{^{\Stacksymbols{+}{\tau}{0}{1}}}{_{\Stacksymbols{\sigma}{+}{6}{1}{\Stacksymbols{\nu}{-}{6}{1}}{\Stacksymbols{\lambda}{-}{6}{1}};\tau}}\\
&=&-\mathbf{W}^{\sigma\nu}\tR{_{\Stacksymbols{\sigma}{+}{6}{1}{\Stacksymbols{\nu}{-}{6}{1}};\lambda}}
   +(\mathbf{W}^{\sigma\Stacksymbols{-}{\nu}{0}{1}}\tR{_{\sigma{\Stacksymbols{\lambda}{+}{6}{1}}}}){_{;\nu}}
   -(\mathbf{W}^{\sigma\nu}\tR{^{\Stacksymbols{+}{\tau}{0}{1}}}{_{\sigma\nu{\Stacksymbols{\lambda}{-}{6}{1}}}}){_{;\tau}}\nonumber\\
&&~~~~~~~~~~~~~~~~~~~~~~
 -\mathbf{W}^{\Stacksymbols{+}{\sigma}{0}{1}\Stacksymbols{-}{\nu}{0}{1}}{_{;\nu}}\tR_{\sigma\lambda}
 +\mathbf{W}^{\Stacksymbols{+}{\sigma}{0}{1}\Stacksymbols{-}{\nu}{0}{1}}{_{;\tau}}\tR^\tau{_{\sigma\nu\lambda}}\\
&=&-\mathbf{W}^{\sigma\nu}\tR{_{\Stacksymbols{\sigma}{+}{6}{1}{\Stacksymbols{\nu}{-}{6}{1}};\lambda}}
 +(\mathbf{W}^{\sigma\Stacksymbols{-}{\nu}{0}{1}}\tR{_{\sigma{\Stacksymbols{\lambda}{+}{6}{1}}}}){_{;\nu}}\nonumber\\
&&+\left(\mathbf{W}^{\Stacksymbols{+}{\tau}{0}{1}\sigma}\tR_{\Stacksymbols{\lambda}{-}{6}{1}\sigma}
-\frac{4\pi(n\!-\!2)}{(n\!-\!1)}(\hbj^{\Stacksymbols{+}{\tau}{0}{1}}{_{,\Stacksymbols{\lambda}{-}{6}{1}}}
\!+\!\tGam^{\Stacksymbols{+}{\tau}{0}{1}}_{\sigma\Stacksymbols{\lambda}{-}{6}{1}}\hbj^\sigma
\!-\!\tGam^\sigma_{\sigma\Stacksymbols{\lambda}{-}{6}{1}}\hbj^{\Stacksymbols{+}{\tau}{0}{1}})\right){_{\!;\,\tau}}\nonumber\\
&&-\frac{4\pi}{(n\!-\!1)}(\hbj^{\nu}\delta^{\sigma}_\nu-\hbj^{\sigma}\delta^{\nu}_\nu)\tR_{\sigma\lambda}
 +\frac{4\pi}{(n\!-\!1)}(\hbj^{\nu}\delta^{\sigma}_\tau-\hbj^{\sigma}\delta^{\nu}_\tau)\tR^\tau{_{\sigma\nu\lambda}}\\
%\end{eqnarray}
%\begin{eqnarray}
&=&-\mathbf{W}^{\sigma\nu}\tR{_{\Stacksymbols{\sigma}{+}{6}{1}{\Stacksymbols{\nu}{-}{6}{1}};\lambda}}
 +(\mathbf{W}^{\sigma\Stacksymbols{-}{\nu}{0}{1}}\tR{_{\sigma{\Stacksymbols{\lambda}{+}{6}{1}}}}){_{;\nu}}
+(\mathbf{W}^{\Stacksymbols{+}{\nu}{0}{1}\sigma}\tR_{\Stacksymbols{\lambda}{-}{6}{1}\sigma}){_{;\nu}}\nonumber\\
&&-\frac{4\pi(n\!-\!2)}{(n\!-\!1)}(\hbj^{\Stacksymbols{+}{\tau}{0}{1}}{_{,\Stacksymbols{\lambda}{-}{6}{1}}}
\!+\!\tGam^{\Stacksymbols{+}{\tau}{0}{1}}_{\sigma\Stacksymbols{\lambda}{-}{6}{1}}\hbj^\sigma
\!-\!\tGam^\sigma_{\sigma\Stacksymbols{\lambda}{-}{6}{1}}\hbj^{\Stacksymbols{+}{\tau}{0}{1}}){_{;\tau}}\nonumber\\
&&+\frac{4\pi(n\!-\!2)}{(n\!-\!1)}\hbj^{\sigma}\tR_{\sigma\lambda}
 +\frac{4\pi}{(n\!-\!1)}\hbj^{\nu}\tR^\sigma{_{\sigma\nu\lambda}}
\end{eqnarray}
\begin{eqnarray}
\label{manyterms}
&=&-\mathbf{W}^{\sigma\nu}(\tR{_{\sigma\nu,\lambda}}
-\tGam^\alpha_{\sigma\lambda}\tR{_{\alpha\nu}}
-\tGam^\alpha_{\lambda\nu}\tR{_{\sigma\alpha}})\nonumber\\
&&+(\mathbf{W}^{\sigma\nu}\tR{_{\sigma\lambda}}){_{,\nu}}
+\tGam^\nu_{\nu\alpha}\mathbf{W}^{\sigma\alpha}\tR{_{\sigma\lambda}}
-\tGam^\alpha_{\lambda\nu}\mathbf{W}^{\sigma\nu}\tR{_{\sigma\alpha}}
-\tGam^\alpha_{\alpha\nu}\mathbf{W}^{\sigma\nu}\tR{_{\sigma\lambda}}\nonumber\\
&&+(\mathbf{W}^{\nu\sigma}\tR_{\lambda\sigma}){_{,\nu}}
+\tGam^\nu_{\alpha\nu}\mathbf{W}^{\alpha\sigma}\tR_{\lambda\sigma}
-\tGam^\alpha_{\nu\lambda}\mathbf{W}^{\nu\sigma}\tR_{\alpha\sigma}
-\tGam^\alpha_{\alpha\nu}\mathbf{W}^{\nu\sigma}\tR_{\lambda\sigma}\nonumber\\
 && -\frac{4\pi(n\!-\!2)}{(n\!-\!1)}[\hbj^\tau{_{,\lambda,\tau}}
\!+\!\tGam^\tau_{\sigma\lambda,\tau}\hbj^\sigma
\!+\!\tGam^\tau_{\sigma\lambda}\hbj^\sigma_{,\tau}
\!-\!\tGam^\sigma_{\sigma\lambda,\tau}\hbj^\tau
\!-\!\tGam^\sigma_{\sigma\lambda}\hbj^\tau_{,\tau}\nonumber\\
&&~~~~~~~~~~~~~~~~~
+\tGam^\tau_{\alpha\tau}(\hbj^\alpha{_{,\lambda}}
\!+\!\tGam^\alpha_{\sigma\lambda}\hbj^\sigma
\!-\!\tGam^\sigma_{\sigma\lambda}\hbj^\alpha)\nonumber\\
&&~~~~~~~~~~~~~~~~~
-\tGam^\alpha_{\tau\lambda}(\hbj^\tau{_{,\alpha}}
\!+\!\tGam^\tau_{\sigma\alpha}\hbj^\sigma
\!-\!\tGam^\sigma_{\sigma\alpha}\hbj^\tau)\nonumber\\
&&~~~~~~~~~~~~~~~~~
-\tGam^\alpha_{\alpha\tau}(\hbj^\tau{_{,\lambda}}
\!+\!\tGam^\tau_{\sigma\lambda}\hbj^\sigma
\!-\!\tGam^\sigma_{\sigma\lambda}\hbj^\tau)\nonumber\\
&&~~~~~~~~~~~~~~~~~
-\hbj^{\sigma}(\tR_{\sigma\lambda}-\tGam^\alpha_{\alpha[\sigma,\lambda]})
 -\hbj^{\sigma}(\tGam^\alpha_{\alpha\sigma,\lambda}\!-\!\tGam^\alpha_{\alpha\lambda,\sigma})]
\end{eqnarray}
With the $\hbj^\sigma$ terms of (\ref{manyterms}), 4C-6A,4D-8D,5A-7A,5B-7B,5C-7C all cancel,
4A and 4E are zero because $\hbj^\nu_{,\nu}\!=\!0$ from (\ref{continuity}),
and 4B,6B,6C,8C cancel the Ricci tensor term 8A,8B.
With the $\mathbf{W}^{\tau\sigma}$ terms of (\ref{manyterms}), all those
with a $\tGam^\alpha_{\nu\mu}$ factor cancel, which are the terms
1C-2C,1B-3C,2B-2D,3B-3D. Doing the cancellations and using (\ref{Mdef}) we get
\begin{eqnarray}
\label{simpleBianchi}
0&=&(\rmN N^{\dashv\nu\sigma}\tR{_{\sigma\lambda}}
+\rmN N^{\dashv\sigma\nu}\tR_{\lambda\sigma}){_{,\nu}}
-\rmN N^{\dashv\nu\sigma}\tR{_{\sigma\nu,\lambda}}.
\end{eqnarray}
Equation (\ref{simpleBianchi}) is a simple generalization of the ordinary contracted Bianchi
identity $2(\rmg\,R^\nu{_\lambda})_{,\nu}\!-\!\rmg\,g^{\nu\sigma}\!R_{\sigma\nu,\lambda}\!=\!0$,
and it applies even when $j^\tau\!\ne0$.
Because $\tGam^\alpha_{\nu\mu}$ has cancelled out of (\ref{simpleBianchi}),
the Christoffel connection $\Gamma^\alpha_{\nu\mu}$ would also cancel, so
a manifestly tensor
relation can be obtained by replacing the ordinary derivatives with covariant
derivatives done with $\Gamma^\alpha_{\nu\mu}$,
\begin{eqnarray}
\label{simpleBianchi2}
0&=&(\rmN N^{\dashv\nu\sigma}\tR{_{\sigma\lambda}}
+\rmN N^{\dashv\sigma\nu}\tR_{\lambda\sigma}){_{;\nu}}
-\rmN N^{\dashv\nu\sigma}\tR{_{\sigma\nu;\lambda}}.
\end{eqnarray}
%\bigskip\\
Rewriting the identity in terms of $g^{\rho\tau}$ and $\hf^{\rho\tau}$
as defined by (\ref{Mdef},\ref{notation0}) gives,
\begin{eqnarray}
0&=&(\rmg\,(g^{\sigma\nu}\!+\!\hf^{\sigma\nu})\tR{_{\sigma\lambda}}
\!+\rmg\,(g^{\nu\sigma}\!+\!\hf^{\nu\sigma})\tR_{\lambda\sigma}){_{;\nu}}
\!-\!\rmg\,(g^{\sigma\nu}\!+\!\hf^{\sigma\nu})\tR{_{\sigma\nu;\lambda}}\\
%&+&(\rmg\,f^{\sigma\nu}\tR{_{\sigma\lambda}}
%\!+\rmg\,f^{\nu\sigma}\tR_{\lambda\sigma}){_{;\nu}}\rmt\Lambda_b^{\!-1/2}
%\!-\!\rmg\,f^{\sigma\nu}\tR{_{\sigma\nu;\lambda}}\rmt\Lambda_b^{\!-1/2}\\
&=&\rmg\,[2\tR^{(\nu}{_{\lambda);\nu}}
-\tR^\sigma_{\sigma;\lambda}]
+\rmg\,[2(\hf^{\nu\sigma}\tR{_{[\lambda\sigma]}}){_{;\nu}}
\!+\!\hf^{\nu\sigma}\tR{_{[\sigma\nu];\lambda}}]\\
&=&\rmg\,[2\tR^{(\nu}{_{\lambda)}}{_{;\nu}}
-\tR^\sigma_{\sigma;\lambda}]
+\rmg\,[3\hf^{\nu\sigma}\tR{_{[\sigma\nu,\lambda]}}
\!+2\hf^{\nu\sigma}{_{;\nu}}\tR{_{[\lambda\sigma]}}].
\end{eqnarray}
Dividing by $2\rmg$ gives another form of the generalized contracted Bianchi identity
\begin{eqnarray}
\left(\tR^{(\nu}{_{\lambda)}}
-\frac{1}{2}\delta^\nu_\lambda\tR^\sigma_\sigma\right){_{\!;\,\nu}}
=\frac{3}{2}\hf^{\nu\sigma}\tR{_{[\nu\sigma,\lambda]}}
+\hf^{\nu\sigma}{_{\!;\nu}}\tR{_{[\sigma\lambda]}}.
\end{eqnarray}
From (\ref{notation0},\ref{genEinstein}) we get the final result (\ref{contractedBianchi}).
\fi
\end{appendix}

%\newpage
%\bibliography{npshifflett}% Produces the bibliography via BibTeX.
%\section*{References}
%\begin{thebibliography}{68}



\end{document}